\documentclass[aps,floats]{revtex4-2}
\usepackage{amsmath,amssymb}
\usepackage{graphicx,epsfig}
\usepackage[greek,english]{babel}
\usepackage{bbold}

\begin{document}
\bibliographystyle {plain}

\pdfoutput=1
\def\oppropto{\mathop{\propto}} 
\def\opsimeq{\mathop{\simeq}}
\def\opoverderline{\mathop{\overline}}
\def\operarrow{\mathop{\longrightarrow}}
\def\opsim{\mathop{\sim}}

\def\opmin{\mathop{\min}} 
\def\opmax{\mathop{\max}} 
\def\oplim{\mathop{\lim}}

%%%%%%%%%%%%%%%%%%%%%%%%%%%%%%%%%%%%%%%%%%%%%%%%%%%%%%%%%%%%%%%%%%%%%%%%%%%%
\title{Explicit dynamical properties of the Pelikan random map in the chaotic region \\
 and at the intermittent critical point towards the non-chaotic region  } 

%%%%%%%%%%%%%%%%%%%%%%%%%%%%%%%%%%%%%%%%%%%%%%%%%%%%%%%%%%%%%%%%%%%%%%%%%%%%

\author{C\'ecile Monthus}
\affiliation{Universit\'e Paris-Saclay, CNRS, CEA, Institut de Physique Th\'eorique, 91191 Gif-sur-Yvette, France}

%%%%%%%%%%%%%%%%%%%%%%%%%%%%%%%%%%%%%%%%%%%%%%%%%%%%%%%%%%%%%%%%%%%%%%%%%%%%

\begin{abstract}
The Pelikan random trajectories $x_t \in [0,1[$ are generated by choosing the chaotic doubling map $x_{t+1}=2 x_t  [\mathrm {mod} 1]$ with probability $p$ and the non-chaotic half-contracting map $x_{t+1}=\frac{x_t}{2}$ with probability $(1-p)$. We compute various dynamical observables as a function of the parameter $p$ via two perspectives. In the first perspective, we focus on the closed dynamics within the subspace of probability densities that remain constant on the binary-intervals $x \in [ 2^{-n-1}, 2^{-n}[$ partitioning the interval $x \in [0,1[$: the dynamics for the weights $\pi_t(n)$ of these intervals corresponds to a biased random walk on the half-infinite lattice $n \in \{0,1,2,..+\infty\}$ with resetting occurring with probability $p$ from the origin $n=0$ towards any site $n$ drawn with the distribution $2^{-n-1}$. In the second perspective, we study the Pelikan dynamics for any initial condition $x_0$ via the binary decomposition $x_t = \sum_{l=1}^{+\infty} \frac{\sigma_l (t)}{2^l} $, where the dynamics for the half-infinite lattice $l=1,2,..$ of the binary variables $\sigma_l(t) \in \{0,1\}$ can be rephrased in terms of two global variables: $z_t$ corresponds to a biased random walk on the half-infinite lattice $z \in \{0,1,2,... +\infty\}$ that may remain at the origin $z=0$ with probability $p$, while $F_t \in \{0,1,2,...,t\}$ counts the number of time-steps $\tau \in [0,t-1]$ where $z_{\tau+1}=0=z_{\tau}$ and represents the number of the binary coefficients of the initial condition that have been erased. We discuss typical and large deviations properties in the chaotic region $\frac{1}{2}<p<1 $ as well as at the intermittent critical point $p_c=\frac{1}{2}$ towards the non-chaotic region $0<p<\frac{1}{2}$.

\end{abstract}

\maketitle

%%%%%%%%%%%%%%%%%%%%%%%%%%%%%%%%%%%%%%%%%%

\section{ Introduction}

The area of random dynamical systems \cite{kifer,kapitaniak,arnold} 
is at the interface between the field of deterministic dynamical systems \cite{ott,beck,schuster,dorfman}
and the field of random processes \cite{gardiner,vankampen,risken}.
One of the simplest example is the Pelikan random map on the interval $x \in [0,1[$
 that has been much studied since its introduction by Pelikan in 1984 \cite{Pelikan}  
 (see the recent study \cite{Ruffo} with references therein
 for a broader perspective and detailed discussions on the relations with other studies):
at each time-step, one chooses either the chaotic doubling map $x_{t+1}=2 x_t  [\mathrm {mod} 1]$ with probability $p$ or the non-chaotic half-contracting map $x_{t+1}=\frac{x_t}{2}$ with probability $(1-p)$. Besides the steady state that is explicitly known as a function of the parameter $p$ in the whole chaotic region $\frac{1}{2}<p<1 $ \cite{Pelikan,Ruffo}, it is interesting to study the convergence properties towards this steady state, and to better understand what happens at the intermittent critical point $p_c=\frac{1}{2}$ towards the non-chaotic region $0<p<\frac{1}{2}$,
as stressed in the recent study \cite{Ruffo} that has motivated the present work.
Our goal will be to compute explicit results for various dynamical observables of the Pelikan map,
without recalling the broader mathematical framework needed to rigorously analyze random dynamical systems
that can be found in the Pelikan paper \cite{Pelikan}.

Let us stress that we will use two different perspectives: 
 
$\bullet$ In the first perspective, we focus on the closed dynamics within the subspace of probability densities that remain constant on the binary-intervals $x \in [ 2^{-n-1}, 2^{-n}[$ partitioning the interval $x \in [0,1[$. The dynamics for the weights $\pi_t(n)$ of these intervals corresponds to a biased random walk on the half-infinite lattice $n \in \{0,1,2,..+\infty\}$ with resetting occurring with probability $p$ from the origin $n=0$ towards any site $n$ drawn with the distribution $2^{-n-1}$. This Markov chain on the half-infinite lattice $n \in \{0,1,2,..+\infty\}$ thus belongs to the broad field of stochastic processes with resetting that have attracted a lot of interest recently (see the recent reviews \cite{review_reset,reset_india} as well as the recent special issue \cite{reset_special}
with references therein), either in relation with intermittent search strategies (see the review \cite{review_search} and references therein) or in relation with the field of non-equilibrium statistical physics, where the language of large deviations \cite{oono,ellis,review_touchette} has been essential to achieve a unified picture (see the reviews with different scopes \cite{derrida-lecture,harris_Schu,searles,harris,mft,sollich_review,lazarescu_companion,lazarescu_generic,jack_review}, the PhD Theses \cite{fortelle_thesis,vivien_thesis,chetrite_thesis,wynants_thesis} 
 and the HDR Thesis \cite{chetrite_HDR}).

$\bullet$ In the second perspective, we study the Pelikan dynamics for any initial condition $x_0$ via the decomposition $x_t = \sum_{l=1}^{+\infty} \frac{\sigma_l (t)}{2^l} $ in terms of the binary variables $\sigma_l(t) \in \{0,1\}$.
The dynamics for the half-infinite lattice $l=1,2,..$ of these binary variables $\sigma_l(t) $ will be rephrased in terms of two global variables: $z_t$ is a biased random walk on the half-infinite lattice $z \in \{0,1,2,... +\infty\}$ that may remain at the origin $z=0$ with probability $p$, while $F_t \in \{0,1,2,...,t\}$ counts the number of time-steps $\tau \in [0,t-1]$ where $z_{\tau+1}=0=z_{\tau}$ and represents the number of the binary coefficients of the initial condition that have been erased.  At the technical level, $F_t$ is thus an additive observable 
of the Markov trajectory $z_{0 \leq \tau \leq t}$. 
Such additive observables of Markov processes have been much studied recently via the appropriate deformations of the Markov generators
 \cite{peliti,derrida-lecture,sollich_review,lazarescu_companion,lazarescu_generic,jack_review,vivien_thesis,lecomte_chaotic,lecomte_thermo,lecomte_formalism,lecomte_glass,kristina1,kristina2,jack_ensemble,simon1,simon2,tailleur,simon3,Gunter1,Gunter2,Gunter3,Gunter4,chetrite_canonical,chetrite_conditioned,chetrite_optimal,chetrite_HDR,touchette_circle,touchette_langevin,touchette_occ,touchette_occupation,garrahan_lecture,Vivo,c_ring,c_detailed,chemical,derrida-conditioned,derrida-ring,bertin-conditioned,touchette-reflected,touchette-reflectedbis,c_lyapunov,previousquantum2.5doob,quantum2.5doob,quantum2.5dooblong,c_ruelle,lapolla,c_east,chabane,us_gyrator,duBuisson_gyrator,c_largedevpearson} or via the contraction principle from higher levels of large deviations that can be written explicitly for the various types of Markov processes,  including discrete-time Markov chains 
\cite{fortelle_thesis,fortelle_chain,review_touchette,c_largedevdisorder,c_reset,c_inference,c_microcanoEnsembles,c_diffReg},
continuous-time Markov jump processes
\cite{fortelle_thesis,fortelle_jump,maes_canonical,maes_onandbeyond,wynants_thesis,chetrite_formal,BFG1,BFG2,chetrite_HDR,c_ring,c_interactions,c_open,barato_periodic,chetrite_periodic,c_reset,c_inference,c_LargeDevAbsorbing,c_microcanoEnsembles,c_susyboundarydriven,c_diffReg,c_inverse},
diffusion processes
\cite{wynants_thesis,maes_diffusion,chetrite_formal,engel,chetrite_HDR,c_lyapunov,c_inference,c_susyboundarydriven,c_diffReg,c_inverse}, 
and jump-diffusion or jump-drift processes \cite{c_reset,c_runandtumble,c_jumpdrift,c_SkewDB}.
The same large deviations framework 
has been applied recently to various deterministic dynamical systems,
 either in continuous time \cite{tailleur_thesis,tailleur_Lyapunov,laffargue}
 or in discrete time for chaotic non-invertible maps \cite{naftali,spain,c_chaos}, including in particular the doubling map
 that enters the definition of the Pelikan map.

The paper is organized as follows.
After recalling the main properties of the Pelikan random map in section \ref{sec_general}, 
we first focus on the closed dynamics within
the subspace of probability densities that remain constant
on the binary-intervals $x \in [ 2^{-n-1}, 2^{-n}[$ partitioning the interval $x \in [0,1[$
in order to describe various properties in sections
\ref{sec_markovBinary}, \ref{sec_propagator} and \ref{sec_excursions}.
We then study the Pelikan map for any initial condition $x_0$ 
via the binary decomposition $x_t = \sum_{l=1}^{+\infty} \frac{\sigma_l (t)}{2^l} $
in section \ref{sec_spins}.
Our conclusions are summarized in section \ref{sec_conclusion},
while three appendices contain more detailed computations.

%%%%%%%%%%%%%%%%%%%%%%%%%%%%%%%%%%%%%%%%%%

\section{ Reminder on the Pelikan random map on the interval $x \in [0,1[$}

\label{sec_general}

In this section, we recall the basic properties of the Pelikan random map 
(see the recent study \cite{Ruffo} and references therein for more details and discussions)
and we introduce the notations that will be useful in the whole paper.

\subsection{ Pelikan random trajectories $x_{t=0,1,2,..}$ on the interval $x \in [0,1[$ }

The Pelikan random map \cite{Pelikan} of parameter $p \in ]0,1[$ defined on the interval $ [0,1[$ 
\begin{eqnarray}
x_{t+1} =  f_{\mathrm {Pelikan}}(x_t) \equiv  
\begin{cases}
f_2(x_t) \equiv 2 x_t  [\text{mod} 1] \ \ \ \ \ \ \ \ \ \text{ with probability $p$ }  
 \\
f_{1/2}(x_t) \equiv \frac{1}{2} x_t  \ \ \ \ \ \ \ \ \ \  \ \ \ \ \text{ with probability $(1-p)$ }
\end{cases}
\label{pelikan}
\end{eqnarray}
involves a random mixture of two deterministic maps with the following properties:

(i) The doubling map $f_2$ 
\begin{eqnarray}
x_{t+1} = f_2(x_t) \equiv 2 x_t  [\text{mod} 1] \equiv  
\begin{cases}
2 x_t \ \ \ \ \ \ \ \ \ \text{  for $0 \leq x_t < \frac{1}{2}$}  
 \\
(2x_t-1) \ \  \text{ for $\frac{1}{2} \leq x_t < 1$}
\end{cases}
\label{doubling}
\end{eqnarray}
is the simplest example of chaotic dynamics,  characterized by the uniform positive Lyapunov exponent 
\begin{eqnarray}
\lambda^{[f_2]} = \ln \left\vert \frac{ df_2(x)}{dx} \right\vert =\ln 2  >0
\label{Lyapunovf2}
\end{eqnarray}
 responsible for the exponential divergence in time of the separation between two nearby trajectories,
while the non-invertibility of the doubling map,
where each value $x_{t+1} \in [0,1[$ has two pre-images 
\begin{eqnarray}
x_t^- && = \frac{x_{t+1}}{2} \in \left[0,\frac{1}{2}\right[
\nonumber \\
x_t^+ && = \left( \frac{1}{2}+\frac{x_{t+1}}{2}\right) \in \left[\frac{1}{2},1 \right[
\label{twopreimages}
\end{eqnarray}
yields that some information about the past is lost at each time-step.
Any smooth initial density $\rho_{t=0}(x_0)$ will converge via the doubling map $f_2$ towards the uniform invariant density
\begin{eqnarray}
 \rho^{[f_2]}_*(  x) =1
\label{perronFrobsteadydoubling}
\end{eqnarray}
The explicit spectral properties governing the convergence towards this uniform steady state
are analyzed in \cite{gaspard,linas,Hu22} and references therein.

(ii) The half-contracting map $f_{1/2}$ 
\begin{eqnarray}
x_{t+1} = f_{1/2}(x_t) \equiv \frac{1}{2} x_t 
\label{half-contracting}
\end{eqnarray}
is characterized by the uniform negative Lyapunov exponent 
\begin{eqnarray}
\lambda^{[f_{1/2}]} = \ln \left\vert \frac{ df_{1/2}(x)}{dx} \right\vert = \ln \left(  \frac{1}{2}\right) = -\ln 2  <0
\label{Lyapunovf1sur2}
\end{eqnarray}
responsible for the exponential convergence in time of the separation between two nearby trajectories,
and any initial condition $x_0$ converges exponentially in time as $ \frac{x_0}{2^t}$
towards the stable fixed-point $x=0$,
so that the invariant density is the singular delta function at the origin
\begin{eqnarray}
 \rho^{[f_{1/2}]}_*(  x) = \delta(x)
\label{perronFrobsteadycontracting}
\end{eqnarray}

%%%%%%%%%%%%%%%%%%%%%%%%%%%%%

\subsection{ Distribution of the finite-time Lyapunov exponent on the time-window $[0,T]$ }

Among the $T$ steps of the time-window $[0,T]$, if there are $T_+ \in \{0,1,..,T\}$ steps where the doubling map $f_2$ is chosen, and $T_-=T-T_+$ where the half-contracting map $f_{1/2}$ is chosen,
then the time-averaged Lyapunov exponent is
\begin{eqnarray}
\lambda_T = \frac{ T_+\lambda^{[f_2]}   + T_- \lambda^{[f_{1/2}]}}{T} 
= \frac{ (2T_+ -T) }{T} \ln 2
\label{FiniteTimeLyapunov}
\end{eqnarray}
As a consequence, the probability distribution of $\lambda_T$ is governed by the following binomial distribution 
\begin{eqnarray}
B_T(T_+)  =  \frac{ T!}{ T_+ ! (T-T_+)!} p^{T_+} (1-p)^{T-T_+} \ \ \ \text{ for } \ \ T_+\in \{0,1,..,T\}
\label{Binomial}
\end{eqnarray}
For large time $T \to + \infty$, the Stirling formula yields that
the empirical fraction $p_+ \equiv \frac{T_+}{T}$ of the steps where the doubling map is chosen
displays the large deviation form
 \begin{eqnarray}
B_T(T_+=T p_+)  =  \frac{ T!}{ (T p_+) ! (T[1-p_+])!} p^{T p_+} (1-p)^{T(1-p_+)} 
\opsimeq_{T \to + \infty} e^{ - T \left[ p_+ \ln \left(  \frac{p_+}{p}\right) + (1-p_+) \ln \left(  \frac{1-p_+}{1-p}\right)\right] }
\label{Binomiallargedev}
\end{eqnarray}
around the typical value $p_+^{\mathrm {typ}}=p$ which is the only value where the rate function in the exponential vanishes.

As a consequence, the large deviations properties of the time-averaged Lyapunov exponent 
 are governed by Eq. \ref{Binomiallargedev} 
after the trivial change of variables $\lambda_T = (2 p_+ -1) \ln 2 $ of Eq. \ref{FiniteTimeLyapunov}.
The change of sign of the typical value $\lambda^{\mathrm {typ}}_T  =(2 p_+^{\mathrm {typ}} -1) \ln 2 =(2p-1) \ln 2$  
that is the only value that survives in the limit $T \to + \infty$ determines the critical point $p_c=\frac{1}{2}$
between the chaotic and the non-chaotic regions
(see the recent study \cite{Ruffo} and references therein for more details)
\begin{eqnarray}
 \lambda_T^{\mathrm {typ}}  =  (2p-1) \ln 2
 \begin{cases}
>0 \ \  \text{ in the chaotic region } \frac{1}{2} < p \leq 1 
 \\
 =0 \ \ \text{ at the intermittent critical point } p_c = \frac{1}{2} 
  \\
 <0 \ \ \text{ in the non-chaotic region } 0 \leq p < \frac{1}{2} 
\end{cases}
\label{LyapunovTyp}
\end{eqnarray}

%%%%%%%%%%%%%%%%%%%%%%%%%%%%%%%%

\subsection{ Dynamics for the
probability density $\rho_t(  x)  $ on the real space interval $x \in [0,1[$ }

The forward kernel $w({\tilde x} \vert x) $
associated to the Pelikan map of Eq. \ref{pelikan} 
\begin{eqnarray}
w({\tilde x} \vert x)  
\equiv \delta({\tilde x}-f_{\mathrm {Pelikan}}(x) )
&& =
p \left[  \delta \left({\tilde x} - 2x \right) \theta \left( 0 \leq x < \frac{1}{2}\right) 
+\delta \left({\tilde x} - (2x-1) \right) \theta \left( \frac{1}{2} \leq x < 1 \right) \right]
+ (1-p)  \delta \left({\tilde x} - \frac{x}{2} \right)
\nonumber \\
&& 
= p \left[  \frac{ \delta(x- \frac{ {\tilde x}}{2} ) + \delta(x-\frac{ {\tilde x}+1}{2} ) }{2 } \right]
+2 (1-p) \delta \left(x- 2{\tilde x}  \right)  \theta \left( 0 \leq {\tilde x} < \frac{1}{2} \right)
\label{Wpelikan}
\end{eqnarray}
governs the Frobenius-Perron dynamics 
of the probability density $\rho_t(\cdot)  $
\begin{eqnarray}
  \rho_t({\tilde x}) = \int_0^1 dx w({\tilde x} \vert x) \rho_{t-1}(x)
= p \frac{  \rho_{t-1}( \frac{ {\tilde x}}{2} ) +  \rho_{t-1}( \frac{ {\tilde x}+1}{2} ) }{2 }
+2 (1-p) \rho_{t-1} ( 2{\tilde x}  )   \theta \left( 0 \leq {\tilde x} < \frac{1}{2} \right)
\label{FP}
\end{eqnarray}

As expected from the discussion of Eq. \ref{LyapunovTyp} concerning the Lyapunov exponent, 
the dynamics of Eq. \ref{FP}
for the probability density $\rho_t(\cdot)  $ will converge towards an invariant density $\rho_*(\cdot)$ defined on the whole interval $x \in [0,1[$ only in the chaotic region $\frac{1}{2} < p \leq 1$, while it will be attracted  
towards the singular delta function at the origin of Eq. \ref{perronFrobsteadycontracting}
in the whole non-chaotic region $0 \leq p < \frac{1}{2}$ 
(see the recent study \cite{Ruffo} and references therein for more details)
\begin{eqnarray}
\rho_t(  x)  \opsimeq_{t \to + \infty} 
 \begin{cases}
\rho_*(x) \ \ \  \text{ in the chaotic region } \frac{1}{2} < p \leq 1 
  \\
\delta(x) \ \ \text{ in the non-chaotic region } 0 \leq p < \frac{1}{2} 
\end{cases}
\label{steadyordelta}
\end{eqnarray}
The invariant density $\rho_*(\cdot)$ in the chaotic region $\frac{1}{2} < p \leq 1$ 
is explicit \cite{Pelikan,Ruffo}
and will be recalled in Eq. \ref{BinaryDensitysteady} of the next section, 
where the dynamics of Eq. \ref{FP} 
within a simple subspace of probability densities is mapped onto a Markov chain on the semi-infinite lattice.

%%%%%%%%%%%%%%%%%%%%%%%%%%%%%%%

\section{ Markov chain dynamics in the simplest subspace of probability densities }

\label{sec_markovBinary}

In the present section, as well as in the two next sections \ref{sec_propagator} and \ref{sec_excursions},
we focus on the subspace of probability densities $\rho_t(x)$ that remain constant
on the binary-intervals $x \in [ 2^{-n-1}, 2^{-n}[$ partitioning the interval $x \in [0,1[$
\begin{eqnarray}
  \rho^{[\mathrm {Partition}]}_t(x) && = \sum_{n=0}^{+\infty} 2^{n+1} \pi_t(n) \theta \left( 2^{-n-1} \leq x < 2^{-n} \right)
\nonumber \\
&& =  2 \pi_t(0) \theta \left( \frac{1}{2} \leq x < 1 \right)
+4 \pi_t(1) \theta \left( \frac{1}{4} \leq x < \frac{1}{2} \right)
+8 \pi_t(2) \theta \left( \frac{1}{8} \leq x < \frac{1}{4} \right)
+...
\label{BinaryDensity}
\end{eqnarray}
where the weights $\pi_t(n) \in [0,1]$ of the binary-intervals $x \in [ 2^{-n-1}, 2^{-n}[$ of widths $2^{-n-1}$
\begin{eqnarray}
  \pi_t(n) \equiv \int_{2^{-n-1} }^{2^{-n}} dx  \rho^{[\mathrm {Partition}]}_t(x)  
  \label{defpi}
\end{eqnarray}
satisfy the normalisation
\begin{eqnarray}
 1  = \int_0^1 dx  \rho^{[\mathrm {Partition}]}_t(x)  = \sum_{n=0}^{+\infty} \pi_t(n) 
\label{normapi}
\end{eqnarray}

%%%%%%%%%%%%%%%%%%%%%%%%%%%%%%%%%%%%%%%%%

\subsection{ Closed dynamics for the weights $\pi_t(n)$ of the binary intervals $x \in [ 2^{-n-1}, 2^{-n}[$ }

Plugging the form $ \rho^{[\mathrm {Partition}]}_t(x)  $ of Eq. \ref{BinaryDensity} into the dynamics of Eq. \ref{FP} yields
 the following closed dynamics for the weights $\pi_t(\cdot)$
of the binary intervals
\begin{eqnarray}
 \pi_t({\tilde n})  && = p   \pi_{t-1}({\tilde n}+1) + (1-p)   \pi_{t-1}({\tilde n}-1)  \theta( {\tilde n} \geq 1 )
 + p 2^{-{\tilde n}-1} \pi_{t-1}(0) 
 \nonumber \\
 && \equiv \sum_{n=0}^{+\infty} M({\tilde n} \vert n) \pi_{t-1}(n)
\label{FPan}
\end{eqnarray}
where the Markov matrix $M({\tilde n} \vert n) $ defined on the semi-infinite lattice $n=0,1,2,..+\infty$ 
\begin{eqnarray}
M({\tilde n} \vert n)  = p   \delta_{n,{\tilde n}+1} \theta( n \geq 1 )+ (1-p)   \delta_{n,{\tilde n}-1}  \theta( {\tilde n} \geq 1 )
 + p 2^{-{\tilde n}-1} \delta_{n,0}
\label{MarkovMatrixM}
\end{eqnarray}
satisfying the normalization
\begin{eqnarray}
\sum_{{\tilde n}=0}^{+\infty} M({\tilde n} \vert n)  = 1 \ \ \text{ for any } \ \ n=0,1,2,..
\label{MarkovMatrixMnorma}
\end{eqnarray}
can be interpreted as a biased random walk with resetting as follows (see Figure 1): 

(i) the random walker at position $n \geq 1$ 
can either jump towards the left neighbor ${\tilde n}=(n-1)$
with probability $p$ or towards the right neighbor ${\tilde n}=(n+1)$ with probability $(1-p)$;

(ii) the random walker at the origin $n =0$ 
can jump towards the right neighbor ${\tilde n}=1$ with probability $(1-p)$,
while with the complementary probability $p$,
the impossible jump towards the left neighbor ${\tilde n}=-1$ 
 is replaced by a reset towards any site of the lattice ${\tilde n}=0,1,..$ drawn with the resetting probability 
\begin{eqnarray}
  \pi^{[\mathrm {Uniform}]}_{{\tilde n} } \equiv 2^{-{\tilde n}-1} \ \ \text{ with the normalization } \ \ \sum_{{\tilde n}=0}^{+\infty} \pi^{[\mathrm {Uniform}]}_{{\tilde n} }=1
  \label{defpireset}
\end{eqnarray}
that corresponds to the uniform probability on the interval $x \in [0,1[$ via Eq. \ref{BinaryDensity}
\begin{eqnarray}
  \rho^{[\mathrm {Uniform}]}(x)  \equiv \sum_{n=0}^{+\infty} 2^{n+1} \pi^{[\mathrm {Uniform}]}_n \theta \left( 2^{-n-1} \leq x < 2^{-n} \right)
 =  \sum_{n=0}^{+\infty}  \theta \left( 2^{-n-1} \leq x < 2^{-n} \right)
 =  \theta \left( 0 \leq x < 1 \right)
\label{UniformBinaryDensity}
\end{eqnarray}

\begin{figure}[h]
  \centering
  \includegraphics[width=.9\textwidth]{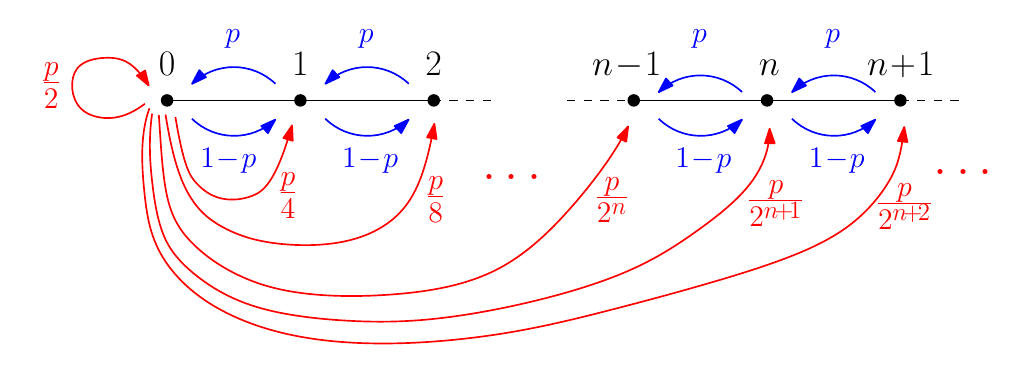}
  \caption{The Markov chain for the weights $\pi_t(n)$ on the half-infinite lattice $n=0,1,2,...+\infty$ corresponds to a biased random walk that jumps to the left with probability $p$ and to the right with probability $(1-p)$, while at the origin $n=0$, the impossible jump to the left is replaced by a reset towards any site $n=0,1,..,+\infty$ with the probability $p 2^{-n-1}$.}
  \label{fig:reset}
\end{figure}

%%%%%%%%%%%%%%%%%%%%%%%%%%%%%%%%%%%%%%%%

\subsection{  Steady state $\pi_*(n) $ in the chaotic region $\frac{1}{2} <p \leq 1$  }

\label{subsec_noneqsteady}

\subsubsection{  Derivation as the non-equilibrium steady state $\pi_*(n) $ of the biased random walk with resetting from the origin  }

The steady state $\pi_*(\cdot)$ of the Markov chain of Eq. \ref{FPan}
\begin{eqnarray}
 \pi_*({ \tilde n})  = p   \pi_*({ \tilde n}+1) + (1-p)   \pi_*({ \tilde n}-1)  \theta( n \geq 1 )
 + p 2^{-{ \tilde n}-1} \pi_*(0) 
  \equiv \sum_{n=0}^{+\infty} M({ \tilde n} \vert n) \pi_*(n)
\label{MarkovSteady}
\end{eqnarray}
can be obtained directly from the solution of a recurrence 
with constant coefficients and inhomogeneous terms, as done already in the Pelikan paper \cite{Pelikan},
and as will be done in Appendix \ref{app_laplace} to analyze more complicated observables.
However from a physical point of view,
it is more instructive here  
to compute $\pi_*({ \tilde n})$ as a non-equilibrium steady state together  
with the corresponding steady link-currents $J_*({ \tilde n}-1/2)$ flowing  
from the sites ${ \tilde n} $ towards their left neighbors $ ({ \tilde n}-1)$
\begin{eqnarray}
J_* ({ \tilde n}-1/2) \equiv p    \pi_*({ \tilde n})- (1- p)  \pi_*({ \tilde n-1})
\label{Jstarbond}
\end{eqnarray}
and with the steady reset-currents $J_*^{Reset} ({\tilde n}) $ 
flowing from the origin $n=0$ towards the sites ${ \tilde n} =0,1,2,..$
\begin{eqnarray}
J_*^{Reset} ({\tilde n}) \equiv p  2^{-{ \tilde n}-1} \pi_*(0) 
\label{Jstarreset}
\end{eqnarray}

Then Eq. \ref{MarkovSteady} for the steady state $\pi_*(\cdot) $ can be rewritten
as the vanishing of the sum of steady currents arriving at the bulk nodes ${ \tilde n}=1,2,..  $
\begin{eqnarray}
0   = J_* ({ \tilde n}+1/2)- J_* ({ \tilde n}-1/2) + J_*^{Reset} ({\tilde n})
\label{MarkovMatrixMringsteadyjring}
\end{eqnarray}
and arriving at the boundary node ${ \tilde n} =0$ 
\begin{eqnarray}
0   =  J_* (1/2) - \sum_{{\tilde n}=0}^{+\infty}  J_*^{Reset} ({\tilde n}) 
 + J_*^{Reset} (0)
  =  J_*(1/2) - \sum_{{\tilde n}=1}^{+\infty}  J_*^{Reset} ({\tilde n}) 
\label{MarkovMatrixMringsteadyjringzero}
\end{eqnarray}
These two equations yield that all the steady link-currents $J_*({ \tilde n}-1/2)$ can be computed in terms of the steady reset-currents $J_*^{Reset} (\cdot) $ of Eq. \ref{Jstarreset} to obtain
\begin{eqnarray}
 J_* (n-1/2) =  \sum_{{\tilde n}=n}^{+\infty}  J_*^{Reset} ({\tilde n})
 = \sum_{{\tilde n}=n}^{+\infty}  p  2^{-{ \tilde n}-1} \pi_*(0) =  p  2^{-n} \pi_*(0) 
\label{Jsteadylink}
\end{eqnarray}
Now that all the steady currents have been 
rewritten in terms of the steady density $\pi_*(0)$ at the origin $n=0$,
one can plug Eq. \ref{Jsteadylink} into Eq. \ref{Jstarbond} 
in order to obtain the following recurrence for the steady density
\begin{eqnarray}
     \pi_*(n) && =  \frac{J_* (n-1/2) }{p}  +\frac{1-p}{p}   \pi_*(n-1) 
     =  2^{-n} \pi_*(0)  +\frac{1-p}{p}   \pi_*(n-1) 
\label{Jstarbondrec}
\end{eqnarray}
The iteration of this recurrence yields the steady density $   \pi_*(n) $ for any $n$ 
\begin{eqnarray}
     \pi_*(n) && = 2^{-n} \pi_*(0)  +\frac{1-p}{p}  \left[ 2^{-(n-1)} \pi_*(0)  +\frac{1-p}{p}   \pi_*(n-2) \right] 
     = ...= \pi_*(0) \sum_{m=0}^{n} 2^{-(n-m)} \left(\frac{1-p}{p} \right)^m 
     \nonumber \\     && 
    = \pi_*(0) 2^{-n} \sum_{m=0}^{n}  \left( 2 \frac{1-p}{p} \right)^m 
 %         = \pi_*(0) 2^{-n} \frac{1-\left(2\frac{1-p}{p} \right)^{n+1}}{1-\left(2 \frac{1-p}{p} \right)}
 %    \nonumber \\     && 
     = \pi_*(0) 2 p \frac{ 2^{-n-1} -\left(\frac{1-p}{p} \right)^{n+1}}{ 3p-2}
\label{Jstarbondrecsol}
\end{eqnarray}
in terms of $  \pi_*(0) $ that is determined by the normalization condition of Eq. \ref{normapi}
\begin{eqnarray}
1 && =\sum_{n=0}^{+\infty} \pi_*(n)  
= \frac{ \pi_*(0) 2 p}{ 3p-2}
\sum_{n=0}^{+\infty} \left[  2^{-n-1} -\left(\frac{1-p}{p} \right)^{n+1} \right]
 =\frac{ \pi_*(0) 2 p}{ 3p-2} \left[  1  - \frac{1-p}{2p-1}  \right]
% \nonumber \\ &&
 =\frac{ \pi_*(0) 2 p}{ 2p-1}
\label{BinaryDensitynormasteady}
\end{eqnarray}
Plugging the solution
\begin{eqnarray}
\pi_*(0) =\frac{ 2p-1}{  2 p}
\label{solusteadyzero}
\end{eqnarray}
into Eq. \ref{Jstarbondrecsol} yields the steady state $\pi_* (n) $ in the chaotic region $1/2<p \leq 1 $
\begin{eqnarray}
  \pi_* (n) = \frac{2p-1}{3p-2} \left[ 2^{-n-1} - \left( \frac{1-p}{p} \right)^{n+1}\right] 
\label{BinarySteadySOL}
\end{eqnarray}
that can be plugged into Eq. \ref{BinaryDensity}
to obtain the steady density $ \rho_*(x)  $ as a function of $x \in [0,1[$
\begin{eqnarray}
  \rho_*(x) && = \sum_{n=0}^{+\infty} 2^{n+1} \pi_*(n) \theta \left( 2^{-n-1} \leq x < 2^{-n} \right)
\nonumber \\
&& =   \frac{2p-1}{3p-2} 
 \sum_{n=0}^{+\infty} \left[ 1 - \left( 2 \frac{1-p}{p} \right)^{n+1}\right]  \theta \left( 2^{-n-1} \leq x < 2^{-n} \right)
\label{BinaryDensitysteady}
\end{eqnarray}
in agreement with the initial derivation \cite{Pelikan}.

%%%%%%%%%%%%%%%%%%%%%%%%%%%%%%%%%%%%%%%%%%%%

\subsubsection{ Reminder on the properties of the steady state $\rho_*(x) $ as a function of the parameter 
$\frac{1}{2} < p \leq 1 $ }

As stressed in \cite{Ruffo}, the properties of the steady state $\rho_*(x) $ change as follows
as a function of the parameter $p$ within the chaotic region $\frac{1}{2} < p \leq 1 $.

The value of the steady state $\rho_*(x)$ on the interval $\frac{1}{2} \leq x <1 $ corresponding to the value $n=0$
\begin{eqnarray}
\text{for } \frac{1}{2} \leq x <1 :   \rho_*(x) &&  =  2 \pi_*(n=0) = \frac{2p-1}{3p-2} 
 \left[ 1 -  2 \frac{1-p}{p} \right] =   \frac{2p-1}{p} 
\label{BinaryDensitysteady1}
\end{eqnarray}
vanishes linearly near the critical point $p \to \frac{1}{2}^+$.

The behavior of the steady state $\rho_*(x)$ near the origin $x \to 0^+$
corresponding to $n \to + \infty$ 
displays the following properties 
as a function of the parameter $p$ within the chaotic region $1/2<p \leq 1 $
(see the recent study \cite{Ruffo} and references therein for more details):

(i) the value of the steady density $ \rho_*(x)  $ at the origin $x \to 0^+$ 
remains finite only for $\left( 2 \frac{1-p}{p} \right) <1 $ i.e. only in the region $\frac{2}{3} < p \leq 1$
\begin{eqnarray}
\lim_{x \to 0^+ }\rho_*(  x)  = \lim_{n \to + \infty} \left( \frac{2p-1}{3p-2} 
 \left[ 1 - \left( 2 \frac{1-p}{p} \right)^{n+1}\right] \right) =
 \begin{cases}
 \frac{2p-1}{3p-2}  \ \ \  \text{ for } \frac{2}{3} < p \leq 1 
  \\
+\infty \ \ \text{ for } \frac{1}{2} < p <  \frac{2}{3}
\end{cases}
\label{steadyatxeq0}
\end{eqnarray}

(ii) at the point $p=\frac{2}{3}$, the computation of Eq. \ref{Jstarbondrecsol} valid for $p \ne \frac{2}{3}$
should be changed to take into account $\left( 2 \frac{1-p}{p} \right) \to 1$
\begin{eqnarray}
\text{for  }   p=\frac{2}{3} : \ \      \pi_*(n) && 
   = \pi_*(0) 2^{-n} \sum_{m=0}^{n}  \left( 2 \frac{1-p}{p} \right)^m 
          = \pi_*(0) 2^{-n} (n+1)
\label{Jstarbondrecsolspecial23}
\end{eqnarray}
while the normalization condition of Eq. \ref{BinaryDensitynormasteady} becomes
\begin{eqnarray}
1 && =\sum_{n=0}^{+\infty} \pi_*(n)  
=\pi_*(0)  \sum_{n=0}^{+\infty} 2^{-n} (n+1) = 4 \pi_*(0)
\label{BinaryDensitynormaspecial23}
\end{eqnarray}
so that the steady state of Eq. \ref{BinaryDensitysteady} becomes
\begin{eqnarray}
\text{for  }   p=\frac{2}{3} : \ \    \rho_*(x) && = \sum_{n=0}^{+\infty} 2^{n+1} \pi_*(n) \theta \left( 2^{-n-1} \leq x < 2^{-n} \right)
 =   \sum_{n=0}^{+\infty} \frac{n+1}{2}  \theta \left( 2^{-n-1} \leq x < 2^{-n} \right)
\label{BinaryDensitysteadyspecial23}
\end{eqnarray}
that can be interpreted as a logarithmic divergence near the origin $x_n \equiv 2^{-n-1} \to 0^+$ 
where $n+1= - \frac{\ln x_n}{\ln 2} \to + \infty$
\begin{eqnarray}
\text{for  }   p=\frac{2}{3} : \ \    \rho_*(x_n=2^{-n-1})  = \frac{n+1}{2}  =  - \frac{\ln (x_n)}{2 \ln 2}
\label{BinaryDensitysteadyspecial23log}
\end{eqnarray}

(iii) In the region $\frac{1}{2} < p <  \frac{2}{3}$ where $\left( 2 \frac{1-p}{p} \right)>1$,
the leading behavior of the steady density of Eq. \ref{BinaryDensitysteady}
at $x_n \equiv 2^{-n-1}$ can be interpreted as a normalizable power-law divergence 
near the origin $x_n =2^{-n-1} \to 0^+$ 
where $n+1= - \frac{\ln x_n}{\ln 2} \to + \infty$
\begin{eqnarray}
\text{ for }  \frac{1}{2} < p <  \frac{2}{3} :  \rho_*(x_n=2^{-n-1}) && \opsimeq_{n \to + \infty}  
 \frac{2p-1}{2-3p}   \left( 2 \frac{1-p}{p} \right)^{n+1} = \left( \frac{2p-1}{2-3p}  \right) x_n^{\mu(p) -1 }
\label{BinaryDensitysteadyoriginpowerlaw}
\end{eqnarray}
where the exponent
\begin{eqnarray}
\mu(p) \equiv \frac{ \ln \left(  \frac{p}{1-p} \right)}{\ln 2} \in ]0,1[ \ \ \text{for} \ \ \frac{1}{2} < p <  \frac{2}{3}
\label{mup}
\end{eqnarray}
grows from the vanishing value $\mu \to 0^+$ near the critical point $p \to \frac{1}{2}^+ $
that corresponds to the normalizability limit
for the power-law divergence with exponent $(\mu-1) $ near the origin
\begin{eqnarray}
 \mu(p) \opsimeq_{p \to \frac{1}{2}^+} 0^+
\label{mup1demi}
\end{eqnarray}
towards the unity value $\mu \to 1^-$ for $p \to \frac{2}{3}^- $ 
\begin{eqnarray}
 \mu(p) \opsimeq_{p \to \frac{2}{3}^-} 1^- 
\label{mup23}
\end{eqnarray}
where the divergence with the power-law $(\mu-1) $ is replaced by the logarithmic divergence of Eq. \ref{BinaryDensitysteadyspecial23}.

%%%%%%%%%%%%%%%%%%%%%%%%%%%%%%%%%%%%%%%%

\subsection{  Discussion  }

Besides the existence and the properties of the steady state $\rho_*(x)$ recalled above,
it is interesting to analyze the dynamical properties
 when the initial density $\rho_{t=0}(x)$ belongs to the subspace 
 of Eq. \ref{BinaryDensity} for any value of the parameter $p$:

(i) in the chaotic region $\frac{1}{2} < p \leq 1$ 
characterized by the presence of the steady density $\rho_*(x)$
of Eq. \ref{BinaryDensitysteady}, 
it is important to analyze the convergence properties towards the steady state $\rho_*(x) $.

(ii) at the critical point $p_c=\frac{1}{2}$ and in the non-chaotic region $ 0 \leq p < \frac{1}{2} $
where there is no steady state, it is interesting to study the properties
of the transient dynamics, for instance when the initial condition $x_0$ is drawn with
the uniform density $\rho^{[\mathrm {Uniform}]}(x_0)=1 $ of Eq. \ref{UniformBinaryDensity}.

To analyze the general dynamical properties with the subspace 
 of Eq. \ref{BinaryDensity}, we will consider as initial condition
the normalized flat distribution on the single interval $x \in [ 2^{-n_0-1} , 2^{-n_0}]$
\begin{eqnarray}
  \rho^{[n_0]}_{t=0}(x) \equiv  2^{n_0+1}  \theta \left( 2^{-n_0-1} \leq x < 2^{-n_0} \right)
  \ \ \ \ \text{corresponding to the single weight } \ \ \pi^{[n_0]}_{t=0}(n)=\delta_{n,n_0}
\label{BinaryDensityn0}
\end{eqnarray} 
and we will study the properties of the 
finite-time propagator $\pi_t(n \vert n_0)$ in the next section.

%%%%%%%%%%%%%%%%%%%%%%%%%%%%%%%%%%%%%%%%

\section{  Properties of the finite-time propagator $\pi_t(n \vert n_0)$ for the weights    }

\label{sec_propagator}

In this section, we focus on the finite-time propagator $ \pi_t(n \vert n_0) $ associated to the 
Markov matrix $M$ of Eq. \ref{MarkovMatrixM}
\begin{eqnarray}
 \pi_t(n \vert n_0)  \equiv \langle n \vert M^t \vert n_0 \rangle
\label{propagatordef}
\end{eqnarray}
that satisfies the forward dynamics of Eq. \ref{FPan}
\begin{eqnarray}
 \pi_t(n \vert n_0)  && 
 = \sum_m \langle n \vert M \vert m \rangle \langle m \vert M^{t-1} \vert n_0 \rangle
 = \sum_m M (n  \vert m)  \pi_{t-1}(m \vert n_0)
 \nonumber \\
 && = p   \pi_{t-1}(n+1\vert n_0) + (1-p)   \pi_{t-1}(n-1\vert n_0)  \theta(n \geq 1 )
 + p 2^{-n-1} \pi_{t-1}(0\vert n_0) 
\label{FPpropagator}
\end{eqnarray}
and the initial condition $\pi_{t=0}(n \vert n_0)=\delta_{n,n_0} $.

%%%%%%%%%%%%%%%%%%%%%%%%%%%%%%%%%%%%%%%%%%

\subsection{  Explicit time-Laplace-transform ${\tilde \pi}_s(n \vert n_0) $ 
of the propagator $\pi_t(n \vert n_0)$     }

\label{subsec_laplacepropagator}

The time-Laplace-transform ${\tilde \pi}_s(n \vert n_0) $ of the propagator $\pi_t(n \vert n_0)$
for $s \in ]0,+\infty[$
\begin{eqnarray}
{\tilde \pi}_s(n \vert n_0) \equiv \sum_{t=0}^{+\infty} e^{-s t} \pi_t(n \vert n_0) 
\label{piLaplace}
\end{eqnarray}
satisfies the following recurrence using Eq. \ref{FPpropagator}
\begin{eqnarray}
{\tilde \pi}_s(n \vert n_0) && = \pi_{t=0}(n \vert n_0) +\sum_{t=1}^{+\infty} e^{-s t} 
\left[ p   \pi_{t-1}(n+1\vert n_0) + (1-p)   \pi_{t-1}(n-1\vert n_0)  \theta(n \geq 1 )
 + p 2^{-n-1} \pi_{t-1}(0\vert n_0)  \right]
 \nonumber \\
&& =\delta_{n,n_0} +
 p e^{-s }  {\tilde \pi}_s(n+1\vert n_0) + (1-p) e^{-s }  {\tilde \pi}_s(n-1\vert n_0)  \theta(n \geq 1 )
 + p e^{-s }2^{-n-1} {\tilde \pi}_s(0\vert n_0)  
\label{piLaplacerec}
\end{eqnarray}
and the normalization over $n$
\begin{eqnarray}
\sum_{n=0}^{+\infty} {\tilde \pi}_s(n \vert n_0) = \sum_{t=0}^{+\infty} e^{-s t} \left[\sum_{n=0}^{+\infty} \pi_t(n \vert n_0) \right]
=\sum_{t=0}^{+\infty} e^{-s t}= \frac{1}{1-e^{-s} }
\label{piLaplacenorma}
\end{eqnarray}

The computation described in Appendix \ref{app_laplace} leads to the explicit expressions of Eqs \ref{recapexpanded}
and \ref{BilanSolutionnzerovanish} that can be summarized by
\begin{eqnarray}
\text{ for } \ \ n \geq n_0 \geq 0 : \ \ \ 
{\tilde \pi}_s(n \vert n_0)   
&&  =    (r_+  +    r_-)  \left[  \frac{r_-^{n+1}  (  r_-^{-n_0-1}-  r_+^{-n_0-1}   ) } {     r_+  -    r_- }  
+   \frac{ (2^{-n-1}    - r_-^{n+1}  )  }{    ( r_+ -1  ) (1-2 r_-)} r_+^{-n_0-1}    \right]
\nonumber \\
\text{ for } \ \ 0 \leq n \leq n_0 : \ \ \ 
{\tilde \pi}_s(n \vert n_0)   
&& =   (r_+  +    r_-)  \left[  \frac{  ( r_+^{n+1} - r_-^{n+1}  ) } {     r_+  -    r_- }  
+   \frac{ (2^{-n-1}    - r_-^{n+1}  )  }{    ( r_+ -1  ) (1-2 r_-)}   \right] r_+^{-n_0-1} 
 \label{recap}
\end{eqnarray}
in terms of the notations introduced in Eq. \ref{2dsol}
\begin{eqnarray}
r_{\pm}(s) \equiv \frac{ e^s }{2p } \left[ 1 \pm \sqrt{ 1 - 4  p(1-p) e^{-2s}} \right]
\label{2dsolmain}
\end{eqnarray}
where we have added the explicit dependence with respect to the Laplace parameter $s$ 
in order to write their series expansions near the origin $s \to 0^+$ 
\begin{eqnarray}
r_{\pm}(s) && = \frac{ [1+s+O(s^2) ] }{2p } \left[ 1 \pm \sqrt{ 1 - 4  p(1-p) (1-2s+O(s^2) )} \right]
\nonumber \\
&& = \frac{ [1+s+O(s^2) ] }{2p } \left[ 1 \pm \sqrt{ (1-2p)^2 \left( 1+ \frac{8  p(1-p) }{(1-2p)^2} s+O(s^2) \right)} \right]
\nonumber \\
&& = \frac{ [1+s+O(s^2) ] }{2p } \left[ 1 \pm \vert 1-2p \vert  \left( 1+ \frac{4  p(1-p) }{(1-2p)^2} s+O(s^2) \right) \right]
\label{2dsolzero}
\end{eqnarray}
that will be useful in the following subsections to analyze the asymptotic behavior of the propagator 
$\pi_t(n \vert n_0)$
for large time $t \to + \infty$ in the various regions of the parameter $p$. 

%%%%%%%%%%%%%%%%%%%%%%%%%%%%%%%%%%%%%%

\subsubsection{ Asymptotic analysis in the chaotic region $\frac{1}{2} < p \leq 1$}

In the chaotic region $\frac{1}{2} < p \leq 1$ where 
the propagator $ \pi_t(n \vert n_0) $ converges towards the steady state $\pi_*(n)$ of Eq. \ref{BinarySteadySOL}
\begin{eqnarray}
 \pi_t(n \vert n_0)  \opsimeq_{t \to + \infty} \pi_*(n)
\label{Cvsteady}
\end{eqnarray}
the time-Laplace-transform of Eq. \ref{piLaplace} will display the following divergence for $s \to 0^+$ 
\begin{eqnarray}
{\tilde \pi}_s(n \vert n_0) \equiv \sum_{t=0}^{+\infty} e^{-s t} \pi_t(n \vert n_0)  
\opsimeq_{s \to 0^+} \frac{\pi_*(n)}{s}
\label{piLaplaceCVsteady}
\end{eqnarray}

The series expansions of Eqs \ref{2dsolzero} in the chaotic region $\frac{1}{2} < p \leq 1$
\begin{eqnarray}
r_+(s) && =  [1+s+O(s^2) ] \left[ 1 + \frac{2(1-p) }{2p-1} s+O(s^2)  \right]
= 1 + \frac{1 }{2p-1} s+O(s^2)
\nonumber \\
r_-(s) && =  [1+s+O(s^2) ]  \left[ \frac{1-p}{p} - \frac{2  (1-p) }{(2p-1)} s+O(s^2)  \right]
= \frac{1-p}{p} - \frac{  (1-p) }{p (2p-1)}s+O(s^2)
\label{2dsolchaos}
\end{eqnarray}
yield that the only terms of order $\frac{1}{s}$ in Eq. \ref{recap}
are the terms containing $\frac{1}{(r_+-1)}$ leading to
\begin{eqnarray}
\pi_*(n) && = \lim_{s \to 0^+} \left[ s{\tilde \pi}_s(n \vert n_0) \right]
 =  \frac{   r_+(0)  +    r_-(0)   }{  \frac{1 }{2p-1}    [1-2 r_-(0)]}  (  2^{-n-1}   -  [r_-(0)]^{n+1}  ) [r_+(0)]^{-n_0-1} 
\nonumber \\
&& =   \frac{  (2p-1) }{ (3p-2) }  \left[  2^{-n-1}   -  \left( \frac{1-p}{p} \right)^{n+1}  \right] 
\label{piLaplaceCVsteadychaotic}
\end{eqnarray}
in agreement with Eq. \ref{BinarySteadySOL} as it should for consistency.

In order to compare with the following subsections, it is useful to mention
that the convergence of Eq. \ref{Cvsteady} means
that the total time spent as position $n$ in the time-window $[0,T]$ diverges 
linearly in time for large time $T \to + \infty $
\begin{eqnarray}
\sum_{t=0}^T \pi_t(n \vert n_0)  \opsimeq_{T \to + \infty} T \pi_*(n)
\label{Cvsteadytime}
\end{eqnarray}

%%%%%%%%%%%%%%%%%%%%%%%%%%%%%%

\subsubsection{ Asymptotic analysis in the non-chaotic region $0<p < \frac{1}{2}$}

In the non-chaotic region $0<p < \frac{1}{2}$,
 the series expansions of Eqs \ref{2dsolzero}
\begin{eqnarray}
r_+(s) && =  [1+s+O(s^2) ]  \left[ \frac{1-p}{p}  + \frac{2(1-p) }{1-2p} s+O(s^2)  \right]
= \frac{1-p}{p}  + \frac{(1-p) }{p(1-2p)} s+O(s^2)
\nonumber \\
r_-(s) && = [1+s+O(s^2) ]  \left[ 1  - \frac{2(1-p) }{1-2p} s+O(s^2)  \right]
= 1 - \frac{1 }{1-2p} s+O(s^2)
\label{2dsolzerononchaos}
\end{eqnarray}
yield that ${\tilde \pi}_{s}(n \vert n_0)$ does not contain any singular term of order $\frac{1}{s}$
in contrast to Eq. \ref{piLaplaceCVsteadychaotic}
\begin{eqnarray}
\pi_*(n)  = \lim_{s \to 0^+} \left[ s{\tilde \pi}_s(n \vert n_0) \right] =0 
\label{piLaplaceCVsteadynonchaos}
\end{eqnarray}
but takes a finite value ${\tilde \pi}_{s=0}(n \vert n_0)  $ for $s=0$
that represents the total time spent at $n$ when starting at $n_0$
\begin{eqnarray}
{\tilde \pi}_{s=0}(n \vert n_0) = \sum_{t=0}^{+\infty}  \pi_t(n \vert n_0) 
 =\begin{cases}
\frac{  1  } {  1-2p }  \left[ 1     -       2^{-n-1}  \left( \frac{p}{1-p}  \right)^{n_0+1} \right]
\ \ \ \ \ \ \ \ \ \ \ \ \ \  \text{ for } \ \ n \geq n_0 \geq 0
 \\
 \\
\frac{  1  } {  1-2p }  \left[  \left( \frac{p}{1-p}  \right)^{n_0-n} -   2^{-n-1}   \left( \frac{p}{1-p}  \right)^{n_0+1} \right]
 \text{ for } \ \ 0 \leq n \leq n_0
\end{cases}
\label{pis0totaltimefinite}
\end{eqnarray}

%%%%%%%%was %%%%%%%%%%%%%%%%%%%%%%

\subsubsection{ Asymptotic analysis at the critical point $p_c=\frac{1}{2}$}

At the critical point $p_c=\frac{1}{2}$,  the series expansions of Eqs \ref{2dsolzero} 
involve the following singular contributions in $\sqrt{s}$ 
\begin{eqnarray}
r_{\pm}(s)  =  [1+s+O(s^2) ]  \left[ 1 \pm \sqrt{ 2s+O(s^2) } \right]
=1 \pm \sqrt{ 2s} +O(s)
\label{2dsolcriti}
\end{eqnarray}
As a consequence, ${\tilde \pi}_{s}(n \vert n_0)$ does not contain any singular term of order $\frac{1}{s}$
\begin{eqnarray}
\pi_*(n)  = \lim_{s \to 0^+} \left[ s{\tilde \pi}_s(n \vert n_0) \right] =0 
\label{piLaplaceCVsteadynoncriti}
\end{eqnarray}
but diverges as $\frac{1}{\sqrt{s}}$ in contrast to the finite value of Eq. \ref{pis0totaltimefinite} concerning the region $p<1/2$
\begin{eqnarray}
{\tilde \pi}_{s}(n \vert n_0) = \frac{ \sqrt{ 2}  (1-  2^{-n-1}     )}{\sqrt{s} } +O(1)
\label{pis0totaltimecriti}
\end{eqnarray}
This corresponds to the following asymptotic behavior for large time $t$
\begin{eqnarray}
 \pi_t(n \vert n_0) \opsimeq_{t \to + \infty}  \frac{ \sqrt{ 2}  (1-  2^{-n-1}     )}{\sqrt{\pi t} } 
\label{pis0timecritit}
\end{eqnarray}
so that the total time spent as position $n$ in the time-window $[0,T]$ diverges as $\sqrt{T}$ for large time $T \to + \infty $
\begin{eqnarray}
\sum_{t=0}^T \pi_t(n \vert n_0) \opsimeq_{T \to + \infty}  \frac{ 2 \sqrt{ 2}  (1-  2^{-n-1}     )}{\sqrt{\pi } } \sqrt{T}
\label{pis0totaltimecritit}
\end{eqnarray}
in contrast to the linear divergence of Eq. \ref{Cvsteadytime} for $p>1/2$
and to the finite value of Eq. \ref{pis0totaltimefinite} for $p<1/2$.

%%%%%%%%%%%%%%%%%%%%%%%%%%%%%%%%%%%%%%%%%%%%%%%%

\subsection{ Probability distribution $F_{t_1}(n \vert n_0)$ of the first-passage time $t_1$ at $n$ when starting at $n_0$ }

\label{subsec_firstpassagetime}

The propagator $\pi_t(n \vert n_0)$ can be decomposed with respect to 
the probability distribution $F_{t_1}(n \vert n_0)$ of the first-passage time $t_1$ at $n$ when starting at $n_0$
\begin{eqnarray}
\pi_t(n \vert n_0) = \sum_{t_1=0}^t \pi_{t-t_1}(n \vert n_0) F_{t_1}(n \vert n_0)
\label{propagatorAndFirstPassage}
\end{eqnarray}
This convolution in time translates into the following simple product for the Laplace transforms
\begin{eqnarray}
{\tilde \pi}_s(n \vert n_0) = \sum_{t=0}^{+\infty} e^{-st} \pi_t(n \vert n_0) 
= {\tilde \pi}_s(n \vert n)  {\tilde F}_s(n \vert n_0)
\label{propagatorAndFirstPassageLaplace}
\end{eqnarray}
so that the Laplace transform $ {\tilde F}_s(n \vert n_0) $ can be computed from 
the Laplace transform ${\tilde \pi}_s(n \vert .) $
\begin{eqnarray}
{\tilde F}_s(n \vert n_0) \equiv \sum_{t=0}^{+\infty} e^{-st} F_t(n \vert n_0)  
= \frac{{\tilde \pi}_s(n \vert n_0) } {{\tilde \pi}_s(n \vert n) }
  \label{PDFLaplace}
\end{eqnarray}
where the value unity for $n=n_0$ corresponds to $F_{t}(n_0 \vert n_0) =\delta_{t,0}$.

The explicit result of Eq. \ref{recap}
for ${\tilde \pi}_s(n \vert n_0) $ yields
\begin{eqnarray}
\text{ for } \ \ n \geq n_0 \geq 0 : \ \ \ 
{\tilde F}_s(n \vert n_0)   
&&  =    \frac{ \frac{  ( r_-^{n-n_0}- r_-^{n+1} r_+^{-n_0-1}   ) } {     r_+  -    r_- }  
+   \frac{ (2^{-n-1}    - r_-^{n+1}  )  }{    ( r_+ -1  ) (1-2 r_-)} r_+^{-n_0-1}    }
{   \frac{  ( 1 - r_-^{n+1}  r_+^{-n-1}  ) } {     r_+  -    r_- }  
+   \frac{ (2^{-n-1}    - r_-^{n+1}  )  }{    ( r_+ -1  ) (1-2 r_-)}   r_+^{-n-1}  } 
\nonumber \\
\text{ for } \ \ 0 \leq n \leq n_0 : \ \ \ 
{\tilde F}_s(n \vert n_0)   && =  r_+^{n-n_0} 
 \label{PDFLaplacerecap}
\end{eqnarray}
 The huge simplification of the second line concerning the region $0 \leq n < n_0 $ can be understood as follows:
 the first-passage at $n$ from any bigger position $n_0$ is the same as for the simple walk involving
  jumps towards nearest-neighbors since the reset events occurring only from the origin $n=0$
  have not been yet possible.
  This is in contrast to the region $n_0 <n$ where the reset events are responsible
  for the more complicated result of the first line of Eq. \ref{PDFLaplacerecap}.

   %%%%%%%%%%%%%%%%%%%%%%%%%%%%%%%%%%%%%%

\subsubsection{ Normalization ${\tilde F}_{s=0}(n \vert n_0)  $ of the First-Passage-Time distribution $F_t (n \vert n_0) $
in the non-chaotic region $0 \leq p <\frac{1}{2} $}

  In the non-chaotic region $0 \leq p <\frac{1}{2} $ where $ {\tilde \pi}_{s=0}(n \vert n_0)$ of Eq. \ref{pis0totaltimefinite}
  is finite, the normalization ${\tilde F}_{s=0}(n \vert n_0)  $ of Eq. \ref{PDFLaplace}
  of the First-Passage-Time distribution $F_t (n \vert n_0)  $
  reads
  \begin{eqnarray}
{\tilde F}_{s=0}(n \vert n_0) && \equiv \sum_{t=0}^{+\infty}  F_t(n \vert n_0)  
= \frac{{\tilde \pi}_0(n \vert n_0) } {{\tilde \pi}_0(n \vert n) }
% \nonumber \\ &&
  =\begin{cases}
\frac{   1     -       2^{-n-1}  \left( \frac{p}{1-p}  \right)^{n_0+1}  } 
{  1     -   \left( \frac{p}{ 2(1-p)}  \right)^{n+1}   }  
\ \ \  \text{ for } \ \ n > n_0 \geq 0
 \\
 \\
\left( \frac{p}{1-p}  \right)^{n_0-n} 
\ \ \  \ \ \  \ \ \ \text{ for } \ \ 0 \leq n < n_0
\end{cases}
  \label{PDFLaplacenormanonchaos}
\end{eqnarray}
  which is complementary to the strictly positive probability of never visiting $n$ when starting at $n_0$
    \begin{eqnarray}
1-{\tilde F}_{s=0}(n \vert n_0) &&
% \nonumber \\ &&
  =\begin{cases}
\frac{   2^{-n-1} \left[ \left( \frac{p}{1-p}  \right)^{n_0+1}   -   \left( \frac{p}{ 1-p}  \right)^{n+1}   \right]              } 
{  1     -   \left( \frac{p}{ 2(1-p)}  \right)^{n+1}   }  >0
\ \ \  \text{ for } \ \ n > n_0 \geq 0
 \\
 \\
1- \left( \frac{p}{1-p}  \right)^{n_0-n} >0
\ \ \ \ \ \ \ \ \  \text{ for } \ \ 0 \leq n < n_0
\end{cases}
  \label{PDFLaplacenormanonchaosNEVER}
\end{eqnarray}

  %%%%%%%%%%%%%%%%%%%%%%%%%%%%%%%%%%%%%%

\subsubsection{ Mean-First-Passage-Time $\tau(n \vert n_0) $ at $n$ when starting at $n_0$ in the chaotic region $\frac{1}{2} < p \leq 1$}

In the chaotic region $\frac{1}{2} < p \leq 1$ where the asymptotic behavior of ${\tilde \pi}_s(n \vert n_0) $ for $s \to 0^+$ 
is given by Eq. \ref{piLaplaceCVsteady}, 
the normalization ${\tilde F}_{s=0}(n \vert n_0)  $ of Eq. \ref{PDFLaplace}
  of the First-Passage-Time distribution $F_t (n \vert n_0)  $
  is unity in contrast to Eq. \ref{PDFLaplacenormanonchaos}
  \begin{eqnarray}
{\tilde F}_{s=0}(n \vert n_0) && \equiv \sum_{t=0}^{+\infty}  F_t(n \vert n_0)  
= \lim_{s \to 0^+} \left( \frac{{\tilde \pi}_s(n \vert n_0) } {{\tilde \pi}_s(n \vert n) } \right) =1
  \label{PDFLaplacenormachaos}
\end{eqnarray}
and the series expansion at first order in $s$
\begin{eqnarray}
{\tilde F}_s(n \vert n_0) = 1-s \tau(n \vert n_0) +o(s)
  \label{PDFLaplaceseries}
\end{eqnarray}
is useful to compute the Mean-First-Passage-Time $\tau(n \vert n_0) $ 
at $n$ when starting at $n_0$ 
  \begin{eqnarray}
\tau(n \vert n_0) && \equiv \sum_{t=0}^{+\infty} t  F_t(n \vert n_0)  
  =\begin{cases}
 \frac{  1 - \left(  \frac{1-p}{p} \right)^{n-n_0} } {    (2p-1) \pi_*(n)  } 
  + \frac{ n_0- n}{ 2p-1 }
\ \ \  \text{ for } \ \ n > n_0 \geq 0
 \\
 \\
\frac{ n_0- n}{ 2p-1 }
\ \ \  \ \ \ \ \ \ \ \ \ \ \ \ \ \ \ \ \text{ for } \ \ 0 \leq n < n_0
\end{cases}
  \label{MFPTchaos}
\end{eqnarray}
where $\pi_*(n)  $ is the steady state given in Eq. \ref{BinarySteadySOL}.

One can check that the MFPT $\tau(n \vert n_0) $ of Eq. \ref{MFPTchaos}
satisfies the standard recurrence with respect to the starting point associated to the Markov matrix $M$ of Eq. \ref{MarkovMatrixM}
\begin{eqnarray}
 \tau(n \vert n_0)  && = 1 + \sum_{m_0}  \tau(n \vert m_0) M(m_0,n_0 )
 \nonumber \\
 && = 1 +    (1-p) \tau(n \vert n_0+1) +   p \tau(n \vert n_0-1) \theta(n_0 \geq 1)  + p  \delta_{n_0,0} \left[ \sum_{m_0=0}^{+\infty} \tau(n \vert m_0)  2^{-m_0-1} \right] 
 \label{MFPTbackward}
\end{eqnarray}
in the two regions $n_0>n$ and $n_0<n$ with the boundary condition  $\tau(n \vert n)=0$.

 %%%%%%%%%%%%%%%%%%%%%%%%%%%%%%%%%%%%%%

\subsubsection{ First-Passage-Time distribution $F_t (n \vert n_0) $ at $n$ when starting at $n_0$
at the critical point $p_c=\frac{1}{2}$}

At the critical point $p_c=\frac{1}{2}$ where the asymptotic behavior of ${\tilde \pi}_s(n \vert n_0) $ for $s \to 0^+$ 
is given by Eq. \ref{pis0totaltimecriti}, 
the normalization ${\tilde F}_{s=0}(n \vert n_0)  $ of Eq. \ref{PDFLaplace}
  of the First-Passage-Time distribution $F_t (n \vert n_0)  $
  is unity 
  \begin{eqnarray}
{\tilde F}_{s=0}(n \vert n_0) && \equiv \sum_{t=0}^{+\infty}  F_t(n \vert n_0)  
= \lim_{s \to 0^+} \left( \frac{{\tilde \pi}_s(n \vert n_0) } {{\tilde \pi}_s(n \vert n) } \right) =1
  \label{PDFcriti}
\end{eqnarray}
but the expansion of $ {\tilde F}_s(n \vert n_0)$ for $s \to 0^+$ contains singular contributions of order $\sqrt{s}$ 
in contrast to Eq. \ref{PDFLaplaceseries} involving finite MFPT $\tau(n \vert n_0)$ for $p>1/2$.
Using 
\begin{eqnarray}
{\tilde \pi}_{s}(n \vert n_0)  
=\begin{cases}
 \frac{ \sqrt{ 2}  (1-  2^{-n-1}     )}{\sqrt{s} } 
 - (2n-1) +2^{-n-1} (2 n_0-1) +O(\sqrt{s})
\ \ \  \ \ \ \ \ \ \ \ \ \ \ \  \text{ for } \ \ n \geq n_0 \geq 0
 \\
 \\
\frac{ \sqrt{ 2}  (1-  2^{-n-1}     )}{\sqrt{s} } 
+ (-1+  2^{-n-1}     ) (2 n_0-1) +O(\sqrt{s})
\ \ \  \ \ \ \ \ \ \ \ \ \ \ \ \ \ \ \ \text{ for } \ \ 0 \leq n \leq n_0
\end{cases}
\label{pis0totaltimecritiseries}
\end{eqnarray}
one obtains the expansions
 \begin{eqnarray}
{\tilde F}_{s}(n \vert n_0) && 
=\begin{cases}
 1- \sqrt{2s} \frac{ (n-n_0)}{ 2^{n+1}-1} +O(s)
\ \ \  \ \ \ \ \ \ \ \ \ \ \ \ \ \ \ \text{ for } \ \ n \geq n_0 \geq 0
 \\
 \\
1- \sqrt{2s}  (n_0-n) +O(s)
\ \ \  \ \ \ \ \ \ \ \ \ \ \ \ \ \ \ \ \text{ for } \ \ 0 \leq n \leq n_0
\end{cases}
  \label{PDFcritis}
\end{eqnarray}
that can be translated into the following power-law decays as $t^{-\frac{3}{2}}$ for 
the First-Passage-Time distribution $F_t (n \vert n_0) $
 \begin{eqnarray}
F_t(n \vert n_0) && 
\opsimeq_{t \to + \infty} \begin{cases}
 \frac{\sqrt{2} (n-n_0)}{ (2^{n+1}-1) \sqrt{\pi} t^{\frac{3}{2}}} 
\ \ \  \ \ \ \ \ \ \ \ \ \  \ \ \text{ for } \ \ n \geq n_0 \geq 0
 \\
 \\
\frac{\sqrt{2} (n_0-n)}{\sqrt{\pi} t^{\frac{3}{2}}}  
\ \ \  \ \ \ \ \ \ \ \ \ \ \ \ \ \ \ \ \text{ for } \ \ 0 \leq n \leq n_0
\end{cases}
  \label{PDFcritispower}
\end{eqnarray}

%%%%%%%%%%%%%%%%%%%%%%%%%%%%%%%%%%%%%%%%%%%%%%%%%

\subsection{ Explicit spectral decomposition of the propagator $ \pi_t (n \vert  n_0) $ }

\label{subsec_propagatorspectral}

Besides all the properties that have been described above in relation with the 
explicit time-Laplace-transform ${\tilde \pi}_s(n \vert 0) $,
it is useful to have another perspective
via the explicit spectral decomposition of the propagator $ \pi_t (n \vert  n_0) $
 discussed in Appendix \ref{app_spectral}
\begin{eqnarray}
\pi_t (n \vert  n_0) = \pi_*(n) \theta\left( \frac{1}{2} <p \leq 1\right)
+ \int_{0}^{ \pi} \frac{dq}{ \pi} \bigg[ \lambda (q) \bigg]^t \langle n \vert R_q \rangle  \langle L_q \vert n_0 \rangle 
  \label{spectralreset}
\end{eqnarray}
where the normalizable steady state $\pi_*(n) $ of Eq. \ref{BinarySteadySOL}
exists only in the chaotic region $\frac{1}{2} <p \leq 1 $,
while the continuum spectrum of eigenvalues $\lambda(q)$ parametrized by the Fourier momentum $q \in [0, \pi]$
\begin{eqnarray}
\lambda(q) \equiv \sqrt{ p (1-p) } (e^{iq}+e^{-iq} ) = \sqrt{ 4 p (1-p) } \cos q 
\label{lambdaqmomentum}
\end{eqnarray}
is associated to the right eigenvector
\begin{eqnarray}
R_q(n) \equiv \langle n \vert R_q \rangle  
&& =   \frac{1}{\sqrt 2} \left(\frac{1-p}{p} \right)^{\frac{n+1}{2}}
\bigg[    e^{iq(n+1)} -  
 \frac{\bigg(\sqrt{ \frac{1-p}{p} } e^{- iq}-1\bigg)\bigg(2 \sqrt{ \frac{1-p}{p} } e^{ iq}-1\bigg)}
 {\bigg(\sqrt{ \frac{1-p}{p} } e^{ iq}-1\bigg) \bigg(2 \sqrt{ \frac{1-p}{p} } e^{- iq} -1 \bigg)}
  e^{-iq(n+1)}
\nonumber \\
&& -  \frac{ i \sin q}
 {\bigg(\sqrt{ \frac{1-p}{p} } e^{ iq}-1\bigg) \bigg(2 \sqrt{ \frac{1-p}{p} } e^{- iq} -1 \bigg)}
    \left( 2 \sqrt{ \frac{1-p}{p} } \right)^{- n}  \bigg]
 \label{Rqnq}
\end{eqnarray}
and to the left eigenvector
\begin{eqnarray}
L_q(n_0) \equiv \langle L_q \vert n_0 \rangle   = \frac{1}{\sqrt 2} \left(\frac{1-p}{p} \right)^{-\frac{n_0+1}{2}} 
\left[ e^{-iq(n_0+1)} 
-  \frac{\bigg(\sqrt{ \frac{1-p}{p} } e^{ iq}-1\bigg) \bigg(2 \sqrt{ \frac{1-p}{p} } e^{- iq} -1 \bigg)}
{\bigg(\sqrt{ \frac{1-p}{p} } e^{- iq}-1\bigg)\bigg(2 \sqrt{ \frac{1-p}{p} } e^{ iq}-1\bigg)}  e^{iq(n_0+1)} \right]
 \label{Lqnq}
\end{eqnarray}
of the Markov matrix $M$ of Eq. \ref{MarkovMatrixM}
\begin{eqnarray}
M \vert R_q \rangle && = \lambda_q \vert R_q \rangle
\nonumber \\
\langle L_q \vert M && = \lambda_q \langle L_q \vert
\label{MarkovMatrixMeigenq}
\end{eqnarray}

For large time $t \to + \infty$, the integral over the momentum $q$ in the spectral decomposition of Eq. \ref{spectralreset}
will be dominated by the vicinity of zero-momentum $q \simeq 0$
corresponding to the highest eigenvalue $ \lambda(q=0) =  \sqrt{ 4 p (1-p) } $.
As a consequence outside the critical point $p_c=\frac{1}{2}$,
the integral is governed by the leading exponential time-decay
\begin{eqnarray}
\pi_t (n \vert  n_0) - \pi_*(n) \theta\left( \frac{1}{2} <p \leq 1\right)
\oppropto_{t \to + \infty}  \bigg[ \lambda (0) \bigg]^t  = \bigg[ \sqrt{ 4 p (1-p) } \bigg]^t = e^{ - \frac{ t }{t_{\mathrm {corre}}(p) } }
  \label{spectralresettinfty}
\end{eqnarray}
where the correlation-time 
\begin{eqnarray}
t_{\mathrm {corre}}(p) \equiv \frac{1}{\ln \left( \frac{1}{ \sqrt{ 4 p (1-p) } } \right) }   = \frac{1}{\ln \left( \frac{1}{ \sqrt{ 1- (1-2p)^2 } }\right)   }
  \label{tcorre}
\end{eqnarray}
diverges near the critical point $p_c=\frac{1}{2}$ as
\begin{eqnarray}
t_{\mathrm {corre}}(p)  \frac{1}{\ln \left( \frac{1}{ \sqrt{ 1- 4 (p-p_c)^2 } } \right)   } \opsimeq_{p \to p_c} \frac{1}{2 ( p - p_c)^2 }
  \label{tcorrecriti}
\end{eqnarray}

At the critical point $p_c = \frac{1}{2}$, where the right and left eigenvectors of Eqs \ref{Rqnq}
and \ref{Lqnq}
reduce to
\begin{eqnarray}
R_q(n)   
&& =   \frac{1}{\sqrt 2}
\bigg[    e^{iq(n+1)} +  
 \frac{\bigg(2  e^{ iq}-1\bigg)} { \bigg(2  e^{- iq} -1 \bigg)}  e^{-iq(n+2)}
 -  \frac{ e^{- i \frac{q}{2} }  \sin q} {  \sin \left[ \frac{q}{2}\right] \bigg(2  e^{- iq} -1 \bigg)}    2^{- n-1}  \bigg]
    \nonumber \\
L_q(n_0)  &&  = \frac{1}{\sqrt 2}  
\left[ e^{-iq(n_0+1)} + \frac{ \bigg(2 e^{- iq} -1 \bigg)}{\bigg(2  e^{ iq}-1\bigg)}  e^{iq(n_0+2)} \right]    
 \label{RqLqcriti}
\end{eqnarray}
the asymptotic behavior for large time $t \to + \infty$
 can be analyzed via the change of variables $q = \frac{ k }{ \sqrt t }  $ 
in the integral of Eq. \ref{spectralreset}
\begin{eqnarray}
&& \pi_t (n \vert  n_0)  
=  \int_{0}^{ \pi} \frac{dq}{ \pi} \bigg[ \cos (q) \bigg]^t R_q (n) L_q (n_0)
= \int_{0}^{ \pi \sqrt{t} } \frac{dk}{ \pi \sqrt{t}} e^{ t \ln \left[ \cos \left( \frac{ k }{ \sqrt t } \right) \right] } R_{ \frac{ k }{ \sqrt t }} (n) L_{ \frac{ k }{ \sqrt t }} (n_0)
\nonumber \\
&& \opsimeq_{t \to + \infty}
 \int_{0}^{ +\infty} \frac{dk}{  \pi \sqrt{t}} e^{ - \frac{ k^2 }{ 2 }  } 
\bigg(  2 \cos \left[ \frac{ k n}{ \sqrt t } \right]    -      2^{- n}    \bigg)  \cos \left[ \frac{ k n_0}{ \sqrt t } \right]  
\oppropto_{t \to + \infty}  \frac{1}{ \sqrt{t}}
  \label{spectralresetcriti}
\end{eqnarray}
to obtain the power-law decay $\frac{1}{ \sqrt{t}} $  that is standard for the non-biased random walk 
with scaling $ n \propto \sqrt t$. In addition, the appearance of the cosines $\cos \left[ \frac{ k n}{ \sqrt t } \right] $ 
and $\cos \left[ \frac{ k n_0}{ \sqrt t } \right]$ in the rescaled right and left eigenvectors 
show that the resetting procedure from the origin towards 
the exponential distribution $2^{-{\tilde n}-1}$
becomes asymptotically equivalent at large time $t$ and large distance $n$ to the reflecting boundary at the origin,
in agreement with intuition.

%%%%%%%%%%%%%%%%%%%%%%%%%%%%%%%%%%%%%%%%%

\section{ Markov trajectories $n(0 \leq t \leq T)$ in terms of excursions between reset events  }

\label{sec_excursions}

In this section, we describe how the Markov trajectories $n(0 \leq t \leq T)$
 generated by the Markov matrix $M$
can be decomposed in terms of the reset events occurring during the time-window $[0,T]$
and in terms of the excursions between them.

\subsection{ Decomposition of the Markov matrix $M$ into two contributions }

The Markov matrix $M$ of Eq. \ref{MarkovMatrixM} can be decomposed into two contributions
\begin{eqnarray}
M({\tilde n} \vert n)  =M_{\mathrm {Abs}}({\tilde n} \vert n) +M_{\mathrm {Reset}}({\tilde n} \vert n) 
\label{MarkovMatrixM2}
\end{eqnarray}
where
\begin{eqnarray}
M_{\mathrm {Reset}}({\tilde n} \vert n)  \equiv p 2^{-{\tilde n}-1} \delta_{n,0}
\label{MarkovMatrixMeset}
\end{eqnarray}
represents the contribution of the reset events from $n=0$ towards all the other sites ${\tilde n} \in \{0,1,..,+\infty\}$
while
\begin{eqnarray}
M_{\mathrm {Abs}}({\tilde n} \vert n)  = 
p   \delta_{n,{\tilde n}+1} \theta( n \geq 1 )+ (1-p)   \delta_{n,{\tilde n}-1}  \theta( {\tilde n} \geq 1 )
\label{MarkovMatrixMabs}
\end{eqnarray}
can be reinterpreted as a biased random walk on the half-line $n \in \{0,1,..,+\infty\}$
with an absorbing boundary condition at $(-1)$ where the particle jumps with probability $p$ when at the site $n=0$,
whose properties are recalled in Appendix \ref{app_RW}.

%%%%%%%%%%%%%%%%%%%%%%%%%%%%%%%%%%%%%%%%%%%%%%%

\subsection{ Decomposition of the trajectories $n(0 \leq t \leq T)$ in terms of the reset events and excursions between them }

Plugging the decomposition of Eq. \ref{MarkovMatrixM2} for the Markov matrix $M$ 
into the probabilities of the trajectories $n(0 \leq t \leq T)$ when the initial condition $n(0)$ 
is drawn with the probability $2^{-n(0)-1} $ corresponding to the uniform density $ \rho^{[\mathrm {Uniform}]}(x_0)  =1$
of Eq. \ref{UniformBinaryDensity}
\begin{eqnarray}
P^{\mathrm {Traj}} [n(0 \leq t \leq T) ] && =2^{-n(0)-1} \prod_{t=1}^T M(n(t) \vert n(t-1) ) 
= 2^{-n(0)-1} \prod_{t=1}^T \left[ M_{\mathrm {Abs}}(n(t) \vert n(t-1) ) + M_{\mathrm {Reset}}(n(t) \vert n(t-1) )\right]
\nonumber \\
&& = 2^{-n(0)-1} \sum_{K=0}^T \sum_{0<T_1<T_2<..<T_K \leq T} 
\left( \prod_{j=1}^K 
\left[ M_{\mathrm {Reset}}(n(T_j) \vert n(T_j-1) ) \prod_{t_j=1+T_{j-1}}^{T_j-1} M_{\mathrm {Abs}}(n(t_j) \vert n(t_j-1) ) \right] \right)
\nonumber \\
&& \times  \left[ \prod_{t_{K+1}=1+T_K}^{T} M_{\mathrm {Abs}}(n(t_{K+1}) \vert n(t_{K+1}-1) ) \right] 
\nonumber \\
&& = 2^{-n(0)-1} \sum_{K=0}^T \sum_{0<T_1<T_2<..<T_K \leq T} 
\left( \prod_{j=1}^K 
\left[  2^{-{n(T_j)}-1} p \delta_{n(T_j-1),0}
 \prod_{t_j=1+T_{j-1}}^{T_j-1} M_{\mathrm {Abs}}(n(t_j) \vert n(t_j-1) ) \right] \right)
\nonumber \\
&& \times  \left[ \prod_{t_{K+1}=1+T_K}^{T} M_{\mathrm {Abs}}(n(t_{K+1}) \vert n(t_{K+1}-1) ) \right] 
\label{Ptraj}
\end{eqnarray}
leads to the decomposition in terms of the number $K \in \{0,1,..,T\}$ of reset events at times 
$T_0 \equiv 0<T_1<T_2<..<T_K \leq T $
with the corresponding positions
\begin{eqnarray}
n(T_j) && = n_j \ \ \ \text{drawn with the reset probability $2^{-1-n_j}$ }
\nonumber \\
n(T_j-1) && =0 \text{ last position of the excursion starting at $n(T_{j-1})=n_{j-1}$ }
\label{spacetimereset}
\end{eqnarray}

The interpretation of Eq. \ref{Ptraj} is the following:

(1) The particle starts at $n(0)$ drawn with $2^{-1-n(0)}$ and evolves with the matrix $M_{\mathrm {Abs}} $
up to the time $(T_1-1)$ where the position vanishes $n(T_1-1) =0$ and where the particle 
jumps towards the absorbing site $n=-1$ at  at time $T_1$ in the language of the Appendix \ref{app_RW}, 
while here this absorption is replaced by a reset
at position $n(T_1)=n_1$ drawn with $2^{-1-n_1}$.

(2) After the reset event towards the position $n(T_1)=n_1$, the particle evolves with the matrix $M_{\mathrm {Abs}} $
up to the time $(T_2-1)$ where the position vanishes $n(T_2-1) =0$ and where the particle 
jumps towards the absorbing site $n=-1$ at  at time $T_2$ in the language of the Appendix \ref{app_RW}, 
while here the absorption is replaced by a reset
at position $n(T_2)=n_2$ drawn with $2^{-1-n_2}$.

This decomposition of the trajectory into reset events and into excursions between them governed by the matrix $M_{\mathrm {Abs}} $
continues up to the last reset event at position $n(T_K)=n_K$: 
then the particle evolves with the matrix $M_{\mathrm {Abs}} $
up to the final time $T$ at some position $n(T)$.

%%%%%%%%%%%%%%%%%%%%%%%%%%%

\subsection{ Statistical properties of the reset events  }

\subsubsection{ Joint probability distribution of the times $T_j$ and of the positions $n_j$ of the reset events }

If one keeps only the information on the number $K$ of reset events,
on the corresponding resetting times $T_j$ and on the corresponding reset positions $n(T_j)=n_j$ for $j=0,1,..,K$, 
the joint probability of these variables
corresponds to the integration of Eq. \ref{Ptraj} over the positions $n(t)$ at times $t \ne (T_1,..,T_K)$,
i.e. over the positions during the excursions between the reset events
\begin{eqnarray}
P_T^{\mathrm {Resets}} [K; (T_.,n_.) ] && = 
\left[ \sum_{n(T)=0}^{+\infty} \langle n(T) \vert M_{\mathrm {Abs}}^{T-T_K} \vert n_K \rangle 2^{-n_K-1} \right]
\left[ \prod_{j=0}^{K-1} p  \langle 0 \vert M_{\mathrm {Abs}}^{T_{j+1}-1-T_j} \vert n_j \rangle 2^{-n_j-1} \right] 
\label{Probaresetjointtn}
\end{eqnarray}
with the following interpretation of the building blocks:

(i) the term $j \in \{0,..,K-1\}$ involves the reset probability distribution $2^{-n_j-1} $ of the position $n_j=n(T_j)$
and the probability $A(t_j \vert n_j ) $ of the absorption time $t_j=T_{j+1}-T_j$ when starting at position $n_j$
\begin{eqnarray}
 p  \langle 0 \vert M_{\mathrm {Abs}}^{T_{j+1}-1-T_j} \vert n_j \rangle  
 = p P^{Abs}_{T_{j+1}-1-T_j}(0 \vert n_j )
 = A(T_{j+1}-T_j \vert n_j )
\label{traductionAn}
\end{eqnarray}
whose properties are described in detail in Appendix \ref{app_RW} starting at Eq. \ref{probaAbst}.

(ii) the remaining term involving the sum over the final position $n(T)$ at time $T$
of the propagator associated to the matrix $M_{\mathrm {Abs}}$ 
and starting at the position $n_K$ at $T_K$ of the last reset event 
\begin{eqnarray}
\sum_{n(T)=0}^{+\infty} \langle n(T) \vert M_{\mathrm {Abs}}^{T-T_K} \vert n_K \rangle 
= \sum_{n(T)=0}^{+\infty}P^{Abs}_{T-T_K}(n(T) \vert n_K ) = S(T-T_K \vert n_K)
\label{traductionslastexcurstion}
\end{eqnarray}
corresponds to the survival probability  introduced in Eq. \ref{SurvivalPtimagesevol} of Appendix \ref{app_RW}.

With these notations, the joint probability distribution of Eq. \ref{Probaresetjointtn} reads
\begin{eqnarray}
P_T^{\mathrm {Resets}} [K; (T_.,n_.)  ]  = S(T-T_K \vert n_K)
\left[ \prod_{j=0}^{K-1}  A(T_{j+1}-T_j \vert n_j ) 2^{-n_j-1} \right] 
\label{Probaresetjointtnfinal}
\end{eqnarray}

%%%%%%%%%%%%%%%%%%%%%%%%%%%

\subsubsection{ Joint probability distribution of the times $T_j$  of the reset events }

If one sums Eq. \ref{Probaresetjointtnfinal}
 over the reset positions $n_j$,
one obtains that the probability to see $K$ reset events at times $(T_1,..,T_K)$ during the time-window $[0,T]$
\begin{eqnarray}
P^{\mathrm {Resets}}_T [K; (T_1,..,T_K)  ] && = \left[ \prod_{j=0}^{K-1} \left( \sum_{n_j=0}^{+\infty} \right) \right] \sum_{n(T)=0}^{+\infty}  
P_T^{\mathrm {Resets}} [K; (T_.,n_.)  ]  
\nonumber \\ 
&& = S(T-T_K)
\left[ \prod_{j=0}^{K-1} A(T_{j+1}-T_j  )  \right] 
\label{Probareset}
\end{eqnarray}
involves the absorption probability $A(t ) $ and the survival probability $S(t)$
introduced in Eqs \ref{Abdistrireset} and \ref{survivalprobareset} of Appendix \ref{app_RW}.

%%%%%%%%%%%%%%%%%%%%%%%%%%%

\subsubsection{ Probability distribution of the number $K$ of reset events }

Finally, the sum of Eq. \ref{Probareset}
over the reset times $0<T_1<T_2<..<T_K \leq T$ gives the probability to see $K$ reset events
 during the time-window $[0,T]$
\begin{eqnarray}
P^{\mathrm {Resets}}_T [K ] && \equiv \sum_{0<T_1<T_2<..<T_K \leq T}P^{\mathrm {Resets}}_T [K; (T_1,..,T_K)  ] 
\nonumber \\ 
&& = \sum_{0<T_1<T_2<..<T_K \leq T} S(T-T_K)
\left[ \prod_{j=0}^{K-1} A(T_{j+1}-T_j  )  \right] 
\label{Probaresetnumber}
\end{eqnarray}
In particular, the probability to see $K=0$ reset events reduces to the survival probability $S(T)$
whose asymptotic decay for large $T$ has been discussed in Eqs \ref{Survivalinfinityreset},
\ref{survivalresetchaos} and
\ref{survivalresetcriti}
 of Appendix \ref{app_RW}
\begin{eqnarray}
P^{\mathrm {Resets}}_T [K=0 ]   = S(T)
\opsimeq_{T \to + \infty}   \begin{cases}
S( \infty) =   \frac{ 2(1-2p)  }{ 2 - 3 p } >0  \ \ \ \ \ \ \ \ \ \ \ \ \ \ \ \ \ \ \ \ \ \ \ \ \ \ \ \ \ \ \ \ \ \ \ \  \text{for } 0 \leq p < \frac{1}{2}  
 \\ 
4 \sqrt{\frac{2}{\pi T} } \ \ \ \ \ \ \ \ \ \ \ \  \ \ \ \ \ \ \ \ \ \ \ \ \ \ \ \ \ \ \ \ \ \ \ \ \ \ \ \ \ \ \ \ \ \ \ \  \ \ \ \ \ \ \ \   \text{for } p_c=\frac{1}{2}
 \\
 \frac{p 8 \sqrt{2}    }{ \sqrt \pi\left( 2- \sqrt{ \frac{p}{1-p} } \right)^2 \left[ \ln \left(\frac{1}{4p(1-p) } \right) \right] T^{3/2}}
 e^{-T \frac{ \ln \left(\frac{1}{4p(1-p) } \right) }{2} }  \ \ \ \ \ \ \ \ \ \  \text{for } \frac{1}{2} < p \leq 1 
\end{cases}
\label{Probaresetnumberzero}
\end{eqnarray}
i.e. it remains finite in the non-chaotic region $0 \leq p < \frac{1}{2} $,
it decays exponentially in time in the chaotic region $\frac{1}{2} < p \leq 1 $,
and it decays only with the power-law $1/\sqrt{T} $ at the critical point $p_c=\frac{1}{2} $.

In order to analyze the probabilities of other values of $K \ne 0$, 
it is convenient to introduce the average of Eq. \ref{Probaresetnumber} 
over the normalized distribution $(1-e^{-s} ) e^{-s T} $ for the time $T=0,1,2,..+\infty$ 
to obtain the following simple explicit result 
in terms of the Laplace transforms ${\tilde A}(s ) $ and $ {\tilde S}(s )$ 
of Eqs \ref{Absinireset} and \ref{survivalLaplacereset}
\begin{eqnarray}
{\tilde P}^{\mathrm {Resets}}_s [K ] && \equiv \sum_{T=0}^{+\infty} (1-e^{-s} ) e^{-s T} P^{\mathrm {Resets}}_T [K ] = 
(1-e^{-s} )  {\tilde S}(s ) \big[{\tilde A}(s ) \big]^K 
\nonumber \\
&& = \left( 1- {\tilde A}(s ) \right) \big[{\tilde A}(s ) \big]^K
= 2 [r_+(s)-1] \bigg( 2 r_+(s)-1\bigg)^{-K-1}
\label{ProbaresetnumberLaplace}
\end{eqnarray}

In particular, one can use the result of Eq. \ref{probaAbstnormareset} concerning ${\tilde A}(s=0 ) $
and the following discussion to obtain the properties of the number of resets in the limit $T \to + \infty$:

(i) in the non-chaotic region $ 0 \leq p < \frac{1}{2} $ where the forever-survival probability of Eq. \ref{Survivalinfinityreset}
is strictly positive 
\begin{eqnarray}
S( \infty) \equiv 1- {\tilde A}(s=0 ) =   \frac{ 2(1-2p)  }{ 2 - 3 p } >0
\label{Survivalinfinityresettext}
\end{eqnarray}
one obtains that in the limit $T \to + \infty$, the number $K$ of reset events remains a finite random variable 
distributed with the geometric distribution 
\begin{eqnarray}
P^{\mathrm {Resets}}_{T=\infty} [K ] = {\tilde P}^{\mathrm {Resets}}_{s=0} [K ]  = \left( 1- {\tilde A}(0 ) \right) \big[{\tilde A}(0 ) \big]^K
= S( \infty) \big[ 1- S( \infty)\big]^K\ \ \text{for } \ \ K=0,1,2,...,+\infty
\label{ProbaresetnumberLaplacesofinite}
\end{eqnarray}

(ii) in the chaotic region $ \frac{1}{2} < p \leq 1$ where
the forever-survival probability vanishes $S( \infty) =0$, and where the normalized probability distribution $A(t ) $
has the finite moment given by Eq. \ref{MFABsreset}
\begin{eqnarray}
\langle t \rangle \equiv \sum_{t=1}^{+\infty} t A(t ) = \frac{ 2}{ 2p-1}  
\label{MFABsresettext}
\end{eqnarray}
then the leading behavior of Eq. \ref{ProbaresetnumberLaplace} for $s \to 0$ reads
\begin{eqnarray}
{\tilde P}^{\mathrm {Resets}}_{s} [K ] && \opsimeq_{s \to 0} \left( s \langle t \rangle +O(s^2) \right) \big[1- s \langle t \rangle +O(s^2)\big]^K
% \nonumber \\ &&  
\opsimeq_{s \to 0}  s \langle t \rangle  e^{ - K s \langle t \rangle}
\label{ProbaresetnumberLaplaces0}
\end{eqnarray}
This means that the number $K(T)$ of resets during the time-window $[0,T]$ grows extensively in $T$ for large time $T \to + \infty$
\begin{eqnarray}
K(T) \opsimeq_{T \to + \infty} \frac{T}{\langle t \rangle} =\left( p - \frac{ 1} { 2} \right) T
\label{Kextensive}
\end{eqnarray}

(iii) at the critical point $p_c=\frac{1}{2}$, the forever survival probability of Eq. \ref{Survivalinfinityresettext}
vanishes $S(\infty) =0$, 
but the first moment of Eq. \ref{MFABsresettext} of the normalized probability distribution $A(t ) $ diverges as a consequence of the power-law decay $t^{-3/2}$ of $A(t)$ of Eq. \ref{stirlingcritireset}.
The leading behavior of Eq. \ref{ProbaresetnumberLaplace} for $s \to 0$ using Eq. \ref{Absiniresetcriti}
\begin{eqnarray}
{\tilde P}^{\mathrm {Resets}}_{s} [K ] && \opsimeq_{s \to 0}  \left( 2 \sqrt{2s} +O(s) \right) \big[1-2 \sqrt{2s} +O(s) \big]^K
 \opsimeq_{s \to 0}   2 \sqrt{2s}  e^{- K 2 \sqrt{2s} }
\label{ProbaresetnumberLaplaces0criti}
\end{eqnarray}
means that for large time $T \to + \infty$, the probability distribution $P^{\mathrm {Resets}}_{T} [K ]  $ 
is governed by the following scaling form in terms of the appropriate rescaled variable $k=\frac{K}{ \sqrt{T}} $ 
\begin{eqnarray}
P^{\mathrm {Resets}}_T [K ]  \opsimeq_{T \to + \infty}  \frac{1}{\sqrt T} \Upsilon \left( \frac{K}{ \sqrt{T}}\right)
\label{Probaresetnumbercriti}
\end{eqnarray}
In conclusion, the number $K(T)$ of resets during the time-window $[0,T]$ grows as $\sqrt{T}$
\begin{eqnarray}
K(T) \oppropto_{T \to + \infty} \sqrt{T}
\label{Ksubextensive}
\end{eqnarray}
in contrast to the extensive growth of Eq. \ref{Kextensive} for $p>1/2$ 
and in contrast to the finite values of Eq. \ref{ProbaresetnumberLaplacesofinite} for $p<1/2$.

%%%%%%%%%%%%%%%%%%%%%%%%%%%%%%%%%%%%%%%%%%%%%%

\subsection{ Link with the finite-time propagator $\pi_T(n \vert n_0)$ studied in the previous section \ref{sec_propagator}  }

The probability $\pi_T(n \vert n_0) $ to be at $n(T)=n$ at time $T$ when starting at $n(0)=n_0$ 
that was studied in section \ref{sec_propagator}
can be reproduced from the summation of Eq. \ref{Ptraj} over the other variables
in order to obtain the following decomposition with respect to the number $K$ of reset events during the time-window $[0,T]$
\begin{eqnarray}
\pi_T(n \vert n_0) = P_T^{Abs}(n \vert n_0)
+ \sum_{K=1}^T \sum_{0<T_1<T_2<..<T_K \leq T}
P_{T-T_K}^{Abs}(n ) \left[ \prod_{j=1}^{K-1} A(T_{j+1}-T_j) \right] A(T_1 \vert n_0)
\label{PropagatorResumresetEvents}
\end{eqnarray}
where the first term $P_T^{Abs}(n \vert n_0) $ is the contribution of 
trajectories with $K=0$ reset events.

The time-Laplace-transform of Eq. \ref{PropagatorResumresetEvents} reduces to
\begin{eqnarray}
{\tilde \pi}_s (n \vert n_0)\equiv \sum_{T=0}^{+\infty} e^{-s T} \pi_T(n \vert n_0) 
&& = {\tilde P}_s^{Abs}(n \vert n_0)
+ \sum_{K=1}^{+\infty} 
{\tilde P}_s^{Abs}(n ) \left[ {\tilde A}(s) \right]^{K-1} {\tilde A}(s \vert n_0)
\nonumber \\
&& = {\tilde P}_s^{Abs}(n \vert n_0)
+ \frac{ {\tilde P}_s^{Abs}(n )  {\tilde A}(s \vert n_0) }
{ 1- {\tilde A}(s)}
\label{PropagatorResumresetEventsLaplace}
\end{eqnarray}
where one can plug 
${\tilde P}^{Abs}_s(n \vert n_0)$ of \ref{sPabspm},
${\tilde A}(s \vert n_0) $ of Eq. \ref{probaAbstLaplace},
${\tilde P}_s^{Abs}(n )  $ of Eq. \ref{pabsreset},
and ${\tilde A}(s ) $ of Eq. \ref{Absinireset}
to obtain
\begin{eqnarray}
\text{for }  n \geq n_0 : \ \ {\tilde \pi}_s (n \vert n_0)
&& =\frac{   [ r_+(s)+r_-(s)]  }{    [r_+(s)- r_-(s)]} 
[r_-(s)]^{n+1} \bigg( [r_-(s)]^{-n_0-1}- [ r_+(s) ]^{-n_0-1} \bigg)
\nonumber \\
&&+ \frac{   [ r_+(s)+r_-(s)]  }
{    [r_+(s)-1][1- 2 r_-(s)]}
\left( 2^{-1-n}  - [r_-(s)]^{n+1}\right)   [ r_+(s)]^{-n_0-1}
\nonumber \\
\text{for } 0 \leq n \leq n_0 : \ \ {\tilde \pi}_s (n \vert n_0)
&& = \frac{   [ r_+(s)+r_-(s)]  }{    [r_+(s)- r_-(s)]}  [r_+(s)]^{-n_0-1} \bigg( [r_+(s)]^{n+1}- [ r_-(s) ]^{n+1} \bigg)
+\nonumber \\
&&+ \frac{   [ r_+(s)+r_-(s)]  }
{    [r_+(s)-1][1- 2 r_-(s)]}
\left( 2^{-1-n}  - [r_-(s)]^{n+1}\right)   [ r_+(s)]^{-n_0-1}
\label{PropagatorResumresetEventsLaplaceCalcul}
\end{eqnarray}
in agreement with Eq. \ref{recap} of the previous section.

So Eqs \ref{PropagatorResumresetEvents} and \ref{PropagatorResumresetEventsLaplace} give a simple interpretation 
of the propagator $\pi_T(n \vert n_0)$ in terms of the reset events and excursions between them.

%%%%%%%%%%%%%%%%%%%%%%%%%%%%%%%%%%%%%%%%%%%%%%

\subsection{ Link with the steady state $\pi_*(n)$ in the chaotic region $\frac{1}{2}<p\leq 1$ }

In the chaotic region $\frac{1}{2} < p \leq 1$ where the number $K$ of resets grows extensively in $T$ as Eq. \ref{Kextensive}, the steady state $\pi_*(n)$ can be recovered
if one divides the total occupation time at position $n$ before absorption of Eq. \ref{pabsresetzero}
by the average duration of an excursion given in Eq. \ref{MFABsreset}
\begin{eqnarray}
\frac{ \displaystyle \sum_{t=0}^{+\infty} P^{Abs}_t(n )   }{ \displaystyle \sum_{t=1}^{+\infty} t A(t ) }
  =     \frac{2p-1}{3p-2} \left[2^{-1-n}  - \left( \frac{1-p}{p}\right)^{n+1} \right] = \pi_*(n)
\label{steadyfromExcursions}
\end{eqnarray}
This gives a simple interpretation of the steady state in terms of the excursions between reset events.

%%%%%%%%%%%%%%%%%%%%%%%%%%%%%%%%%

\section{ Pelikan dynamics for the binary decomposition $x_t =\displaystyle \sum_{l=1}^{+\infty} \frac{\sigma_l (t)}{2^l} $  }

\label{sec_spins}

After focusing on the closed dynamics for the special subspace of probability densities $\rho^{[\mathrm {Partition}]}_t(x) $ 
of Eq. \ref{BinaryDensity} in the previous sections, we now return to
 the general case of an arbitrary single initial condition $x_0$
or an arbitrary distribution $\rho_{t=0}(x_0)$ of initial conditions $x_0$.

%%%%%%%%%%%%%%%%%%%%%%%%%%%%%%%%%%%%%%%%%%%%%%%%%%

\subsection{ Pelikan dynamics for the spins $\sigma_l \in \{0,1\}$ parametrizing
the binary decomposition $x =\displaystyle \sum_{l=1}^{+\infty} \frac{\sigma_l }{2^l} $ }

\label{subsec_spins}

When $x_t \in [0,1[$ is represented by its binary coefficients $\sigma_l(t)=0,1$ with $l=1,2,..+\infty$
\begin{eqnarray}
x_t  = \sum_{l=1}^{+\infty} \frac{\sigma_l(t)}{2^l} =  \frac{\sigma_1(t)}{2}+ \frac{\sigma_2(t)}{4}+ \frac{\sigma_3(t)}{8}+...
\label{binary}
\end{eqnarray}
the Pelikan dynamics of Eq. \ref{pelikan}
translates into the following global shift-dynamics to the right or to the left for the half-infinite spin chain 
$[\sigma_1(t),\sigma_2(t),...]$
\begin{eqnarray}
{\tilde \sigma }_l(t+1) =
\begin{cases}
\sigma_{l+1}(t)  \ \ \ \ \ \ \ \ \ \ \ \ \ \ \ \ \ \ \ \ \ \ \ \text{ with probability $p$ }  
 \\
\sigma_{l-1}(t) \theta(l \geq 2) \ \ \ \ \ \ \ \ \ \text{ with probability $(1-p)$ }
\end{cases}
\label{binaryimage}
\end{eqnarray}
where it is important to stress what happens to the first spin $l=1$ 
\begin{eqnarray}
\text{ with probability $p$ } : && \text{ the value $\sigma_1(t)$ is erased forever and the new first spin is ${\tilde \sigma }_1(t+1)=\sigma_2(t) $} 
\nonumber \\
\text{ with probability $(1-p)$ } : && \text{ a zero is injected ${\tilde \sigma }_1(t+1)=0 $ and the new second spin is  ${\tilde \sigma }_2(t+1)=\sigma_1(t) $} 
\label{FirstSpin}
\end{eqnarray}
In summary, the real-space kernel $w(. \vert .) $ of Eq. \ref{Wpelikan} translates for the spins
into the kernel
\begin{eqnarray}
{\cal W}( {\tilde \sigma }_. \vert \sigma_.)  \equiv 
p \prod_{l=1}^{+\infty} \delta_{{\tilde \sigma }_l,\sigma_{l+1}}
  +(1-p) \delta_{{\tilde \sigma }_1,0} \prod_{l=2}^{+\infty} \delta_{{\tilde \sigma }_l,\sigma_{l-1}}
\label{Wpelikansigma}
\end{eqnarray}

The correspondence between the probability density $\rho_t(x)$ on the interval $[0,1[$ 
and the probability ${\cal P}_t(   \sigma_.) $ of the binary variables reads
\begin{eqnarray}
\rho_t(x)   
= \sum_{\sigma_.}  {\cal P}_t(  \sigma_.)  \delta \left( x-\displaystyle \sum_{l=1}^{+\infty} \frac{\sigma_l}{2^l}\right)
\label{rhoxPsigma}
\end{eqnarray}
In particular, the uniform density $\rho^{[\mathrm {Uniform}]}(x)=1$ of Eq. \ref{UniformBinaryDensity} corresponds to the factorized distribution for the spins,
where each spin can take the two values $\sigma_l=0,1$ with the equal probabilities $(1/2,1/2)$
\begin{eqnarray}
  {\cal P}^{[\mathrm {Uniform}]}(\sigma_.)  
 =  \prod_{l=1}^{+\infty} \left( \frac{\delta_{ \sigma _l,0} +  \delta_{\sigma _l,1} }{2} \right)
\label{UniformDensitysigma}
\end{eqnarray}
while the special subspace $\rho^{[\mathrm {Partition}]}_t(x) $ of Eq. \ref{BinaryDensity} 
studied in the previous sections
translates into
\begin{eqnarray}
  {\cal P}^{[\mathrm {Partition}]}_t(\sigma_.) && 
 = \sum_{n=0}^{+\infty}  2^{n+1} \pi_t(n) \left[ \prod_{m=1}^n \delta_{\sigma_m,0} \right]\delta_{\sigma_{n+1},1}\prod_{l=n+2}^{+\infty} \left( \frac{\delta_{ \sigma _l,0} +  \delta_{\sigma _l,1} }{2} \right)
\label{BinaryDensitysigma}
\end{eqnarray}

%%%%%%%%%%%%%%%%%%%%%%%%%%%%%%%%%%%%%%%

\subsection{ Pelikan dynamics in terms of  two global variables $z_t \in \{0,1,2,..,+\infty\}$  and $F_t \in \{0,1,2,..,+\infty\}$ }

\label{subsec_zF}

Since the dynamics of Eq. \ref{binaryimage} involves only global shifts of the spin chain,
the state at time $t$ of Eq. \ref{binary}
can be summarized by two global variables $z_t \in \{0,1,2..,+\infty\}$  and $F_t \in \{0,1,2,..,+\infty\}$ 
that are sufficient to reconstruct all the spins $\sigma_l(t)$ via
\begin{eqnarray}
\sigma_l(t) && = 0 \ \ \ \ \ \ \ \ \ \ \ \ \ \ \ \ \ \ \text{ for } \ \ l =1,2,..,z_t
\nonumber \\
\sigma_{l=z_t+m} (t) && =   \sigma_{ F_t+m} (0)   \ \ \ \ \ \ \ \ \text{ for } \ \ m =1,2,..,+\infty
\label{binarydigmazf}
\end{eqnarray}
corresponding to
\begin{eqnarray}
x_t  = \sum_{l=1}^{+\infty} \frac{\sigma_l(t)}{2^l} 
=  \sum_{l=1}^{z_t} \frac{0}{2^l} + \sum_{l=z_t+1}^{+\infty} \frac{ \sigma_{l -z_t + F_t} (0) }{2^l}
= \sum_{m=0}^{+\infty}  \frac{ \sigma_{ F_t +m} (0) }{2^{z_t+m} }
\label{binaryzf}
\end{eqnarray}
with the following interpretation based on the dynamics of Eq. \ref{FirstSpin} concerning the first spin $l=1$:

$\bullet$ the variable  $z_t \in \{0,1,2,..,t \}$ represents the number of zeros at time $t$ that remain from 
the previous injections of zeros during the dynamics of the previous $t$ time-steps;

$\bullet$ the variable $F_t \in \{0,1,2,..,t \}$ represents the number of spins of the initial condition at $t=0$
that have been erased during the dynamics of the previous $t$ time-steps, so that the initial spins $  \sigma_l(0)$
that are still surviving at time $t$ have indices $l \geq F_t+1$.

The two global variables $(z_t,F_t)$ evolves with the following dynamics during one time-step:

(i) with probability $(1-p)$, the injection of a new zero of Eq. \ref{FirstSpin}
translates into the growth $z_{t+1}=z_t+1$ 
and into the stable value $F_{t+1}=F_t$ ;

(ii) with probability $p$, the erasure of the first spin $\sigma_{l=1}(t)$ of Eq. \ref{FirstSpin} can have two effects 
depending on the number $z_t$ of zeroes:

(ii-a) if $z_t \geq 1$, the erasure of the first zero translates into the decay $z_{t+1}=z_t-1$ 
and into the stable value $F_{t+1}=F_t$;

(ii-b) if $z_t=0$, the erasure of the first spin $\sigma_{l=1}(t)$ corresponds to the erasure of the spin $\sigma_{F_t+1}(0)$ of the initial condition at $t=0$, so this increases by one the number of erased spins $F_{t+1}=F_t+1$ while 
the number of zeros remains $z_{t+1}=0$.

The Markov matrix summarizing the dynamics of these two variables 
\begin{eqnarray}
W( {\tilde z } , {\tilde F } \vert z,F )  
= (1-p) \delta_{{\tilde z }, z+1}  \delta_{{\tilde F },F}
+ p \theta(z \geq 1)  \delta_{{\tilde z }, z-1}  \delta_{{\tilde F },F} + p \delta_{z,0}\delta_{{\tilde z }, 0}  \delta_{{\tilde F },F+1}
\label{Wzf}
\end{eqnarray}
governs the evolution of their joint probability distribution $P_t(z,F) $
\begin{eqnarray}
P_{t+1}( {\tilde z } , {\tilde F }) && = \sum_z \sum_F W( {\tilde z } , {\tilde f } \vert z,F )  P_t(z,F)
\nonumber \\
&& = (1-p) P_t( {\tilde z }-1 , {\tilde F })  \theta({\tilde z } \geq 1)
+ p P_t( {\tilde z }+1 , {\tilde F}) + p \delta_{{\tilde z }, 0} P_t( 0 , {\tilde F }-1)
\label{evolPzf}
\end{eqnarray}
while the initial condition at $t=0$ corresponds to $z_{t=0}=0$ et $F_{t=0}=0$
 \begin{eqnarray}
P_{t=0}  (z,F)= \delta_{z,0} \delta_{F,0}
\label{evolPzfini}
\end{eqnarray}
The propagator for the spins $\sigma_{l=1,2,..,+\infty}$ can be rewritten in terms of the joint distribution $P_t(z,F)$ 
\begin{eqnarray}
  {\cal P}_t \left[ \sigma_. \vert \sigma_.(0) \right]   = \sum_{z=0}^{+\infty}  \sum_{F=0}^{+\infty} P_t(z,F)
   \left[ \prod_{l=1}^{z_t} \delta_{\sigma_l,0} \right] 
   \left[\prod_{m=1}^{+\infty}  \delta_{\sigma _{z_t+m},\sigma_{F_t+m}(0)}  \right]
\label{Ptsigmafromzf}
\end{eqnarray}
to analyze the Pelikan dynamics starting from an arbitrary initial condition $x_0$ parametrized by its binary coefficients
$\sigma_l(t=0)$, or from an arbitrary initial probability distribution $\rho_{t=0}(x_0)$
after its translation into the probability distribution ${\cal P}_{t=0} \left[ \sigma_.(0) \right] $ for the binary coefficients
\begin{eqnarray}
{\cal P}_t \left[ \sigma_.  \right] && = \sum_{\sigma_.(0) }  {\cal P}_t \left[ \sigma_. \vert \sigma_.(0) \right]   {\cal P}_{t=0} \left[ \sigma_.(0) \right]
 \nonumber \\
 && = \sum_{z=0}^{+\infty}  \sum_{F=0}^{+\infty} P_t(z,F) \left[ \prod_{l=1}^{z_t} \delta_{\sigma_l,0} \right] 
\left(  \sum_{\sigma_.(0) } {\cal P}_{t=0} \left[ \sigma_.(0) \right]
   \left[\prod_{m=1}^{+\infty}  \delta_{\sigma _{z_t+m},\sigma_{F_t+m}(0)}  \right] \right)
\label{Ptsigmafromzfconvol}
\end{eqnarray}

%%%%%%%%%%%%%%%%%%%%%%%
 
 \subsection{ Closed dynamics for the probability distribution $P_t(z)$ of $z $ alone with its consequences }
 
 \label{subsec_closedz}
 
 The probability distribution $P_t(z)$ of $z $ alone after summing the joint distribution $P_t(z,F) $ 
 over the variable $F$
\begin{eqnarray}
 P_t(z)  \equiv \sum_{F=0}^{+\infty} P_t(z,F)
\label{zalone}
\end{eqnarray}
satisfies the following closed dynamics using Eq. \ref{evolPzf}
\begin{eqnarray}
 P_{t+1}({\tilde z})  =  (1-p)   P_t({\tilde z}-1)  \theta( {\tilde z} \geq 1 )
 + p   P_t({\tilde z}+1) 
 + p \delta_{{\tilde z},0} P_t(0) \equiv \sum_{z=0}^{+\infty} W( {\tilde z }  \vert z )  P_t(z)
\label{dynz}
\end{eqnarray}
with the Markov matrix
\begin{eqnarray}
W( {\tilde z }  \vert z )  
= (1-p) \delta_{{\tilde z }, z+1} 
+ p \theta(z \geq 1)  \delta_{{\tilde z }, z-1}   + p \delta_{z,0}\delta_{{\tilde z }, 0}  
\label{Wz}
\end{eqnarray}
The comparison with the Markov matrix $M({\tilde n} \vert n) $ of Eq. \ref{MarkovMatrixM} 
shows that the only difference is that 
the resets from the origin $n=0$ towards $\tilde n$ distributed with $2^{-{\tilde n}-1}$
have been replaced by a 'reset' from the origin $z=0$ towards itself ${\tilde z}=0$,
i.e. by a stay at the origin.
As a consequence, the dynamics for the variable $z$ alone is even simpler than the dynamics for the variable $n$
that has been discussed in detail in the previous sections.

%%%%%%%%%%%%%%%%%%%%%%%%%%%%%%%%

  \subsubsection{ Steady state $P_*(z) $ in the chaotic region $\frac{1}{2} <p \leq 1 $ }
  
  In the chaotic region $\frac{1}{2} <p \leq 1 $, while the steady state $\pi_*(n) $ for the variable $n$ is out-of-equilibrium
  with non-vanishing steady currents as described in subsection \ref{subsec_noneqsteady},
   the steady state $P_*(z) $ for the variable $z$ is at equilibrium
  with vanishing steady current on each link flowing  
from ${ \tilde z} $ towards its left neighbor $ ({ \tilde z}-1)$
  \begin{eqnarray}
0 && = J_* ({ \tilde z}-1/2) \equiv p    P_*({ \tilde z})- (1- p)  P_*({ \tilde z-1})
\label{Jstarbondz}
\end{eqnarray}
so that the equilibrium steady state reduces to the following geometric distribution
\begin{eqnarray}
 P_*(z)  = \frac{2p-1}{p} \left( \frac{1-p}{p} \right)^z \ \ \ \text{ for } z=0,1,2,...
\label{steadyz}
\end{eqnarray}
On the contrary in the non-chaotic region $0 \leq p <\frac{1}{2} $, 
the number $z_t$ of zeroes will grow extensively in time $t$, 
while at the critical point $p_c=\frac{1}{2}$, the number $z_t$ of zeroes will scale as $\sqrt{t}$,
so that it is useful to consider the finite-time probability distribution $P_t(z) $ in the next subsection.

%%%%%%%%%%%%%%%%%%%%%%%%%%%%%%%%

  \subsubsection{ Finite-time probability distribution $P_t(z) =P_t(z \vert z_0=0)$ starting from the initial condition $z_0=0$  }

 From the comparison between the dynamics in $z$ and in $n$ discussed after Eq. \ref{dynz},
 one obtains that the time-Laplace-transform ${\tilde P}_s(z) $ of the finite-time probability distribution $P_t(z) =P_t(z \vert z_0=0)$
 \begin{eqnarray}
{\tilde P}_s(z) \equiv \sum_{t=0}^{+\infty} e^{-s t} P_t(z)
\label{pszLaplace}
\end{eqnarray}
satisfies the following simpler counterpart of Eq. \ref{PropagatorResumresetEventsLaplace}
\begin{eqnarray}
{\tilde P}_s (z ) 
&& = {\tilde P}_s^{Abs}(z \vert 0)
+ \sum_{K=1}^{+\infty} 
{\tilde P}_s^{Abs}(z \vert 0) \left[ {\tilde A}(s \vert 0) \right]^{K-1} {\tilde A}(s \vert 0)
= {\tilde P}_s^{Abs}(z \vert 0)
+ \frac{ {\tilde P}_s^{Abs}(z \vert 0)  {\tilde A}(s \vert 0) }
{ 1- {\tilde A}(s\vert 0)}
\nonumber \\
&& = \frac{ {\tilde P}_s^{Abs}(z \vert 0)  }
{ 1- {\tilde A}(s\vert 0)}
\label{PropagatorResumresetEventsLaplacezcalcul}
\end{eqnarray}
where one can plug 
${\tilde P}^{Abs}_s(z \vert n_0=0)$ of \ref{sPabspm},
and ${\tilde A}(s \vert n_0=0) $ of Eq. \ref{probaAbstLaplace}
to obtain the explicit time-Laplace-transform ${\tilde P}_s(z) $
\begin{eqnarray}
{\tilde P}_s (z ) 
  = \frac{ \frac{r_+(s)+r_-(s)}{ r_+(s)-r_-(s) } [r_-(s)]^{z+1} \bigg( [r_-(s)]^{-1}- [ r_+(s) ]^{-1} \bigg)  }
{ 1- [ r_+(s)]^{-1}}
  =  \frac{r_+(s)+r_-(s)}{ r_+(s)-1 } [r_-(s)]^z 
\label{PropagatorResumresetEventsLaplacez}
\end{eqnarray}
in terms of the two roots $r_{\pm}(s)$ of Eq. \ref{2dsolmain}.

The spectral decomposition of the propagator $P_t(z \vert z_0)$ is also much simpler than Eq. \ref{spectralreset}
for the propagator $\pi_t (n \vert  n_0) $ as a consequence of the similarity transformation
\begin{eqnarray}
P_t(z \vert z_0) = \left( \frac{1-p}{p} \right)^{\frac{z-z_0}{2}} \psi_t(z \vert z_0)
\label{similarity}
\end{eqnarray}
towards the propagator $\psi_t(z \vert z_0) $ satisfying the discrete-time Euclidean lattice quantum  dynamics obtained from Eq. \ref{dynz}
\begin{eqnarray}
\text{ for } z \geq 1 : \ \ \   
&& \psi_t(z \vert z_0)  
  =  \sqrt{p (1-p) } \left[ \psi_{t-1}(z-1 \vert z_0)   +  \psi_{t-1}(z+1 \vert z_0) \right] \equiv \sum_y \langle z \vert \Omega \vert y \rangle \psi_{t-1}(y \vert z_0)
 \nonumber \\
 \text{ for } z =0 : \ \ \   
 &&  \psi_t(0 \vert z_0)  
  =   \sqrt{p (1-p) } \psi_{t-1}(1 \vert z_0) + p  \psi_t(0 \vert z_0) \equiv \sum_y \langle 0 \vert \Omega \vert y \rangle \psi_{t-1}(y \vert z_0)
\label{dynzpsi}
\end{eqnarray}
that involves the symmetric matrix $\Omega$ with the off-diagonal elements
\begin{eqnarray}
\langle z \vert \Omega \vert z+1 \rangle = \langle z+1 \vert \Omega \vert z \rangle =  \sqrt{p (1-p) } \ \ \ \text{ for} \ z=0,1,2,..
\label{off-diag}
\end{eqnarray}
and the only diagonal element at the boundary site $z=0$
\begin{eqnarray}
\langle 0 \vert \Omega \vert 0 \rangle =p
\label{diag}
\end{eqnarray}
As a consequence, the spectral decomposition of the quantum propagator $\psi_t(z \vert z_0)$ reduces to
\begin{eqnarray}
\psi_t(z \vert z_0)  && = \langle z \vert \psi_* \rangle  \langle \psi_* \vert z_0 \rangle \theta\left( \frac{1}{2} <p \leq 1\right)
+ \int_{0}^{ \pi} \frac{dq}{ \pi} \bigg[ \lambda (q) \bigg]^t \langle z \vert \phi_q \rangle  \langle \phi_q \vert z_0 \rangle 
\nonumber \\
&& =  \psi_*(z) \psi_*(z_0) \theta\left( \frac{1}{2} <p \leq 1\right)
+ \int_{0}^{ \pi} \frac{dq}{ \pi} \bigg[ \lambda (q) \bigg]^t  \phi_q (z)  \overline{ \phi_q (z_0) }
  \label{spectralpsi}
\end{eqnarray}
where the normalizable quantum ground-state $\psi_*(z) $ associated to the eigenvalue $\lambda=1$ exists only in the chaotic region$\frac{1}{2} <p \leq 1$
in relation with Eq. \ref{steadyz}
\begin{eqnarray}
\psi_*(z)= \sqrt{ P_*(z) }  = \sqrt{ \frac{2p-1}{p} } \left( \frac{1-p}{p} \right)^{\frac{z}{2}} \ \ \ \text{ for } z=0,1,2,...
\label{steadyzpsigs}
\end{eqnarray}
while the continuous spectrum of eigenvalues $\lambda(q) \equiv \sqrt{ p (1-p) } (e^{iq}+e^{-iq} ) = \sqrt{ 4 p (1-p) } \cos q $
of Eq. \ref{lambdaqmomentum} is associated to eigenvectors 
corresponding to linear combinations of the two plane-waves $e^{\pm iq z}$
\begin{eqnarray}
\phi_q(z)= \frac{1}{\sqrt 2} \left[ e^{i q z } + \Gamma(q) e^{- i q z } \right]
\label{eigenphiq}
\end{eqnarray}
that satisfy the eigenvalue equation for $z \geq 1$
\begin{eqnarray}
\text{ for } z \geq 1 : \ \ \   
&& \lambda(q) \phi_q(z)  
  =  \sqrt{p (1-p) } \left[ \phi_q(z-1)   + \phi_q(z+1) \right] 
\label{eigenpsiz1}
\end{eqnarray}
and at the boundary $z=0$
\begin{eqnarray}
0 && = - \lambda(q) \phi_q(0)     +   \sqrt{p (1-p) }  \phi_q(1)  + p  \phi_q(0) 
= - \sqrt{p (1-p) }  \phi_q(-1)  + p  \phi_q(0)
\nonumber \\
&& = - \sqrt{p (1-p) }  \left[ e^{-i q  } + \Gamma(q) e^{ i q  }\right]
+ p  \left[1 + \Gamma(q)  \right]
  \label{eigenpsiz0}
\end{eqnarray}
that deternines the coefficient $\Gamma(q)$ 
\begin{eqnarray}
 \Gamma(q) = -  \frac{\sqrt{ \frac{1-p}{p} } e^{- iq}-1} {\sqrt{ \frac{1-p}{p} } e^{ iq}-1 }
  \label{eigenpsiz0sol}
\end{eqnarray}

Putting everything together, the spectral decomposition of the propagator $P_t(z \vert z_0)$ 
obtained from Eq. \ref{similarity} and Eq. \ref{spectralpsi}
reads
\begin{eqnarray}
P_t(z \vert z_0) && = \left( \frac{1-p}{p} \right)^{\frac{z-z_0}{2}} \psi_t(z \vert z_0)
% \nonumber \\ &&
 = \left( \frac{1-p}{p} \right)^{\frac{z-z_0}{2}} \left[\psi_*(z) \psi_*(z_0) \theta\left( \frac{1}{2} <p \leq 1\right)
+ \int_{0}^{ \pi} \frac{dq}{ \pi} \bigg[ \lambda (q) \bigg]^t  \phi_q (z)  \overline{ \phi_q (z_0) } \right]
\nonumber \\
&& =P_*(z)  \theta\left( \frac{1}{2} <p \leq 1\right)
+ \left( \frac{1-p}{p} \right)^{\frac{z-z_0}{2}}  \int_{0}^{ \pi} \frac{dq}{2 \pi} \bigg[ \lambda (q) \bigg]^t  
\left[ e^{i q z } + \Gamma(q) e^{- i q z } \right]  
\left[ e^{-i q z_0 } + \overline{\Gamma(q) } e^{ i q z_0 } \right]  
\label{similarityspectral}
\end{eqnarray}

At the critical point $p_c=\frac{1}{2}$, the spectral decomposition of Eq. \ref{similarityspectral}
reduces to
\begin{eqnarray}
P^{\mathrm {Criti}}_t(z \vert z_0) &&  = \int_{0}^{ \pi} \frac{dq}{2 \pi} \bigg[ \cos (q) \bigg]^t  
\left[ e^{i q z } +  e^{- i q (z+1) } \right]  
\left[ e^{-i q z_0 } +  e^{ i q (z_0+1) } \right]  
\nonumber \\
&&  = 2  \int_{0}^{ \pi} \frac{dq}{ \pi} \bigg[ \cos (q) \bigg]^t  
 \cos \left[ q \left(z+\frac{1}{2} \right) \right]  
 \cos \left[ q \left(z_0+\frac{1}{2} \right) \right]  
\label{similarityspectralcriti}
\end{eqnarray}
that corresponds to the unbiased random walk on the half-infinite lattice $z=0,1,2,..+\infty$
in the presence of a reflecting boundary at $z=-\frac{1}{2}$,
so that the critical propagator of Eq. \ref{similarityspectralcriti}
can also be rewritten in terms of the free propagator $P^{\mathrm {Free}}_t(m) $
of Eqs \ref{Gtpm} and \ref{Gspectral} of Appendix \ref{app_RW} for the special case $p_c=\frac{1}{2}$
\begin{eqnarray}
P^{\mathrm {FreeCriti}}_t(m)  =
\int_{-\pi}^{ \pi} \frac{dq}{2 \pi} \bigg[ \cos (q) \bigg]^t e^{i q m } 
 = \frac {t!}{ \left( \frac{t+m}{2}\right) ! \left(\frac{t-m}{2}\right) !} 2^{-t} \ \ \text{ for } m \in \{-t,...,+t\}
\label{Gtpmcriti}
\end{eqnarray}
via the method of images with a primary source at $z_0$ and a secondary source at $(-z_0-1)$
\begin{eqnarray}
P^{\mathrm {Criti}}_t(z \vert z_0)     = P^{\mathrm {FreeCriti}}_t(z-z_0)  +  P^{\mathrm {FreeCriti}}_t(z+z_0+1)
\label{similarityspectralcritiImages}
\end{eqnarray}

%%%%%%%%%%%%%%%%%%%%%%%
 
 \subsubsection{ Application to the dynamics starting from the initial uniform distribution $\rho_{t=0}(x_0)=1$ on $x_0 \in [0,1[$ }

If the initial condition at $t=0$ is the uniform distribution $\rho_{t=0}(x_0)=1$ on $x_0 \in [0,1[$
that translates into $ {\cal P}_{t=0}(\sigma_.) =  {\cal P}^{[\mathrm {Uniform}]}(\sigma_.)  $ of Eq. \ref{UniformDensitysigma}
where all the spins have the same random distribution,
then ${\cal P}_t \left[ \sigma_.  \right]  $ of Eq. \ref{Ptsigmafromzfconvol}
 depends only on the variable $z_t$ and not on the variable $F_t$    
\begin{eqnarray}
  {\cal P}_t(\sigma_.)   = \sum_{z=0}^{+\infty} {\cal P}_t(z) \left[ \prod_{l=1}^{z_t} \delta_{\sigma_l,0} \right] 
\left(  \prod_{l=z_t+1}^{+\infty} \left( \frac{\delta_{ \sigma _l,0} +  \delta_{\sigma _l,1} }{2} \right) \right)
\label{UniformDensitysigmadyn}
\end{eqnarray}
The translation towards the continuous variable $x \in [0,1[$ reduces to
\begin{eqnarray}
  \rho_t(x)   = \sum_{z=0}^{+\infty} P_t(z) \frac{ \theta( 0 \leq x \leq 2^{-z} )}{ 2^{-z} }
\label{UniformDensitysigmadyninx}
\end{eqnarray}
that belongs to the subspace $\rho^{[\mathrm {Partition}]}_t(x) $ of Eq. \ref{BinaryDensity}
where the weights
\begin{eqnarray}
  \pi_t(n) && = \int_{2^{-n-1} }^{2^{-n}} dx     \rho_t(x)
  =  \sum_{z=0}^{+\infty} P_t(z) 2^z \int_0^1 dx  \theta( 0 \leq x \leq 2^{-z} ) \theta \left( 2^{-n-1} \leq x < 2^{-n} \right)
\nonumber \\
&&   = 2^{-n-1} \sum_{z=0}^{n} P_t(z) 2^{z}
  \label{defpizt}
\end{eqnarray}
depend only on the finite-time probability distribution $P_t(z) =P_t(z \vert z_0=0)$
discussed in detail in the previous section.

In the chaotic region $\frac{1}{2}<p \leq 1$, the steady state $ P_*(z)  $ of Eq. \ref{steadyz} can be plugged into Eq. \ref{defpizt}
to recover the steady state $\pi_*(\cdot)$ of Eq. \ref{BinarySteadySOL}
\begin{eqnarray}
  \pi_*(n) && = 2^{-n-1}   \sum_{z=0}^{n} P_*(z) 2^{z-n-1}
  =  2^{-n-1} \frac{2p-1}{p} \sum_{z=0}^{n}  \left( 2 \frac{1-p}{p} \right)^z 
  = 2^{-n-1} \frac{2p-1}{p} \left[ \frac{ 1-  \left( 2 \frac{1-p}{p} \right)^{n+1} }{ 1- 2 \frac{1-p}{p} } \right]
  \nonumber \\
  && =   \frac{2p-1}{3p-2} \left[  2^{-n-1} -  \left(  \frac{1-p}{p} \right)^{n+1}  \right]
  \label{defpiztstar}
\end{eqnarray}
This result gives still another interpretation of the steady state $\pi_*(n) $
in terms of the steady distribution $P_*(z) $ of the number $z$ of zeroes.

Since each spin $\sigma_l(0)$ has for average value $\frac{1}{2}$
in the uniform distribution  $  {\cal P}^{[\mathrm {Uniform}]}(\sigma_.)  $ of Eq. \ref{UniformDensitysigma},
the average value of $x_t$ 
can be evaluated 
using Eq. \ref{binaryzf}
\begin{eqnarray}
 \langle x_t  \rangle_{{\cal P}^{[\mathrm {Uniform}]}(\sigma_.)} 
 && = \sum_{\sigma_.(0) } {\cal P}^{Unif} \left[ \sigma_.(0) \right]
 \sum_{z=0}^{+\infty} \sum_{F=0}^{+\infty} P_t(z,F) 
  \left[ \sum_{m=1}^{+\infty} \frac{ \sigma_{ F+m} (0) }{2^{z+m}} \right]
\nonumber \\
&& =    \sum_{z=0}^{+\infty} \sum_{F=0}^{+\infty} P_t(z,F) 
   \sum_{m=1}^{+\infty} \frac{ 1 }{2^{z+m+1}}  =  \sum_{z=0}^{+\infty}  P_t(z) 2^{-z-1}
\label{averageunif}
\end{eqnarray}
In the chaotic region $\frac{1}{2}<p \leq 1$, the steady state $ P_*(z)  $ of Eq. \ref{steadyz} can be plugged into Eq. \ref{averageunif} to obtain the convergence towards
\begin{eqnarray}
 \langle x_{t=+\infty}  \rangle_{{\cal P}^{[\mathrm {Uniform}]}(\sigma_.)} && 
  =  \sum_{z=0}^{+\infty}  P_*(z) 2^{-z-1}
  = \frac{2p-1}{2p} \sum_{z=0}^{+\infty}    \left( \frac{1-p}{ 2p} \right)^z 
  = \frac{2p-1}{3p-1} 
\label{averageunifinfty}
\end{eqnarray}
in agreement with the direct computation of the average of $x$ drawn with the steady state $\rho_*(x)$
of Eq. \ref{BinaryDensitysteady}
\begin{eqnarray}
 \langle x \rangle_{\rho_*(\cdot)} && \equiv  \int_0^1 dx x \rho_*(x)
 = \sum_{n=0}^{+\infty} 2^{n+1} \pi_*(n) \int_{2^{-n-1}}^{2^{-n}} dx x   
 = \sum_{n=0}^{+\infty} 2^{n+1} \pi_*(n)    \frac{ (2^{-n} - 2^{-n-1}) (2^{-n} + 2^{-n-1}) }{2}
  \nonumber \\
 && 
 =  \sum_{n=0}^{+\infty}  \pi_*(n)    \frac{  (2^{-n} + 2^{-n-1}) }{2}
 = \frac{3}{2} \sum_{n=0}^{+\infty}  \pi_*(n)  2^{-n-1} 
 = \frac{3(2p-1)}{2(3p-2)}  \sum_{n=0}^{+\infty}   \left[ 4^{-n-1} - \left( \frac{1-p}{2 p} \right)^{n+1}\right] 
% =  \frac{3(2p-1)}{2(3p-2)}   \left[ \frac{1}{4-1} - \frac{1}{\frac{2 p}{1-p} -1}\right] 
% = \frac{3(2p-1)}{2(3p-2)}   \left[ \frac{1}{3} - \frac{1-p}{3p -1}\right]  
% = \frac{3(2p-1)}{2(3p-2)} \times   \frac{2(3p-2) }{3(3p -1)}
%   \nonumber \\ && 
  = \frac{ 2p-1 }{3p -1}
\label{avxstarsteady}
\end{eqnarray}

%%%%%%%%%%%%%%%%%%%%%%%

  \subsection{ Interpretation of the variable $F_t$ as an additive observable 
  of the Markov trajectory $z_{0 \leq \tau \leq t}$ }
  
  \label{subsec_additive}
  
  In the dynamics for the two variables $(z_t,F_t)$ described before Eq. \ref{Wzf},
  the variable $F_t$ changes only via the incrementation $F_{t+1}=F_t+1$ in the case
   (ii-b) when $z_t=0$ and $z_{t+1}=0$.
  As a consequence, $F_t$ counts the number of times $\tau \in \{0,1,..,t-1 \}$ where  $z_{\tau}=0=z_{\tau+1}$
  \begin{eqnarray}
 F_t= \sum_{\tau=0}^{t-1} \delta_{z_{\tau},0} \delta_{z_{\tau+1},0} 
\label{ftadditive}
\end{eqnarray}
This means that $F_t$ is an additive observable 
  of the Markov trajectory $z_{0 \leq \tau \leq t}$ of the variable $z$ alone
  and can be thus analyzed via the corresponding standard methods.

%%%%%%%%%%%%%%%%%%%%%%%%%%%%%%%%%%%%%%%%%%

 \subsubsection{ Dynamics for the generating function $Z^{[\nu]}_t(z) $ with respect to the variable $F$}
 
 \label{subsec_generatingF}

The generating function with respect to $F$ of the joint distribution $ P_t(z,F)$ 
\begin{eqnarray}
 Z^{[\nu]}_t(z) =  \sum_{F=0}^{+\infty} e^{  \nu F } P_t(z,F)
\label{dynzi}
\end{eqnarray}
satisfies the following closed dynamics using Eq. \ref{evolPzf}
\begin{eqnarray}
 {\cal Z}^{[\nu]}_{t+1}({\tilde z})  = (1-p)   {\cal Z}^{[\nu]}_{t}({\tilde z}-1)  \theta( {\tilde z} \geq 1 )
+ p   {\cal Z}^{[\nu]}_{t}({\tilde z}+1)  
 + p e^{\nu} \delta_{{\tilde z},0} {\cal Z}^{[\nu]}_{t}(0) \equiv \sum_{z=0}^{+\infty} W^{[\nu]}( {\tilde z }  \vert z )  {\cal Z}^{[\nu]}_t(z)
\label{dynzgene}
\end{eqnarray}
where the $\nu$-deformation $W^{[\nu]} $ of the Markov matrix $W=W^{[\nu=0]}$ of Eq. \ref{Wz} 
\begin{eqnarray}
W^{[\nu]}( {\tilde z }  \vert z )  
\equiv (1-p) \delta_{{\tilde z }, z+1} 
+ p \theta(z \geq 1)  \delta_{{\tilde z }, z-1}   + p e^{\nu} \delta_{z,0}\delta_{{\tilde z }, 0}  
\label{Wznudeformed}
\end{eqnarray}
concerns only the amplitude $p e^{\nu} $ of the last term in $\delta_{z,0}\delta_{{\tilde z }, 0} $,
while the initial condition at $t=0$ of Eq. \ref{evolPzfini} translates into
 \begin{eqnarray}
Z^{[\nu]}_{t=0}(z) =  \sum_{F=0}^{+\infty} e^{ - \nu F } P_{t=0}(z,F)
=\sum_{F=0}^{+\infty} e^{ - \nu F }   \delta_{z,0} \delta_{F,0} = \delta_{z,0}
\label{Znufini}
\end{eqnarray}

The sum over $z$ of Eq. \ref{dynzi}
that represents the generating function of the variable $F$ alone
\begin{eqnarray}
Z^{[\nu]}_t \equiv \sum_{z=0}^{+\infty}  Z^{[\nu]}_t(z) =  \sum_{F=0}^{+\infty} e^{  \nu F } \left[ \sum_{z=0}^{+\infty}P_t(z,F)\right]=  \sum_{F=0}^{+\infty} e^{  \nu F } P_t(F)
\label{dynzisum}
\end{eqnarray}
evolves according to
\begin{eqnarray}
Z^{[\nu]}_{t+1} && = \sum_{{\tilde z}=0}^{+\infty} \left[ (1-p)   {\cal Z}^{[\nu]}_{t}({\tilde z}-1)  \theta( {\tilde z} \geq 1 )
+ p   {\cal Z}^{[\nu]}_{t}({\tilde z}+1)  
 + p e^{\nu} \delta_{{\tilde z},0} {\cal Z}^{[\nu]}_{t}(0)  \right]
 \nonumber \\
 && =Z^{[\nu]}_t + p (e^{\nu} -1 ) {\cal Z}^{[\nu]}_{t}(0)
\label{dynzgenesum}
\end{eqnarray}
With respect to $\nu=0$ where the constant value $Z^{[\nu=0]}_t =1$ corresponds to the conserved 
normalization of the probability distribution $P_t(F)$,
the interpretation of Eq. \ref{dynzgenesum}
depends on the sign of $\nu$:

(i) $\nu>0$ corresponds to a population $Z^{[\nu]}_t $ that can only grow
via the reproduction factor $ p (e^{\nu} -1 ) $ for the population ${\cal Z}^{[\nu]}_{t}(z=0) $ at the origin $z=0$;

(ii) $\nu<0$ corresponds to a population $Z^{[\nu]}_t $ that can only decay
via the killing factor $ p (1-e^{\nu}  ) $ for the population ${\cal Z}^{[\nu]}_{t}(z=0) $ at the origin $z=0$.

The asymptotic exponential behavior of the population $Z^{[\nu]}_t $ for large time $t \to + \infty$ 
\begin{eqnarray}
Z^{[\nu]}_t \oppropto_{t \to + \infty} \bigg[ \Lambda(\nu) \bigg]^t
\label{dynzisumtlarge}
\end{eqnarray}
is thus expected to be governed by values $\Lambda(\nu) > 1$ for $\nu>0$
and by values $\Lambda(\nu) < 1$ for $\nu<0$
only if the fraction $ \frac{{\cal Z}^{[\nu]}_{t}(z=0)}{{\cal Z}^{[\nu]}_{t}} $ of the population at
the origin $z=0$ remains finite.

%%%%%%%%%%%%%%%%%%%%%%%%%%%%%%%%%%%%%%%%%%

 \subsubsection{ Explicit results for the Laplace transform $ {\tilde Z}^{[\nu]}_s(z) $ and for $\Lambda(\nu) $ }

For the time-Laplace transform of the generating function $ Z^{[\nu]}_t(z) $ of Eq. \ref{dynzi}
\begin{eqnarray}
 {\tilde Z}^{[\nu]}_s(z) \equiv \sum_{t=0}^{+\infty} e^{-s t} Z^{[\nu]}_t(z)
\label{dynzilaplacedef}
\end{eqnarray}
the only difference with respect to the decomposition of Eq. \ref{PropagatorResumresetEventsLaplacezcalcul} is the additional factor $e^{\nu}$ for each of the $K$ sojourns at the origin
\begin{eqnarray}
 {\tilde Z}^{[\nu]}_s(z) =  \sum_{K=0}^{+\infty} 
{\tilde P}_s^{Abs}(z \vert 0) \left[ e^{-\nu}  {\tilde A}(s \vert 0) \right]^{K} = \frac{ {\tilde P}_s^{Abs}(z \vert 0)  }
{ 1- e^{-\nu}  {\tilde A}(s\vert 0)}
\label{dynzilaplacecalcul}
\end{eqnarray}
where one can plug 
${\tilde P}^{Abs}_s(z \vert n_0=0)$ of \ref{sPabspm},
and ${\tilde A}(s \vert n_0=0) $ of Eq. \ref{probaAbstLaplace}
to obtain the explicit result
\begin{eqnarray}
 {\tilde Z}^{[\nu]}_s(z) =   \frac{r_+(s)+r_-(s)}{ r_+(s)- e^{ \nu} } [r_-(s)]^z 
\label{dynzilaplace}
\end{eqnarray}
in terms of the two roots $r_{\pm}(s)$ of Eq. \ref{2dsolmain},
while the Laplace transform of Eq. \ref{dynzisum} obtained via the sum over $z$ of Eq. \ref{dynzilaplace}
reduces to
\begin{eqnarray}
{\tilde Z}^{[\nu]}_s \equiv \sum_{t=0}^{+\infty} e^{-s t} Z^{[\nu]}_t 
= \sum_{z=0}^{+\infty}  {\tilde Z}^{[\nu]}_s(z) = \frac{r_+(s)+r_-(s)}{ [r_+(s)- e^{ \nu} ]  [1- r_-(s)] } 
\label{dynzisumlaplace}
\end{eqnarray}
The large-time asymptotic behavior of Eq. \ref{dynzisum} means that the Laplace transform ${\tilde Z}^{[\nu]}_s $
converges as long as 
\begin{eqnarray}
 e^{-s}  \Lambda(\nu) <1 \ \ \ \text {i.e. } \ \ s> \ln [\Lambda(\nu) ] 
\label{critere}
\end{eqnarray}
so that $\Lambda(\nu) $ can be found from the value $s= \ln [\Lambda(\nu) ]$ where Eq. \ref{dynzisumlaplace} 
becomes singular.
Let us consider the two factors in the denominator of Eq. \ref{dynzisumlaplace}: 

(a) the factor $[1- r_-(s)] $ vanishes only for $s=0$ in the region $p \leq \frac{1}{2}$

(b) the factor $[r_+(s)- e^{ \nu} ] $ vanishes if
\begin{eqnarray}
2p e^{ \nu-s }  -1 =  \sqrt{ 1 - 4  p(1-p) e^{-2s}}  \ \ \text{ i.e. } 
e^s = p e^{\nu}+(1-p) e^{-\nu} \equiv \Lambda_*(\nu) \ \ \text{ with the condition }  e^{ \nu} \geq \sqrt {\frac{1-p}{p} }
\label{dvsnu}
\end{eqnarray}
The value of $\Lambda_*(\nu)$ when $\nu $ takes the limiting value $\nu =\ln  \sqrt {\frac{1-p}{p} }$ reduces to
\begin{eqnarray}
\Lambda_* \left(  \nu = \ln  \sqrt {\frac{1-p}{p} } \right) = p  \sqrt {\frac{1-p}{p} }
+(1-p)  \sqrt {\frac{p}{1-p} }  = 2 \sqrt{ p(1-p)} = \lambda_{q=0}
\label{dvsnuspecial}
\end{eqnarray}
where one recognizes the maximal value $\lambda_{q=0} $ of the continuum spectrum $\lambda_q$
of Eq. \ref{lambdaqmomentum} associated to Fourier modes that are always present via the square-root present
in $r_{\pm}(s)$.

So one obtains the following discussion as a function of the parameter $p$:

(i) in the chaotic region $\frac{1}{2} < p \leq 1$, 
the asymptotic behavior of Eq. \ref{dynzisumtlarge} is governed by
\begin{eqnarray}
\Lambda(\nu)  =
\begin{cases}
\Lambda_*(\nu) \equiv p e^{\nu}+(1-p) e^{-\nu} \ \ \ \ \ \ \ \ \ \text{ for }  \ \  \nu \geq - \ln \sqrt {\frac{p}{1-p} }
 \\
\lambda_{q=0} =2 \sqrt{ p(1-p)}  \ \ \ \ \ \ \ \ \ \  \ \ \ \ \ \ \text{ for } \ \ \ \nu \leq - \ln \sqrt {\frac{p}{1-p} } <0
\end{cases}
\label{lambdanuchaos}
\end{eqnarray}
with the following interpretation:
the existence of the localized steady state $P_*(z) $ of Eq. \ref{steadyz} for $\nu=0$ 
yields that the reproducing/killing factor at the origin $z=0$ 
is relevant for small $\nu \ne 0$;
this leads to the non-trivial $\Lambda_*(\nu) $ for any reproducing factor corresponding to the region $\nu>0$,
but only for the finite interval $- \ln \sqrt {\frac{p}{1-p} } <\nu<0$ in the killing region,
while for $\nu \leq - \ln \sqrt {\frac{p}{1-p} } $, the killing at the origin becomes too strong with respect to the  
maximal value $\lambda_{q=0} =2 \sqrt{ p(1-p)}$ of the continuum Fourier spectrum $\lambda_q$.

(ii) at the critical point $p_c=\frac{1}{2}$, the asymptotic behavior of Eq. \ref{dynzisumtlarge} is governed by
\begin{eqnarray}
\Lambda(\nu)  =
\begin{cases}
\Lambda_*(\nu) \equiv \frac{e^{\nu}+e^{-\nu}}{2} = \cosh (\nu) \ \ \ \ \ \ \ \ \ \text{ for }  \ \  \nu \geq 0
 \\
1  \ \ \ \ \ \ \ \ \ \  \ \ \ \ \ \ \ \ \ \ \ \ \ \  \ \ \ \ \ \ \ \ \ \ \ \ \ \ \ \ \ \  \text{ for } \ \ \ \nu \leq 0
\end{cases}
\label{lambdanucriti}
\end{eqnarray}
with the following interpretation:
for $\nu=0$, the process $z_{0 \leq \tau \leq t}$ spends a time of order $\sqrt{t}$ at the origin $z=0$.
Any reproducing factor is able to produce the non-trivial $\Lambda_*(\nu) $ for $\nu>0$,
as in one-dimensional quantum mechanics where any localized attractive potential is able to produce a bound state,
while any killing factor for $\nu<0$ is not able to change maximal value $\lambda_{q=0}=1 $ of the continuum Fourier spectrum $\lambda_q$.

(iii) in the non-chaotic region $0<p < \frac{1}{2}$, the asymptotic behavior of Eq. \ref{dynzisumtlarge} is governed by: 
\begin{eqnarray}
\Lambda(\nu)  =
\begin{cases}
\Lambda_*(\nu) \equiv p e^{\nu}+(1-p) e^{-\nu} \ \ \ \ \ \ \ \ \ \text{ for }  \ \  \nu \geq  \ln \frac{1-p}{p} >0
 \\
1  \ \ \ \ \ \ \ \ \ \  \ \ \ \ \ \ \ \ \ \ \ \ \ \ \ \ \ \ \ \ \ \ \ \ \ \ \ \ \ \ \ \  \text{ for } \ \ \ \nu \leq  \ln \frac{1-p}{p} 
\end{cases}
\label{lambdanuchaosnonchaos}
\end{eqnarray}
with the following interpretation: 
for $\nu=0$, the process $z_t$ flows ballistically towards $(+\infty)$, with an exponential time-decay at the origin,
so the reproducing/killing factor at the origin $z=0$ for $\nu \ne 0$ 
is irrelevant for small $\nu$ and is not able to change the value $\Lambda(\nu=0)=1$
 for any killing factor corresponding to the region $\nu<0$,
but only for the finite interval $0\leq \nu \leq \ln \frac{1-p}{p}$ in the reproducing region,
while for $\nu > \ln \frac{1-p}{p}$, the reproducing factor at the origin 
is strong enough to produce the non-trivial $\Lambda_*(\nu) $.

%%%%%%%%%%%%%%%%%%%%%%%%%%%%%%%%%%%%%%%%%%%%%%

 \subsubsection{ Explicit rate function $I(f)$ governing the large deviations of the density $f=\frac{ F_T}{T} \in [0,1]$ of spins erased per unit time   }

The explicit results of the previous section concerning $ \Lambda(\nu) $ are useful to compute 
the rate function $I(f)$ governing the decay for large $T$ of
the probability to see the density $f=\frac{ F_T}{T} \in [0,1]$ of spins erased per unit time on a large time window $[0,T]$
\begin{eqnarray}
 P_T(F=f T) \opsimeq_{T \to + \infty} e^{-T I(f)}
\label{largedevf}
\end{eqnarray}
Plugging this asymptotic form into the generating function of Eq. \ref{dynzisum}
displaying the asymptotic behavior of Eq. \ref{dynzisumtlarge}
\begin{eqnarray}
Z^{[\nu]}_T \opsimeq_{T \to + \infty}  \int_0^1 df   e^{ T [ \nu f - I(f)]}
\oppropto_{T \to + \infty} e^{T \ln [ \Lambda(\nu)] }
\label{saddle}
\end{eqnarray}
yields via the saddle-point evaluation of the integral over $f$
that the rate function $I(f)$ and $\ln [ \Lambda(\nu)] $ are related by the Legendre transform
\begin{eqnarray}
 \nu f - I(f) && = \ln [ \Lambda(\nu)] 
 \nonumber \\
 \nu -I'(f) && =0
\label{legendre}
\end{eqnarray}

In the chaotic region $\frac{1}{2} < p \leq 1 $ with the non-trivial $\Lambda_*(\nu)$ of Eq. \ref{lambdanuchaos},
 the reciprocal Legendre transform reads
\begin{eqnarray}
I(f)  && = \nu f -  \ln [ \Lambda_*(\nu)]  =  \nu f -  \ln [  p e^{\nu}+(1-p) e^{-\nu}]  
 \nonumber \\
 f && = \frac{ d  \ln [ \Lambda_*(\nu)]}{d \nu}  = \frac{ p e^{\nu}-(1-p) e^{-\nu} }{ p e^{\nu}+(1-p) e^{-\nu} }
\label{legendrereci}
\end{eqnarray} 
that can be inverted to obtain $\nu$ as a function of $f$
\begin{eqnarray}
 \nu  = \ln \sqrt{ \frac{(1-p)(1+f)}{p (1-f)}}
\label{nufonctiondef}
\end{eqnarray} 
Plugging this value into the first line of Eq. \ref{legendrereci}
yields the explicit expression of the rate function 
\begin{eqnarray}
I(f)   = f  \ln \sqrt{ \frac{(1-p)(1+f)}{p (1-f)}} -\ln \left[ \sqrt{p(1-p) } \left( \sqrt{ \frac{1+f}{1-f}} +\sqrt{ \frac{1-f}{1+f}} \right)\right]
\label{rateIf}
\end{eqnarray} 
The typical value $f_*$ where the rate function $I(f)$ and its derivative $I'(f)$ vanish 
\begin{eqnarray}
I(f_*)=0=I'(f_*)
\label{ftypderi}
\end{eqnarray} 
corresponds to $\nu=0$ in the Legendre transform of Eq. \ref{legendre}, so that 
the reciprocal Legendre transform of Eq. \ref{legendrereci}
yields that the typical value is
\begin{eqnarray}
 f_* =  2p-1
\label{ftypchaos}
\end{eqnarray} 
while the validity region for $\Lambda_*(\nu) $ in Eq. \ref{lambdanuchaos} using Eq. \ref{nufonctiondef}
\begin{eqnarray}
0 \leq   \nu + \ln \sqrt {\frac{p}{1-p} } = \ln \sqrt{ \frac{1+f}{1-f}} 
\label{lambdanuchaosvali}
\end{eqnarray}
is always satisfied for $f \in [0,1]$.

At the critical point $p_c=\frac{1}{2}$, the typical value of Eq. \ref{ftypchaos} vanishes 
\begin{eqnarray}
f_*=0
\label{ftypcriti}
\end{eqnarray} 
and is thus at the boundary on the domain $f \in [0,1]$, while the rate function of Eq. \ref{rateIf}
reduces to
\begin{eqnarray}
I(f)   = f  \ln \sqrt{ \frac{1+f}{1-f}} -\ln \left[ \frac{ \sqrt{ \frac{1+f}{1-f}} +\sqrt{ \frac{1-f}{1+f}} }{2} \right]
\label{rateIfcriti}
\end{eqnarray} 

%%%%%%%%%%%%%%%%%%%%%%%%%%%%%%%%%%

 \subsubsection{  Spectral decomposition of the generating function $Z^{[\nu]}_t(z) $ }
 
 Since the dynamics of Eq. \ref{dynzgene} for the generating function $ Z^{[\nu]}_t(z\vert z_0) $ of Eq. \ref{dynzi}
differs only by the amplitude $p e^{\nu}$ of the last term with respect to the dynamics of
Eq. \ref{dynz} concerning the probability distribution $P_t(z)$ whose spectral decomposition was discussed in detail between Eqs \ref{similarity} and \ref{similarityspectral},
one can obtain the spectral decomposition of $Z^{[\nu]}_t(z)  $
just by making a few changes as follows.
The similarity transformation analogous to Eq. \ref{similarity}
\begin{eqnarray}
Z^{[\nu]}_t(z\vert z_0)= \left( \frac{1-p}{p} \right)^{\frac{z-z_0}{2}} \psi_t^{[\nu]}(z \vert z_0)
\label{similaritynu}
\end{eqnarray}
is useful to write the spectral decomposition of the $\nu$-deformed quantum propagator 
\begin{eqnarray}
\psi_t^{[\nu]}(z \vert z_0)  && 
= \bigg[ \Lambda_*(\nu)  \bigg]^t \langle z \vert \psi^{[\nu]}_* \rangle  \langle \psi_*^{[\nu]} \vert z_0 \rangle 
\theta\left( \frac{1-p}{p} e^{-2 \nu} <1 \right)
+ \int_{0}^{ \pi} \frac{dq}{ \pi} \bigg[ \lambda (q) \bigg]^t \langle z \vert \phi_q^{[\nu]} \rangle  \langle \phi_q^{[\nu]} \vert z_0 \rangle 
  \label{spectralpsinu}
\end{eqnarray}
where the eigenstates $\phi_q^{[\nu]} $ associated to 
the continuous spectrum of eigenvalues $\lambda(q) \equiv \sqrt{ p (1-p) } (e^{iq}+e^{-iq} ) = \sqrt{ 4 p (1-p) } \cos q $ of Eq. \ref{lambdaqmomentum} have the same form as Eq. \ref{eigenphiq}
\begin{eqnarray}
\phi^{[\nu]}_q(z)= \frac{1}{\sqrt 2} \left[ e^{i q z } + \Gamma^{[\nu]}(q) e^{- i q z } \right]
\label{eigenphiqnu}
\end{eqnarray}
with the $\nu$-deformed coefficients
\begin{eqnarray}
 \Gamma^{[\nu]}(q) = -  \frac{\sqrt{ \frac{1-p}{p} } e^{- iq}-e^{\nu}} {\sqrt{ \frac{1-p}{p} } e^{ iq}-e^{\nu} }
  \label{eigenpsiz0solnu}
\end{eqnarray}
while the normalizable $\nu$-deformed quantum ground-state $\psi_*^{[\nu]}(z) $ 
\begin{eqnarray}
\psi_*^{[\nu]}(z)=  \sqrt{1- \frac{1-p}{p}  e^{-2 \nu}  } \left( \sqrt{\frac{1-p}{p} } e^{-\nu}  \right)^{z} \ \ \ \text{ for } \frac{1-p}{p} e^{-2 \nu} <1
\label{steadyzpsigsnu}
\end{eqnarray}
is associated to the $\nu$-deformed eigenvalue $\Lambda_*(\nu) $ corresponding Eq. \ref{dvsnu}
\begin{eqnarray}
 \Lambda_*(\nu) \equiv  p e^{\nu}+(1-p) e^{-\nu} 
 \label{lambdastarnu}
\end{eqnarray}

Putting everything together, the spectral decomposition of the 
generating function $ Z^{[\nu]}_t(z\vert z_0) $ reads
\begin{eqnarray}
&& Z^{[\nu]}_t(z\vert z_0)  = \left( \frac{1-p}{p} \right)^{\frac{z-z_0}{2}} \psi_t^{[\nu]}(z \vert z_0)
\nonumber \\
&& =\bigg[ \Lambda_*(\nu)  \bigg]^t  \left( \frac{1-p}{p} \right)^{\frac{z-z_0}{2}} \psi^{[\nu]}_* (z)   \psi_*^{[\nu]}( z_0 )
\theta\left( e^{\nu} > \sqrt{\frac{1-p}{p} } \right)
+\left( \frac{1-p}{p} \right)^{\frac{z-z_0}{2}} \int_{0}^{ \pi} \frac{dq}{ \pi} \bigg[ \lambda (q) \bigg]^t  \phi_q^{[\nu]} (z) 
\overline{   \phi_q^{[\nu]} (z_0) } 
\nonumber \\
&& =\bigg[ \Lambda_*(\nu)  \bigg]^t  \left[ 1- \frac{1-p}{p}  e^{-2 \nu}  \right]
 \left( \frac{1-p}{p}  e^{-\nu}  \right)^{z} e^{- \nu z_0} 
\theta\left( e^{\nu} > \sqrt{\frac{1-p}{p} } \right)
% \nonumber \\ &&
+\left( \frac{1-p}{p} \right)^{\frac{z-z_0}{2}} \int_{0}^{ \pi} \frac{dq}{ \pi} \bigg[ \lambda (q) \bigg]^t  \phi_q^{[\nu]} (z) 
\overline{   \phi_q^{[\nu]} (z_0) } 
\nonumber \\
\label{similarityspectralnu}
\end{eqnarray}
with the following leading exponential behavior for large $t$ 
\begin{eqnarray}
&& Z^{[\nu]}_t(z\vert z_0)  \opsimeq_{t \to + \infty}
\begin{cases}
\bigg[ \Lambda_*(\nu)  \bigg]^t  = \bigg[ p e^{\nu}+(1-p) e^{-\nu}  \bigg]^t \ \ \ \ \ \ \ \ \ \text{ for }  \ \  \nu \geq - \ln \sqrt {\frac{p}{1-p} }
\\ 
\\
\bigg[ \lambda (0) \bigg]^t =\bigg[ 2 \sqrt{ p(1-p)} \bigg]^t  \ \ \ \ \ \ \ \ \ \ \ \ \ \ \ \text{ for } \ \ \ \nu \leq - \ln \sqrt {\frac{p}{1-p} } <0
\end{cases}
\label{similarityspectralnutlarge}
\end{eqnarray}

In the chaotic region $\frac{1}{2} < p \leq 1$, the sum over $z$ of Eq. \ref{similarityspectralnu}
involves only convergent series, so that the link between the asymptotic behaviors of Eq. \ref{similarityspectralnutlarge}
and Eq. \ref{lambdanuchaos} is direct.
In the non-chaotic region $0<p < \frac{1}{2}$, the sum over $z$ of Eq. \ref{similarityspectralnu}
are not always convergent series, leading to some differences between the asymptotic behaviors of
Eq. \ref{similarityspectralnutlarge}
and Eq. \ref{lambdanuchaosnonchaos}.

%%%%%%%%%%%%%%%%%%%%%%%%%%%%%%%%%%%%%%%%%%%%%%

 \subsection{ Time-correlation  $C(t)=\langle x_t x_0 \rangle $ when the initial condition $x_{t=0}$ is drawn with the uniform distribution }
 
 \label{subsec_corre}
 
 The product $x_t x_0$ reads in terms of the initial spins $\sigma_.(t=0)$
using Eq. \ref{binaryzf}
\begin{eqnarray}
x_t x_0  = \left[ \sum_{m=1}^{+\infty} \frac{ \sigma_{ F_t+m} (0) }{2^{z_t+m}} \right]
\left[ \sum_{l=1}^{+\infty} \frac{\sigma_l(0)}{2^l}  \right]
\label{correzf}
\end{eqnarray}
When the initial distribution at $t=0$ is the uniform distribution $\rho_{t=0}(x_0)=1$ on $x_0 \in [0,1[$
that translates into $ {\cal P}_{t=0}(\sigma_.(0)) =  {\cal P}^{[\mathrm {Uniform}]}(\sigma_.(0))  $ of Eq. \ref{UniformDensitysigma},
 the correlation between the two spins $\sigma_l(0) $ and $\sigma_j (0) $ reduces to
\begin{eqnarray}
  \sum_{\sigma_.(0) } \sigma_l(0)\sigma_j (0) {\cal P}^{[\mathrm {Uniform}]}_{t=0} \left[ \sigma_.(0) \right] 
  = \delta_{l,j} \frac{1}{2} +(1- \delta_{l,j} ) \frac{1}{4} =  \frac{1+ \delta_{l,j} }{4}
\label{corre2spinuniform}
\end{eqnarray}
The average of Eq. \ref{correzf} over the uniform distribution  ${\cal P}^{[\mathrm {Uniform}]}(\sigma_.(0))  $
for the initial condition yields the time-correlation $C(t)$ in terms of the joint distribution $P_t(z,F)$
\begin{eqnarray}
C(t) && \equiv \langle x_t x_0 \rangle_{{\cal P}^{[\mathrm {Uniform}]}(\sigma_.(0))} 
 =  \sum_{z=0}^{+\infty} \sum_{F=0}^{+\infty} P_t(z,F) 
  \sum_{l=1}^{+\infty} \sum_{m=1}^{+\infty} \frac{ 1 }{2^{z+m+l}} 
\left( \sum_{\sigma_.(0) } \sigma_l(0)\sigma_{ F+m} (0) {\cal P}^{[\mathrm {Uniform}]}_{t=0} \left[ \sigma_.(0) \right] \right)
\nonumber \\
&&
= \sum_{z=0}^{+\infty} 2^{-z-2} \sum_{F=0}^{+\infty} P_t(z,F) 
 \sum_{m=1}^{+\infty} 2^{-m}  \sum_{l=1}^{+\infty} 2^{-l} \left[ 1+ \delta_{l,F+m} \right]
% \nonumber \\&& 
=  \sum_{z=0}^{+\infty}2^{-z-2} \sum_{F=0}^{+\infty} P_t(z,F) 
 \sum_{m=1}^{+\infty} 2^{-m}  \left[ 1+ 2^{-F-m} \right]
 \nonumber \\
&& =  \sum_{z=0}^{+\infty}2^{-z-2} \sum_{F=0}^{+\infty} P_t(z,F) 
   \left[ 1+ \frac{2^{-F} }{3} \right]
    \nonumber \\
&& =  \sum_{z=0}^{+\infty}  2^{-z-2} P_t(z)
    +   \frac{1 }{3}  \sum_{z=0}^{+\infty}  2^{-z-2} Z^{[\nu=-\ln 2]}_t(z)
\label{corretime}
\end{eqnarray}
where the first term involves the probability distribution $P_t(z)$ of the variable $z$ alone
that has been discussed in detail in section \ref{subsec_closedz},
while the second term involves the generating function of Eq. \ref{dynzi}
for the special value $\nu=-\ln 2$
\begin{eqnarray}
 Z^{[\nu=-\ln 2]}_t(z) \equiv  \sum_{F=0}^{+\infty}  P_t(z,F)2^{-F}
\label{dynziln2}
\end{eqnarray}

The time-Laplace-transform ${\tilde C}(s) $ of the time-correlation $C(t)$ 
can be thus computed in terms of ${\tilde P}_s(z) $ of Eq. \ref{PropagatorResumresetEventsLaplacez}
and in terms of $ {\tilde Z}^{[\nu]}_s(z) $ of Eq. \ref{dynzilaplace} for the special value $\nu=-\ln 2$
to obtain the explicit result
\begin{eqnarray}
{\tilde C}(s) \equiv \sum_{t=0}^{+\infty} e^{-s t} C(t) 
&& =  \sum_{z=0}^{+\infty}  2^{-z-2} {\tilde P}_s(z)
    +   \frac{1 }{3}  \sum_{z=0}^{+\infty}  2^{-z-2}  {\tilde Z}^{[\nu]}_s(z)
    \nonumber \\
    && = \frac{[r_+(s)+r_-(s)]}{ 4 [r_+(s)-1] } \sum_{z=0}^{+\infty}  [ 2^{-1} r_-(s)]^z
    +    \frac{[r_+(s)+r_-(s)]}{ 6 [2 r_+(s)-1] } \sum_{z=0}^{+\infty}    [ 2^{-1}r_-(s)]^z 
     \nonumber \\
    && =  \frac{[r_+(s)+r_-(s)] [8r_+(s)-5]}{ 6 [r_+(s)-1] [2 r_+(s)-1] [ 2-r_-(s)] }  
\label{correlaplace}
\end{eqnarray}
in terms of the two roots $r_{\pm}(s)$.

The asymptotic behavior of $C(t)$ for large time $ t \to + \infty$
can be also analyzed by plugging the spectral decomposition of Eq. \ref{similarityspectral} for the special case $z_0=0$
\begin{eqnarray}
P_t(z \vert z_0=0) &&  =P_*(z)  \theta\left( \frac{1}{2} <p \leq 1\right)
+ \left( \frac{1-p}{p} \right)^{\frac{z}{2}}  \int_{0}^{ \pi} \frac{dq}{2 \pi} \bigg[ \lambda (q) \bigg]^t  
\left[ e^{i q z } + \Gamma(q) e^{- i q z } \right]  
\left[ 1 + \overline{\Gamma(q) }  \right]  
\label{similarityspectralz00}
\end{eqnarray}
into the first contribution of Eq. \ref{corretime},
and by plugging the spectral decomposition of Eq. \ref{similarityspectralnu}
for the special case $z_0=0$ and $\nu=- \ln 2$ with $e^{\nu}=1/2$ and $e^{-\nu}=2$
\begin{eqnarray}
 Z^{[\nu= - \ln 2]}_t(z\vert z_0=0) 
 &&   =\bigg[ \frac{4-3p}{2} \bigg]^t  \left[  \frac{5p-4}{p}    \right]
 \left( 2 \frac{1-p}{p}   \right)^{z} 
\theta\left( p>\frac{4}{5} \right)
 \nonumber \\ &&
 +\left( \frac{1-p}{p} \right)^{\frac{z}{2}} \int_{0}^{ \pi} \frac{dq}{ \pi} \bigg[ \lambda (q) \bigg]^t  \phi_q^{[\nu=-\ln 2]} (z) 
\overline{   \phi_q^{[\nu=-\ln 2]} (0) } 
\label{similarityspectralnuln2}
\end{eqnarray}
Let us discuss the results as a function of the parameter $p$ in the following subsections.

%%%%%%%%%%%%%%%%%%%%%%%%%%%%%%%%%%%%

 \subsubsection{ Asymptotic behavior of the time-correlation  $C(t)$ in the chaotic region $\frac{1}{2} <p \leq 1 $}

In the chaotic region $\frac{1}{2} <p \leq 1 $ where the dynamics converges towards the steady state $\rho_*(x)$
independently of the initial condition,
the time-correlation $C(t)$ of Eq. \ref{corretime} is expected to converge towards the 
product of the average value $\langle x_0 \rangle_{\rho^{[\mathrm {Uniform}]}(\cdot)}=1/2$ of the initial condition $x_0$
and the average value $\langle x \rangle_{\rho_*(\cdot)}$ of $x$ in the steady state $\rho_*(x)$
that has been computed in Eq. \ref{avxstarsteady}
\begin{eqnarray}
C(t) && \opsimeq_{t \to \infty} 
C(\infty) =  \langle x \rangle_{\rho_*(\cdot)}
 \langle x_0 \rangle_{\rho^{[\mathrm {Uniform}]}(\cdot)} = \frac{ 2p-1 }{ 2(3p -1)}
 \label{correinfty}
\end{eqnarray}

The time-Laplace-transform of Eq. \ref{correlaplace} is then expected to display the
following asymptotic behavior for $s \to 0^+$ 
\begin{eqnarray}
{\tilde C}(s)  \opsimeq_{s \to 0^+} \frac{C(\infty)}{s}
\label{correschaos}
\end{eqnarray}
The series expansions of Eqs \ref{2dsolchaos} for $r_{\pm}(s)$ in the chaotic region $\frac{1}{2} < p \leq 1$
yield that the only term of order $\frac{1}{s}$ in Eq. \ref{correlaplace}
comes from the factor $[r_+(s)-1]$ in the denominator leading to the asymptotic value
\begin{eqnarray}
C(\infty) && = \lim_{s \to 0^+} \left[ s {\tilde C}(s) \right]
  =  \frac{[r_+(0)+r_-(0)] [8r_+(0)-5]}{ 6 [\frac{1 }{2p-1}] [2 r_+(0)-1] [ 2-r_-(0)] }  
=  \frac{ 2p-1 }{2(3p -1)}
\label{correlaplaces0chaos}
\end{eqnarray}
in agreement with Eq. \ref{correinfty}.

This asymptotic value $C(\infty)$ can be also reproduced by plugging the steady state $P_*(z)$ of Eq. \ref{steadyz}
into the first contribution of Eq. \ref{corretime}
\begin{eqnarray}
  \sum_{z=0}^{+\infty}  P_*(z) 2^{-z-2}
  =  \frac{2p-1}{4p} \sum_{z=0}^{+\infty}     \left( \frac{1-p}{ 2 p} \right)^z
  = \frac{2p-1}{4p}  \times \frac{1}{1- \frac{1-p}{ 2 p} }
  =  \frac{ 2p-1 }{2(3p -1)}
\label{correinitialuniforminftybis}
\end{eqnarray}
so that the connected correlation can be rewritten as 
\begin{eqnarray}
C^{\mathrm {Connected}}(t) && \equiv C(t) - C(\infty)
=  \sum_{z=0}^{+\infty}  2^{-z-2} \left[ P_t(z) - P_*(z)\right]
    +   \frac{1 }{3}  \sum_{z=0}^{+\infty}  2^{-z-2} Z^{[\nu=-\ln 2]}_t(z)
\label{corretimeconnected}
\end{eqnarray}
where the first contribution involving the spectral decomposition of Eq. \ref{similarityspectralz00}
is dominated by the contribution of the largest Fourier eigenvalue $\lambda(q=0)=\sqrt{4p(1-p)}$ for large time
\begin{eqnarray}
  \sum_{z=0}^{+\infty}  2^{-z-2} \left[ P_t(z) - P_*(z)\right] \opsimeq_{t \to + \infty} \bigg[ \lambda (q=0) \bigg]^t  
  = \bigg[ 2 \sqrt{p(1-p)} \bigg]^t  
\label{corretimeconnected1}
\end{eqnarray}
while the second contribution involving Eq. \ref{similarityspectralnuln2}
is dominated by 
\begin{eqnarray}
\begin{cases}
\Lambda_*(\nu=-\ln 2)=\frac{4-3p}{2} \ \ \ \ \ \ \ \ \ \ \ \ \ \ \ \ \ \ \text{ for }  \ \  \frac{4}{5} \leq p \leq 1
 \\
\lambda (q=0) 2 = \sqrt{ p(1-p)}  \ \ \ \ \ \ \ \ \ \  \ \ \ \ \ \ \ \text{ for } \ \  \frac{1}{2}<p \leq \frac{4}{5}
\end{cases}
\label{lambdanuchaosln2}
\end{eqnarray}

As a consequence, the asymptotic exponential decay of the connected correlation of Eq. \ref{corretimeconnected} is
\begin{eqnarray}
C^{\mathrm {Connected}}(t) && \equiv C(t) - C(\infty)
\oppropto_{t \to +\infty}
\begin{cases}
\left[ \frac{4-3p}{2} \right]^t \ \ \ \ \ \ \ \ \ \ \ \ \ \ \ \ \ \ \ \ \ \ \ \ \ \text{ for }  \ \ \ \  \frac{4}{5} \leq p \leq 1
 \\
\left[ 2 \sqrt{ p(1-p)} \right]^t  \ \ \ \ \ \ \ \ \ \  \ \ \ \ \ \ \text{ for } \ \ \ \frac{1}{2}<p \leq \frac{4}{5}
\end{cases}
\label{corretimeconnectedchaos}
\end{eqnarray}

%%%%%%%%%%%%%%%%%%%%%%%%%%%%%%

\subsubsection{ Asymptotic analysis in the non-chaotic region $0<p < \frac{1}{2}$}

In the non-chaotic region $0<p < \frac{1}{2}$,
 the series expansions of $r_{\pm}(s)$ in Eqs \ref{2dsolzerononchaos}
yield that ${\tilde C}(s)$ of Eq. \ref{correlaplace}
takes a finite value for $s=0$ that represents the converging sum of $C(t)$ over time
\begin{eqnarray}
{\tilde C}(s=0)  \equiv  \sum_{t=0}^{+\infty}  C(t) 
=  \frac{[r_+(0)+r_-(0)] [8r_+(0)-5]}{ 6 [r_+(0)-1] [2 r_+(0)-1] [ 2-r_-(0)] }  = \frac{8-13p}{6 (1-2p) (2-3p) }
\label{correlaplacenonchaos}
\end{eqnarray}

The two spectral decompositions of Eq. \ref{similarityspectralz00}
and \ref{similarityspectralnuln2}
yields that the time-correlation $C(t)$ in Eq. \ref{corretime}
is dominated by the exponential decay governed by the largest Fourier eigenvalue $\lambda(q=0)=\sqrt{4p(1-p)}$ 
\begin{eqnarray}
 C(t) \opsimeq_{t \to + \infty} \bigg[ 2 \sqrt{p(1-p)} \bigg]^t  
\label{corretimenonchaotic}
\end{eqnarray}

%%%%%%%%%%%%%%%%%%%%%%%%%%%%%%%%%%%

\subsubsection{ Asymptotic analysis at the critical point $p_c=\frac{1}{2}$}

At the critical point $p_c=\frac{1}{2}$,  the series expansions of $r_{\pm}(s)$ in Eqs \ref{2dsolcriti}
yield that the time-Laplace-transform ${\tilde C}(s) $ of Eq. \ref{correlaplace} diverges as $s^{-\frac{1}{2}}$ for $s \to 0^+$
\begin{eqnarray}
{\tilde C}(s) \opsimeq_{s \to 0}  \frac{1}{  \sqrt{ 2s}  }  
\label{correlaplacecriti}
\end{eqnarray}
that corresponds to the following power-law decay of the time-correlation $C(t)$ for large $t$
\begin{eqnarray}
C(t)  \opsimeq_{t \to + \infty}  \frac{ 1}{\sqrt{2 \pi t} } 
\label{correcrititime}
\end{eqnarray}

%%%%%%%%%%%%%%%%%%%%%%%%%%%%%%%%%%

\subsection{ Discussion } 

In conclusion, when one wishes to study an observable of the Pelikan dynamics that cannot be computed
within the first perspective (i.e. with probability densities that remain constant on the binary-intervals $x \in [ 2^{-n-1}, 2^{-n}[$ partitioning the interval $x \in [0,1[$),
it is useful to work with the decomposition $x_t = \sum_{l=1}^{+\infty} \frac{\sigma_l (t)}{2^l} $
into the binary coefficients $\sigma_l (t) \in \{0,1 \}  $  whose dynamics can be rephrased by the two global variables 
$(z_t,F_t)$. The dependence of the observable with respect to these two variables $(z_t,F_t)$ 
will then determine the level of difficulty of the corresponding computation.
 
Note that even when the initial condition at $t=0$ is the uniform distribution $\rho_{t=0}(x_0)=1$ on $x_0 \in [0,1[$,
one should distinguish between two types of observables: 

(i) one-time observables that can be computed from the probability density $\rho^{[Partition]}_t(x)$ 
\begin{eqnarray}
 \langle O(x_t)  \rangle_{\rho_{t=0}^{\mathrm {Uniform}}}   \equiv \int_0^1 dx O(x) \rho^{[Partition]}_t(x)
\label{Oaverageunifdef}
\end{eqnarray}
can be evaluated either from the distribution $\pi_t(n)$ of the first perspective using Eq. \ref{BinaryDensity}
 or from the distribution $P_t(z)$ of the second perspective using Eq. \ref{UniformDensitysigmadyninx}
\begin{eqnarray}
 \langle O(x_t)  \rangle_{\rho_{t=0}^{\mathrm {Uniform}}} 
 && =\sum_{n=0}^{+\infty} 2^{n+1} \pi_t(n) \int_{2^{-n-1}}^{2^{-n}} dx O(x) \ \ \ \ \text{ in the first perspective }
  \nonumber \\ 
 && =  \sum_{z=0}^{+\infty}  \frac{ P_t(z) }{ 2^{-z} } \int_0^{2^{-z}} dx O(x) \ \ \ \ \text{ in the second perspective }
\label{Oaverageunif}
\end{eqnarray}
  while the direct relation between $\pi_t(n) $ and $P_t(z) $ was written in Eq. \ref{defpizt}.
  We have discussed the simple example of the observable $O(x)=x$ around Eqs \ref{averageunif}, \ref{averageunifinfty}
  and \ref{avxstarsteady}.

(ii) the other observables that cannot be computed from the probability density $\rho^{[Partition]}_t(x)$ alone,
cannot be evaluated within the first perspective, but can be evaluated within the second perspective where the variable $F$ will appear explicitly besides the variable $z$.  In the previous subsection \ref{subsec_corre},
we have discussed in detail the example of the time-correlation 
$C(t)=\langle x_t x_0 \rangle_{\rho_{t=0}^{\mathrm {Uniform}}} $ that cannot be evaluated within the first perspective,
but that can be evaluated within the second perspective via Eq. \ref{corretime} that involves the two global variables $z$ and $F$.

%%%%%%%%%%%%%%%%%%%%%%%%%%%%%%%

 \section{ Conclusions }

 \label{sec_conclusion}
 
 In summary, we have revisited the Pelikan random map via two perspectives:
 
 \vskip 0.5cm

1) In the first perspective considered in the sections \ref{sec_markovBinary}, \ref{sec_propagator} and \ref{sec_excursions}, we have focused on the closed dynamics within the subspace of probability densities that remain constant on the binary-intervals $x \in [ 2^{-n-1}, 2^{-n}[$ partitioning the interval $x \in [0,1[$:  

$\bullet$ In section \ref{sec_markovBinary}, we have explained that the dynamics for the weights $\pi_t(n)$ of these intervals corresponds to a biased random walk on the half-infinite lattice $n \in \{0,1,2,..+\infty\}$ with resetting occurring with probability $p$ from the origin $n=0$ towards any site $n$ drawn with the distribution $2^{-n-1}$. 
We have also recalled the main properties of the corresponding steady state $\pi_*(n)$ existing in the chaotic region
$\frac{1}{2} < p \leq 1 $.

$\bullet$ In section \ref{sec_propagator}, in order to characterize 
 the convergence towards the non-equilibrium steady state in the chaotic region $\frac{1}{2} < p \leq 1 $, 
as well as the transient dynamics in the non-chaotic region $ 0 \leq p < \frac{1}{2} $, and 
 the intermittent dynamics at the critical point $p_c=\frac{1}{2}$,
we have studied the finite-time propagator $\pi_t(n \vert n_0) $ 
via its explicit time-Laplace transform in subsection \ref{subsec_laplacepropagator},
and via its explicit spectral  decomposition in subsection \ref{subsec_propagatorspectral}.
In subsection \ref{subsec_firstpassagetime}, 
we have also analyzed the first-passage-time distribution $F_{t_1}(n \vert n_0)$
via its explicit time-Laplace transform in order to extract its asymptotic behavior
for large time $t$ in the various phases.

$\bullet$ Finally in section \ref{sec_excursions},
 we have discussed the decomposition of the trajectories $n(0 \leq t \leq T)$ with respect to the reset events and the excursions between them in order to characterize their statistics  as a function of the parameter $p$,
 and to have a better understanding of the properties of whole trajectories.

 \vskip 0.5cm

2) In the second perspective considered in the last section \ref{sec_spins}, we have analyzed the Pelikan map for any initial condition $x_0$ via the binary decomposition $x_t = \sum_{l=1}^{+\infty} \frac{\sigma_l (t)}{2^l} $:

$\bullet$ We have explained that the dynamics for the half-infinite lattice $l=1,2,..$ of the binary variables $\sigma_l(t) \in \{0,1\}$ of subsection \ref{subsec_spins}
can be rephrased in terms of two global variables whose physical meaning is explained in subsection \ref{subsec_zF}: $z_t$ corresponds to a biased random walk on the half-infinite lattice $z \in \{0,1,2,..+\infty\}$ that may remain at the origin $z=0$ with probability $p$, while $F_t \in \{0,1,2,...,t\}$ is the additive observable of the stochastic trajectory $z_{0 \leq \tau \leq t-1} $ that counts the number of time-steps $\tau \in [0,t-1]$ where $z_{\tau+1}=0=z_{\tau}$. 

$\bullet$ In subsection \ref{subsec_closedz}, we have discussed the dynamics
 of the distribution $P_t(z)$ of the variable $z $ alone 
and we have explained the link with the dynamics of the weights $\pi_t(n)$ of the first perspective
(see Eq. \ref{defpizt} for the relation between $\pi_t(n) $ and $P_t(z) $).

$\bullet$ In subsection \ref{subsec_additive}, 
we have characterized the joint probability distribution $P_t(z,F)$ via the explicit spectral decomposition
of the generating function $Z^{[\nu]}_t(z) $  
with respect to the variable $F$ as a function of the conjugated parameter  $\nu$.
 In particular, we have written in Eq. \ref{rateIf} the explicit rate function $I(f)$ that governs the large deviations of the density $f=\frac{ F_T}{T} \in [0,1]$ of spins erased per unit time during the large-time window $[0,T]$ in the chaotic region $\frac{1}{2} < p \leq 1 $ and at the critical point $p_c=\frac{1}{2}$. 

$\bullet$ Finally in subsection \ref{subsec_corre}, we have focused on the time-correlation  $C(t)=\langle x_t x_0 \rangle $ when the initial condition $x_0$ is drawn with the uniform distribution : we have computed its explicit time-Laplace transform in Eq. \ref{correlaplace}
and we have written the asymptotic behaviors for large time $t$ as a function of the parameter $p$
(see Eqs \ref{corretimeconnectedchaos}, \ref{corretimenonchaotic} and \ref{correcrititime}).

 %%%%%%%%%%%%%%%%%%%%%%%%%%%%%%%%%%%%%%%%%%%%%%%%%%%

 \appendix

 \section{ Computation of the time-Laplace-transform ${\tilde \pi}_s(n \vert n_0) $ of the propagator $\pi_t(n \vert n_0)$ }

\label{app_laplace}

In this Appendix, we describe how the time-Laplace-transform ${\tilde \pi}_s(n \vert n_0) $
can be computed as the solution of the recurrence of Eq. \ref{piLaplacerec} written in the main text.

\subsection{ Analysis of the general solution of the recurrence }

The general solution of the homogeneous recurrence associated to Eq. \ref{piLaplacerec}
\begin{eqnarray}
\phi_s(n )  =  p e^{-s }  \phi_s(n+1) + (1-p) e^{-s }  \phi_s(n-1)  
\label{piLaplacerechomo}
\end{eqnarray}
can be written as a linear combination involving two arbitrary constants $C_{\pm} $ 
\begin{eqnarray}
\phi_s(n )  =  C_+ r_+^n +C_-  r_-^n  
\label{recGeneralSolhomo}
\end{eqnarray}
where $r_{\pm}^n $ are two independent solutions of the recurrence of Eq. \ref{piLaplacerechomo}
based on the two roots $r_{\pm}$ of the second order equation
\begin{eqnarray}
0=    pr + \frac{1-p}{r} - e^s \equiv f(r)
\label{2d}
\end{eqnarray}
that read
\begin{eqnarray}
r_{\pm} = \frac{ e^s }{2p } \left[ 1 \pm \sqrt{ 1 - 4  p(1-p) e^{-2s}} \right]
\label{2dsol}
\end{eqnarray}
with their sum and their product 
\begin{eqnarray}
r_++r_- && = \frac{ e^s }{  p }
\nonumber \\
r_+r_- && = \frac{1-p}{p}
\label{sumprod}
\end{eqnarray}
It is also useful to rewrite the function $f(r)$ of Eq. \ref{2d}
in the factorized form
\begin{eqnarray}
f(r) =    pr + \frac{1-p}{r} - e^s = \frac{p}{r} (r-r_+)(r-r_-)
\label{frfactorized}
\end{eqnarray}
At $r=1$, the function $f(r)$ takes the negative value for $s>0$
\begin{eqnarray}
f(r=1) =    1 - e^s = p (1-r_+)(1-r_-) = - p (r_+-1)(1-r_-)
\label{frfactorized1}
\end{eqnarray}
so that $r=1$ is between the two positive roots $r_{\pm} $ for $s>0$
\begin{eqnarray}
0 < r_- < 1 < r_+ \ \ \ \ \ \ \ \text{ for } s>0
\label{orderroots}
\end{eqnarray}

When the last term of Eq. \ref{piLaplacerec}
 is added to the homogeneous recurrence of Eq. \ref{piLaplacerechomo}
 to obtain the inhomogeneous recurrence
\begin{eqnarray}
\Phi_s(n )  =  p e^{-s }  \Phi_s(n+1) + (1-p) e^{-s }  \Phi_s(n-1)   +p e^{-s }2^{-n-1} {\tilde \pi}_s(0\vert n_0)  
\label{piLaplacerecinhomo}
\end{eqnarray}
one can write a particular solution behaving as $2^{-n}$ 
\begin{eqnarray}
\Phi_s^{\mathrm {Particular}}(n )  =   B 2^{-n} {\tilde \pi}_s(0\vert n_0)  
\label{piLaplacerecinhomoParticular}
\end{eqnarray}
with the prefactor 
\begin{eqnarray}
B    \equiv  \frac{ p  2^{-1} }{e^s- \frac{p}{2}    - 2(1-p)    }
=  \frac{ p   }{ 2 e^{s }- 4 +3p    }
\label{bsolu}
\end{eqnarray}
that can also be rewritten in terms of the function $f(r)$ of Eq. \ref{frfactorized} 
at the point $r=\frac{1}{2}$
\begin{eqnarray}
f\left(r=\frac{1}{2}\right) && =   \frac{p}{2} + 2(1-p) - e^s =  - \frac{  2e^s -4+3p}{2} 
\nonumber \\
&& = 2 \left(\frac{1}{2}-r_+\right)\left(\frac{1}{2}-r_-\right) = - \frac{ (2 r_+-1) (1-2 r_-)}{2}
\label{frfactorized12}
\end{eqnarray}
to obtain
\begin{eqnarray}
B   = \frac{p  2^{-1} }{ [ - f \left(\frac{1}{2} \right) ]}  = \frac{1}{ (2 r_+-1) (1-2 r_-)}
\label{bsolupm}
\end{eqnarray}

The general solution of Eq. \ref{piLaplacerecinhomo} 
can be then obtained as the sum of the particular solution of Eq. \ref{piLaplacerecinhomoParticular}
and of the general solution of Eq. \ref{recGeneralSolhomo}
\begin{eqnarray}
\Phi_s^{\mathrm {General}}(n )  =  C_+ r_+^n +C_-  r_-^n  +  B 2^{-n} {\tilde \pi}_s(0\vert n_0)  
\label{recGeneralSol}
\end{eqnarray}

Let us now discuss how this general solution can be used to solve Eq. \ref{piLaplacerec} 
for $n_0>0$ and for $n_0=0$ in the two next subsections.
In order to simplify the computations, it is convenient to make a change of variables from 
the two parameters $(p,s)$ towards the two roots $r_{\pm}$ satisfying Eq. \ref{sumprod}
\begin{eqnarray}
 p && = \frac{1}{1+r_+ r_- }
 \nonumber \\
1-p && = \frac{r_+ r_-}{1+r_+ r_- }
\nonumber \\
e^s  && = \frac{r_+ + r_-}{1+r_+ r_- }
\label{sptowardsrpm}
\end{eqnarray}
in order to rewrite Eq. \ref{piLaplacerec}  as
\begin{eqnarray}
0 = (r_+ + r_- ) \left[ \delta_{n,n_0} -  {\tilde \pi}_s(n \vert n_0)  \right]  
+   {\tilde \pi}_s(n+1\vert n_0) + r_+ r_- {\tilde \pi}_s(n-1\vert n_0)  \theta(n \geq 1 )
 + 2^{-n-1} {\tilde \pi}_s(0\vert n_0)  
\label{piLaplacerecrpm}
\end{eqnarray}

%%%%%%%%%%%%%%%%%%%%%%%%%%%%%%%%%%%%%%%%%%%%

\subsection{ Solution ${\tilde \pi}_s(n \vert n_0) $ for any initial condition $n_0>0$ }

In order to satisfy Eq. \ref{piLaplacerecrpm} in the whole region $n>n_0$,
 the solution ${\tilde \pi}_s(n \vert n_0) $ 
should follow the general form of Eq. \ref{recGeneralSol} for $n \geq n_0$, 
but the normalizability at $n \to + \infty$ required by Eq. \ref{piLaplacenorma}
excludes the presence of the root $r_+>1$ of Eq. \ref{orderroots}, so that one is left
with the parametrization in terms of a single constant $C$
\begin{eqnarray}
{\tilde \pi}_s(n \vert n_0)   = C  r_-^n  + B 2^{-n} {\tilde \pi}_s(0\vert n_0) \ \ \text{ for } \ \ n \geq n_0
\label{recGeneralSolAbovedef}
\end{eqnarray}

In order to satisfy Eq. \ref{piLaplacerecrpm} in the whole region $0 < n <n_0$,
the solution ${\tilde \pi}_s(n \vert n_0) $ should
follow the general form of Eq. \ref{recGeneralSol} for $0 \leq n \leq n_0 $ involving two constants $c_{\pm}$
\begin{eqnarray}
{\tilde \pi}_s(n \vert n_0)   =c_+ r_+^n +c_-  r_-^n  + B 2^{-n} {\tilde \pi}_s(0\vert n_0)
\ \ \text{ for } \ \ 0 \leq n \leq n_0
\label{recGeneralSolbelow}
\end{eqnarray}
and should satisfy the following conditions at the two boundaries $n=n_0$ and at $n=0$.

%%%%%%%%%%%%%%%%%%%%%%%%%%%%%%%%%%%%%%%%%

\subsubsection{ Conditions at the boundary $n=n_0$ between the two regions $n \geq n_0$ and $n \leq n_0$ }

At $n=n_0$, the consistency for ${\tilde \pi}_s(n=n_0 \vert n_0) $ 
given by Eq. \ref{recGeneralSolAbovedef}
and by Eq. \ref{recGeneralSolbelow} determines $C$ in terms of $c_{\pm}$
\begin{eqnarray}
 C   =c_+ \left( \frac{r_+}{r_- } \right)^{n_0} +c_-   
\label{recGeneralSolbelown0}
\end{eqnarray}
that can be plugged into Eq. \ref{recGeneralSolAbovedef} to obtain
\begin{eqnarray}
{\tilde \pi}_s(n \vert n_0)   =  c_+ r_+^{n_0} r_-^{n-n_0} +c_-  r_-^n  + B 2^{-n} {\tilde \pi}_s(0\vert n_0) \ \ \text{ for } \ \ n \geq n_0
\label{recGeneralSolAbove}
\end{eqnarray}

Since Eq. \ref{piLaplacerecrpm} for $n=n_0$ involves the supplementary term $\delta_{n,n_0}=1$, 
while the terms involving $c_-$ and $B$ 
are the same in the two regions of Eq. \ref{recGeneralSolbelow} and \ref{recGeneralSolAbove}
and thus satisfy the general recurrence of Eq. \ref{piLaplacerecinhomo},
the writing of Eq. \ref{piLaplacerecrpm} for $n=n_0$ will only involve the contributions of the terms 
with the prefactor $c_+$
that are different in the two regions of Eq. \ref{recGeneralSolbelow} and \ref{recGeneralSolAbove}
\begin{eqnarray}
0 && = (r_+ + r_- ) \left[ 1 -  {\tilde \pi}_s(n_0 \vert n_0)  \right]  
+   {\tilde \pi}_s(n_0+1\vert n_0) + r_+ r_- {\tilde \pi}_s(n_0-1\vert n_0)  \theta(n \geq 1 )
 + 2^{-n_0-1} {\tilde \pi}_s(0\vert n_0)  
 \nonumber \\
 && = (r_+ + r_- ) \left[ 1 -  c_+ r_+^{n_0}  \right]  
+   c_+ r_+^{n_0} r_- + r_+ r_-   c_+ r_+^{n_0-1}
 \nonumber \\
 && = (r_+ + r_- ) -   c_+ r_+^{n_0}  (r_+ - r_- ) 
\label{piLaplacerecrpmzero}
\end{eqnarray}
so that $c_+$ reads 
\begin{eqnarray}
c_+   = r_+^{-n_0}  \frac{   r_+  +    r_- } {     r_+  -    r_- }
\label{solucplus}
\end{eqnarray}

%%%%%%%%%%%%%%%%%%%%%%%%%%%%%%%%%%%%%%%%%

\subsubsection{ Conditions at the boundary $n=0$  }

At $n=0$, the consistency of Eq. \ref{recGeneralSolbelow} 
yields
\begin{eqnarray}
c_- = (1-B) {\tilde \pi}_s(0 \vert n_0)   - c_+  
\label{cmoinspizero}
\end{eqnarray}
while Eq. \ref{piLaplacerecrpm} for ${\tilde \pi}_s(n \vert n_0)  $ at $n=0$ 
where the term $n=-1$ is absent with respect to the general recurrence for $n \geq 1$
will be satisfied only if the continuation of Eq. \ref{recGeneralSolbelow} vanishes for $n=-1$
\begin{eqnarray}
0&& ={\tilde \pi}_s(n=-1 \vert n_0)    =c_+ r_+^{-1} +c_-  r_-^{-1}  + 2B  {\tilde \pi}_s(0\vert n_0)
= c_+ r_+^{-1} +\left[ (1-B) {\tilde \pi}_s(0 \vert n_0)   - c_+ \right]   r_-^{-1}  + 2B  {\tilde \pi}_s(0\vert n_0)
\nonumber \\
&& = c_+ (r_+^{-1} - r_-^{-1} )  + \left[ (1-B)  r_-^{-1} + 2B \right] {\tilde \pi}_s(0\vert n_0)
\label{recGeneralSolbelowAbs}
\end{eqnarray}
so that one obtains using $B$ of Eq. \ref{bsolupm}
and $c_+$ of Eq .\ref{solucplus}
\begin{eqnarray}
{\tilde \pi}_s(0\vert n_0)&& = \frac{ c_+ (r_+ -  r_-)  }{  r_+ \left[ 1-B (1-   2 r_-) \right] }
=  \frac{ r_+^{-n_0} (   r_+  +    r_- )  }{  r_+ \left[ 1+ \frac{1}{ (1-2 r_+) }  \right] }
\nonumber \\
&& =  r_+^{-n_0-1} \frac{ (   r_+  +    r_- ) (2 r_+-1) }{  2  ( r_+ -1  ) }
\label{solupizero}
\end{eqnarray}
with the corresponding product $B {\tilde \pi}_s(0\vert n_0)$ 
\begin{eqnarray}
B {\tilde \pi}_s(0\vert n_0)&& =    r_+^{-n_0-1} \frac{ (   r_+  +    r_- )  }{  2  ( r_+ -1  ) (1-2 r_-)}
\label{solupizeroB}
\end{eqnarray}
and the constant $c_-$ of Eq. \ref{cmoinspizero}
\begin{eqnarray}
c_- && = {\tilde \pi}_s(0 \vert n_0) -B {\tilde \pi}_s(0 \vert n_0)   - c_+  
\nonumber \\
&& =   r_+^{-n_0-1}  (   r_+  +    r_- ) 
\left[ \frac{(2 r_+-1) }{  2  ( r_+ -1  ) }
-   \frac{  1  }{  2  ( r_+ -1  ) (1-2 r_-)}
-   \frac{  r_+ } {     r_+  -    r_- } \right]
\nonumber \\
&& = -  r_+^{-n_0-1}  (   r_+  +    r_- ) r_-
\left[    \frac{  1  }{    ( r_+ -1  ) (1-2 r_-)}
+   \frac{  1 } {     r_+  -    r_- } \right]
\label{cmoinssolu}
\end{eqnarray}

One can check that the solution of Eqs \ref{recGeneralSolAbove} and \ref{recGeneralSolbelow}
satisfies the normalization of Eq. \ref{piLaplacenorma}
for any $n_0$
 \begin{eqnarray}
\sum_{n=0}^{+\infty} {\tilde \pi}_s(n \vert n_0) && 
=  \sum_{n=n_0}^{+\infty} c_+ r_+^{n_0} r_-^{n-n_0} 
+ \sum_{n=0}^{n_0-1} c_+ r_+^n 
+ \sum_{n=0}^{+\infty} \left[  c_- r_-^n  + B 2^{-n} {\tilde \pi}_s(0\vert n_0) \right]
\nonumber \\
&& 
= \frac{c_+ r_+^{n_0}}{1-r_-}
+ \frac{c_+ (1-r_+^{n_0})}{1-r_+}
+  \frac{c_- }{1-r_-}
+ 2 B  {\tilde \pi}_s(0\vert n_0)
 \nonumber \\
 \nonumber \\
&& 
= c_+r_+^{n_0} \left[ \frac{1}{1-r_-} - \frac{1}{1-r_+} \right]
+ \left( \frac{c_+ }{1-r_+}
+  \frac{c_- }{1-r_-}
+ 2 B  {\tilde \pi}_s(0\vert n_0) \right)
 \nonumber \\
&&= \frac{1}{1-e^{-s} } + 0
\label{piLaplacenormacheck}
\end{eqnarray}
using Eqs \ref{solucplus},  \ref{solupizeroB} and \ref{cmoinssolu}.

%%%%%%%%%%%%%%%%%%%%%%%%%%%%%%%%%%%%%%%%%

\subsubsection{ Conclusion  }

Putting everything together, the solution of Eqs \ref{recGeneralSolAbove} and \ref{recGeneralSolbelow} can be written in terms $r_{\pm}$ of Eq. \ref{2dsol}
as
\begin{eqnarray}
\text{ for } \ \ n \geq n_0 >0 : \ \ \ 
{\tilde \pi}_s(n \vert n_0)   
&& =    \left(\frac{   r_+  +    r_- } {     r_+  -    r_- } \right) r_-^{n-n_0} 
 -    (   r_+  +    r_- ) 
\left(    \frac{  1  }{    ( r_+ -1  ) (1-2 r_-)}+   \frac{  1 } {     r_+  -    r_- } \right)  r_-^{n+1}  r_+^{-n_0-1}
\nonumber \\
&& +   \frac{ (   r_+  +    r_- )  }{    ( r_+ -1  ) (1-2 r_-)}    2^{-n-1}   r_+^{-n_0-1}
\nonumber \\
\text{ for } \ \ 0 \leq n < n_0 : \ \ \ 
{\tilde \pi}_s(n \vert n_0)   
&& =   \left( \frac{   r_+  +    r_- } {     r_+  -    r_- } \right) r_+^{n-n_0} 
 -    (   r_+  +    r_- ) 
\left(    \frac{  1  }{    ( r_+ -1  ) (1-2 r_-)}+   \frac{  1 } {     r_+  -    r_- } \right)  r_-^{n+1}  r_+^{-n_0-1}
\nonumber \\
&&+   \frac{ (   r_+  +    r_- )  }{    ( r_+ -1  ) (1-2 r_-)} 2^{-n-1}   r_+^{-n_0-1}
 \label{recapexpanded}
\end{eqnarray}
so that it is convenient to regroup the terms with the same denominators to obtain
the final form given in Eq. \ref{recap} of the main text.
This solution derived above for any initial condition $n_0>0$
is actually still valid for $n_0=0$ as described in the next subsection.

%%%%%%%%%%%%%%%%%%%%%%%%%%%%%%%%%%%%%%%%%%%%%%

\subsection{ Solution ${\tilde \pi}_s(n \vert n_0=0) $ for the special initial condition $n_0=0$ }

For the special initial condition $n_0=0$, the region $0 \leq n \leq n_0$ collapses into the single point $n=0$,
so that the previous analysis can be adapted as follows:
the solution of Eq. \ref{recGeneralSolAbove} involving some constant $C_0$
\begin{eqnarray}
{\tilde \pi}_s(n \vert 0)   = C_0  r_-^n  + B 2^{-n} {\tilde \pi}_s(0\vert 0) \ \ \text{ for } \ \ n \geq 0
\label{recGeneralSolAbovenzerovanish}
\end{eqnarray}
should be consistent for $n=0$ 
\begin{eqnarray}
{\tilde \pi}_s(0 \vert 0)   = C_0   +B {\tilde \pi}_s(0\vert 0) 
\label{recGeneralSolAboven000}
\end{eqnarray}
leading to
\begin{eqnarray}
 C_0   =(1-B) {\tilde \pi}_s(0\vert 0) 
\label{recGeneralSolAboven000c}
\end{eqnarray}
while 
Eq. \ref{piLaplacerecrpm} for $n=0=n_0$ where the term $\delta_{n,n_0}=1$ is present
and where the term $n=-1$ is absent is equivalent
 to the following requirement for the continuation of Eq. \ref{recGeneralSolAbovenzerovanish} to $n=-1$
\begin{eqnarray}
0 && = (r_+ + r_- )  -  (r_+ + r_- ) {\tilde \pi}_s(0 \vert 0)  
+   {\tilde \pi}_s(1\vert 0) + r_+ r_- {\tilde \pi}_s(n-1\vert 0)  \theta(n \geq 1)
 + 2^{-n-1} {\tilde \pi}_s(0\vert 0)  
 \nonumber \\
 &&  = (r_+ + r_- )  - r_+ r_- {\tilde \pi}_s(-1\vert 0) 
 \nonumber \\
 &&  = (r_+ + r_- )  - r_+ r_- \left[ C_0  r_-^{-1}  + B 2 {\tilde \pi}_s(0\vert 0)\right]
  \nonumber \\
 &&  = (r_+ + r_- )  - r_+ r_- \left[ (1-B)   r_-^{-1}  + B 2 \right]{\tilde \pi}_s(0\vert 0) 
 \label{piLaplacerecrpm000}
\end{eqnarray}
 so that one obtains 
using $B$ of Eq. \ref{bsolupm}  
\begin{eqnarray}
{\tilde \pi}_s(0\vert 0)= \frac{ r_+ + r_- }{r_+   \left[ 1-B (1-  2 r_-) \right]}
=  \frac{ r_+ + r_-  }{ r_+    \left[ 1- \frac{1}{ 2 r_+ -1 }  \right]}
= \frac{ (r_+ + r_-) (2 r_+-1)  }{ r_+   2 (r_+-1)}
\label{solupi00}
\end{eqnarray}
with the corresponding product
\begin{eqnarray}
 B {\tilde \pi}_s(0\vert 0) =  \frac{ (r_+ + r_-)   }{ r_+   2 (r_+-1)  (1-2 r_-)} 
\label{Bpi00}
\end{eqnarray}
and $C_0$ of Eq. \ref{recGeneralSolAboven000c}
\begin{eqnarray}
C_0 && = {\tilde \pi}_s(0\vert 0) - B {\tilde \pi}_s(0\vert 0) 
=  \frac{ (r_+ + r_-)   }{ r_+   2 (r_+-1)} \left[ 2 r_+-1- \frac{ 1   }{  1-2 r_-} \right]
\nonumber \\
 && = \frac{ (r_+ + r_-)   }{ r_+   } \left[ 1  - \frac{  r_-   }{ (r_+-1) (1-2 r_-)} \right]
\label{soluC0}
\end{eqnarray}

Putting everything together, 
the solution $ {\tilde \pi}_s(n \vert n_0=0) $ of Eq. \ref{recGeneralSolAbovenzerovanish} 
reads
\begin{eqnarray}
\text{ for } \ \ n \geq n_0 =0 : \ \ \ 
 {\tilde \pi}_s(n \vert 0)  
 && =   \frac{ (r_+ + r_-)   }{ r_+   } \left[ 1  - \frac{  r_-   }{ (r_+-1) (1-2 r_-)} \right]  r_-^n  
 + \frac{ (r_+ + r_-)   }{ r_+   2 (r_+-1)  (1-2 r_-)}   2^{-n} 
 \nonumber \\
 && = \frac{ (r_+ + r_-)   }{ r_+   }r_-^n
 + \frac{ (r_+ + r_-)   }{ r_+    (r_+-1)  (1-2 r_-)}  ( 2^{-n-1} - r_-^{n+1} )
\label{BilanSolutionnzerovanish}
\end{eqnarray}
and actually coincides with the naive replacement $n_0 \to 0$ in the solution of Eq. \ref{recapexpanded} 
computed in the previous subsection for $n_0>0$.

%%%%%%%%%%%%%%%%%%%%%%%%%%%%%%%%%%%%%%%%%%

\section{ Reminder on the biased Random-Walk on the half-line with an absorbing boundary condition }

\label{app_RW}

In this Appendix, we recall the properties of the biased random walk that jumps 
to the left neighbor with probability $p$ and to the right neighbor with probability $(1-p)$,
first on the whole one-dimensional lattice $n \in \{-\infty,...,+\infty \}$, 
and then on the half-lattice $n=0,1,..,+\infty$ with an absorbing boundary condition at $n=-1$.

\subsection{ Reminder on the free propagator $P^{\mathrm {Free}}_t(m=n - n_0)$ on the whole one-dimensional lattice $n \in \{-\infty,...,+\infty \}$ }

The probability $P^{\mathrm {Free}}_t(m)$ of the displacement $m=n-n_0$ between the position $n$ at time $t$ 
and the position $n_0$ at time $t=0$ satisfies the recurrence
\begin{eqnarray}
P^{\mathrm {Free}}_t(m) && = (1-p) P^{\mathrm {Free}}_{t-1}(m-1) + p P^{\mathrm {Free}}_{t+1}(m+1)
\label{eqGfree}
\end{eqnarray}
and the initial condition $P^{\mathrm {Free}}_{t=0}(m)=\delta_{m,0}$.

\subsubsection{ Space-time propagator $P^{\mathrm {Free}}_t(m)$ in terms of the binomial distribution }

The $t$ steps can be decomposed into the number $t_+ \in  \{0,1,..,t\}$ of steps to the right and 
the complementary number $t_-=t-t_+$ of steps to the left, while their difference 
produces the total displacement $m=n-n_0=t_+-t_-=2 t_+ -t $. 
Plugging
\begin{eqnarray}
t_+  = \frac{t+ m}{2}
\label{tpm}
\end{eqnarray}
into the binomial distribution for $t_+$ leads
to the space-time propagator  
\begin{eqnarray}
P^{\mathrm {Free}}_t(m) && = \frac {t!}{ t_+ ! (t-t_+) !} (1-p)^{t_+} p^{t-t_+} 
= \frac {t!}{ \left( \frac{t+m}{2}\right) ! \left(\frac{t-m}{2}\right) !} 
(1-p)^{\frac{t+m}{2}} p^{\frac{t-m}{2}} 
\ \ \text{ for } m \in \{-t,...,+t\}
\label{Gtpm}
\end{eqnarray}

%%%%%%%%%%%%%%%%%%%%%%%%%%%%%%%%%%%

\subsubsection{ Spectral decomposition of the propagator $P^{\mathrm {Free}}_t(m=n-n_0) $}

The spectral decomposition of the propagator $P^{\mathrm {Free}}_t(m) $ of Eq. \ref{Gtpm}
\begin{eqnarray}
P^{\mathrm {Free}}_t(m)  
= \left( \frac{1-p}{p} \right)^{\frac{m}{2} }
\int_{-\pi}^{ \pi} \frac{dq}{2 \pi} \bigg[ \lambda (q) \bigg]^t e^{i q m } 
\label{Gspectral}
\end{eqnarray}
involves the continuous spectrum $\lambda(q)$ parametrized by the Fourier momentum $q \in ]-\pi, \pi]$
\begin{eqnarray}
\lambda(q) \equiv \sqrt{ p (1-p) } (e^{iq}+e^{-iq} ) = \sqrt{ 4 p (1-p) } \cos q 
\label{lambdaq}
\end{eqnarray}
with the maximal value for zero momentum $q=0$
\begin{eqnarray}
 \lambda(q=0) =  \sqrt{ 4 p (1-p) } 
 \  \begin{cases}
=1 \ \ \ \ \ \text{ for  }  p_c=\frac{1}{2}
 \\
<1  \ \ \ \ \  \text{ for } p \ne \frac{1}{2}
\end{cases}
\label{lambdaq0}
\end{eqnarray}

%%%%%%%%%%%%%%%%%%%%%%%%%%%%%%%%%

\subsubsection{ Time-Laplace-transform ${\tilde P}^{\mathrm {Free}}_s(m) $ of the propagator $P^{\mathrm {Free}}_t(m) $}

 The time-Laplace-transform of the propagator of Eq. \ref{Gtpm}
\begin{eqnarray}
{\tilde P}^{\mathrm {Free}}_s(m) \equiv \sum_{t=0}^{+\infty} e^{-s t} P^{\mathrm {Free}}_t(m)  
= \begin{cases}
\frac{r_+(s)+r_-(s)}{ r_+(s)-r_-(s) } [ r_-(s) ]^m  \ \ \ \ \ \ \ \ \ \ \ \  \text{for } m \geq 0 
 \\ %
 \\
\frac{r_+(s)+r_-(s)}{ r_+(s)-r_-(s) } \left[ r_+(s)  \right]^m  \ \ \ \ \ \ \ \ \ \  \text{for } m \leq 0 
\end{cases}
\label{GLaplace}
\end{eqnarray}
involves the two roots $r_{\pm}(s)$ introduced in Eq. \ref{2dsol} of the previous Appendix.

%%%%%%%%%%%%%%%%%%%%%%%%%%%%%%%%%%%%

\subsection{ Propagator $P^{Abs}_t(n \vert n_0)$ with an absorbing boundary condition at $n=-1$ }

\subsubsection{ Main observables in the presence of an absorbing boundary condition at $n=-1$  }

In the presence of an absorbing boundary condition at $n=-1$,
the probability $P^{Abs}_t(n \vert n_0)$ to be surviving at position $n \in \{0,1,2,..+\infty\}$ at time $t$
when starting at position $n_0$ at time $t=0$
satisfies the recurrence
\begin{eqnarray}
P^{Abs}_{t}(n \vert n_0) && = (1-p) P^{Abs}_{t-1}(n-1 \vert n_0) \theta(n \geq 1) + p P^{Abs}_{t-1}(n+1 \vert n_0)  
\label{recabs}
\end{eqnarray}
and the initial condition $P^{Abs}_{t=0}(x \vert x_0)= \delta_{n,n_0}$.

The sum over all the possible positions $n=0,1,..$
of the propagator $P^{Abs}_{t}(n \vert n_0)  $ of Eq. \ref{recabs}
yields that the survival probability $S(t \vert n_0)  $ at time $t$ evolves according to
\begin{eqnarray}
S(t \vert n_0) && \equiv \sum_{n=0}^{+\infty} P^{Abs}_t(n \vert n_0) 
= \sum_{n=0}^{+\infty} \left[ (1-p) P^{Abs}_{t-1}(n-1 \vert n_0) \theta(n \geq 1) + p P^{Abs}_{t-1}(n+1 \vert n_0)  \right]
\nonumber \\
&& =(1-p) S(t -1 \vert n_0) + p \left[S(t -1 \vert n_0) - P^{Abs}_{t-1}(0 \vert n_0) \right]
= S(t -1 \vert n_0) - p P^{Abs}_{t-1}(0 \vert n_0)
\label{SurvivalPtimagesevol}
\end{eqnarray}
i.e. the probability to get absorbed at time $t$ 
\begin{eqnarray}
A(t \vert n_0)  \equiv S(t -1 \vert n_0) - S(t \vert n_0)= p P^{Abs}_{t-1}(0 \vert n_0)  
\label{probaAbst}
\end{eqnarray}
can be computed from the probability $P^{Abs}_{t-1}(0 \vert n_0)   $ to be surviving at position $n=0$ at time $(t-1)$.

The survival probability $S(T \vert n_0) $ at time $T$
can be evaluated from the sum of the absorption probability $A(t \vert n_0) $ of Eq. \ref{probaAbst} over the possible times $t=1,2,..,T$
\begin{eqnarray}
S(T \vert n_0) \equiv 1 - \sum_{t=1}^{T} A(t \vert n_0) 
\label{survivalproba}
\end{eqnarray}
The limit $T \to + \infty$ yields the forever-survival probability $S(T=\infty \vert n_0) $,
i.e. to the probability of escape towards $(+ \infty)$ without ever being absorbed at $(-1)$
\begin{eqnarray}
S(\infty \vert n_0) = 1 - \sum_{t=1}^{+\infty} A(t \vert n_0)  
\label{probaAbstinfinitydef}
\end{eqnarray}

%%%%%%%%%%%%%%%%%%%%%%%%%%%%%%%%%%%%%%%%%%%%%

\subsubsection{ Space-time propagator $P^{Abs}_t(n \vert n_0) $ via the method of images }

In the presence of an absorbing boundary condition at $n=-1$,
the method of images with a primary source at $n_0$ and a secondary source at $n_0'=-n_0-2$
yields that $P^{Abs}_t(n \vert n_0) $ can be written in terms of the free propagator $P^{\mathrm {Free}}_t(m)$ of Eq. \ref{Gtpm} as
\begin{eqnarray}
P^{Abs}_t(n \vert n_0) && = P^{\mathrm {Free}}_t(n - n_0) - \left(\frac{p}{1-p}\right)^{n_0+1} P^{\mathrm {Free}}_t(n +n_0+2)
\nonumber \\
&& 
= \frac {t! [ p (1-p)]^{\frac{t}{2}}  }{ \left( \frac{t+(n-n_0)}{2}\right) ! \left(\frac{t-(n-n_0)}{2}\right) !} 
\left(\frac{1-p}{p}\right)^{\frac{(n-n_0)}{2}} 
 - \left(\frac{p}{1-p}\right)^{n_0+1}
\frac {t! [ p (1-p)]^{\frac{t}{2}}  }{ \left( \frac{t+(n+n_0+2)}{2}\right) ! \left(\frac{t-(n+n_0+2)}{2}\right) !} 
\left(\frac{1-p}{p}\right)^{\frac{(n+n_0+2)}{2}} 
\nonumber \\
&& 
= t! [ p (1-p)]^{\frac{t}{2}} \left(\frac{1-p}{p}\right)^{\frac{(n-n_0)}{2}} 
\left[ \frac {1}{ \left( \frac{t+(n-n_0)}{2}\right) ! \left(\frac{t-(n-n_0)}{2}\right) !} 
 -
\frac { 1 }{ \left( \frac{t+(n+n_0+2)}{2}\right) ! \left(\frac{t-(n+n_0+2)}{2}\right) !} 
\right]
\label{Ptimages}
\end{eqnarray}
where the weight $\left(\frac{p}{1-p}\right)^{n_0+1} $  of the secondary source at $n_0'=-n_0-2$
has been chosen to satisfy the absorbing boundary condition at $n=-1$ at any time $t$
\begin{eqnarray}
P^{Abs}_t(n=-1 \vert n_0) && = P_t(-1 - n_0) - \left(\frac{p}{1-p}\right)^{n_0+1} P_t(n_0+1)
\nonumber \\
&& 
=  t! [ p (1-p)]^{\frac{t}{2}} \left(\frac{1-p}{p}\right)^{\frac{(-1-n_0)}{2}} 
\left[ \frac {1}{ \left( \frac{t-(1+n_0)}{2}\right) ! \left(\frac{t+(1+n_0)}{2}\right) !} 
 -
\frac { 1 }{ \left( \frac{t+(1+n_0)}{2}\right) ! \left(\frac{t-(1+n_0)}{2}\right) !} 
\right] =0 \ \ \ 
\label{Ptimagesabs}
\end{eqnarray}

The absorption probability $A(t +1\vert n_0) $ of Eq. \ref{probaAbst}
can be then computed from $P^{Abs}_t(n=0 \vert n_0) $ 
\begin{eqnarray}
A(t +1\vert n_0) &&= p P^{Abs}_t(0 \vert n_0)  
= p\left[ P_t( -n_0) - \left(\frac{p}{1-p}\right)^{n_0+1} P_t(n_0+2)\right]
\nonumber \\
&& =p  t! [ p (1-p)]^{\frac{t}{2}} \left(\frac{1-p}{p}\right)^{- \frac{n_0}{2}} 
\left[ \frac {1}{ \left( \frac{t-n_0}{2}\right) ! \left(\frac{t+n_0}{2}\right) !} 
 -
\frac { 1 }{ \left( \frac{t+n_0}{2} +1\right) ! \left(\frac{t-n_0}{2}-1\right) !} 
\right]
\nonumber \\
&& 
=(n_0+1)  p \left(\frac{1-p}{p}\right)^{- \frac{n_0}{2}} [ p (1-p)]^{\frac{t}{2}}
\frac{   t!   }
{ \left( \frac{t+n_0}{2} +1\right) ! \left( \frac{t-n_0}{2}\right) !} 
\label{probaAbstimages}
\end{eqnarray}
The Stirling approximation for the factorials yields that the asymptotic behavior for large $t$
\begin{eqnarray}
A(t +1\vert n_0) \opsimeq_{t \to + \infty}  
 \frac{  (n_0+1) p}{\sqrt \pi} \left(\frac{1-p}{p}\right)^{- \frac{n_0}{2}} [ 4 p (1-p)]^{\frac{t}{2}} \left( \frac{2}{t} \right)^{\frac{3}{2}} 
\label{stirling}
\end{eqnarray}
is dominated for $p \ne \frac{1}{2}$ by the exponential decay $[ 4 p (1-p)]^{\frac{t}{2}} = [\lambda(q=0)]^t$ 
where one recognizes the highest eigenvalue $\lambda(q=0)= \sqrt{ 4 p (1-p) }  $ of Eq. \ref{lambdaq0},
while it is governed by the power-law decay as $t^{-3/2}$ for the critical case $p_c=\frac{1}{2}$ 
where the highest eigenvalue $\lambda(q=0)= \sqrt{ 4 p (1-p) }  $ reaches unity.

%%%%%%%%%%%%%%%%%%%%%%%%%%%%%%%%%%%%%%%

\subsubsection{ Spectral decomposition of the propagator $P^{Abs}_t(n \vert n_0)   $ }

Plugging the spectral decomposition of $P_t(\cdot)$ of Eq. \ref{Gspectral}
into the first line of Eq. \ref{Ptimages} yields
\begin{eqnarray}
P^{Abs}_t(n \vert n_0) && = P_t(n - n_0) - \left(\frac{p}{1-p}\right)^{n_0+1} P_t(n +n_0+2)
\nonumber \\
&& =  \left( \frac{1-p}{p} \right)^{\frac{n-n_0}{2} }
\int_{-\pi}^{ \pi} \frac{dq}{2 \pi} \bigg[ \lambda (q) \bigg]^t 
e^{i q (n +1) } \bigg( e^{ - i q ( n_0+1) } -  e^{i q (n_0+1) } \bigg)
\nonumber \\
&& =2 \left( \frac{1-p}{p} \right)^{\frac{n-n_0}{2} }
\int_{0}^{ \pi} \frac{dq}{\pi}  \bigg[ \lambda (q) \bigg]^t 
\sin[ q ( n+1) ] \sin[ q ( n_0+1) ]
\nonumber \\
&& \equiv \int_{0}^{ \pi} \frac{dq}{\pi}  \bigg[ \lambda (q) \bigg]^t 
\langle n \vert R^{Abs}_q \rangle  \langle L_q^{Abs} \vert n_0 \rangle 
\label{Ptimagesspectral}
\end{eqnarray}
where 
\begin{eqnarray}
\langle n \vert R^{Abs}_q \rangle && \equiv \left( \frac{1-p}{p} \right)^{\frac{n}{2} } \sqrt{2} \sin[ q ( n+1) ]
\nonumber \\
\langle L_q^{Abs} \vert n_0 \rangle && \equiv  \left( \frac{1-p}{p} \right)^{-\frac{n_0}{2} }  \sqrt{2} \sin[ q ( n+1) ]
\label{eigenabs}
\end{eqnarray}
represent the right and left eigenvectors associated to   
the eigenvalue $\lambda(q)$ of Eq. \ref{lambdaq} in the presence of the absorbing boundary condition at $n=-1$.

Let us stress the difference:

(i) for the biased random walk on
 the whole line $n \in \{-\infty,...,+\infty\}$,
 the eigenvalue $\lambda (q) =\lambda(-q)$ is doubly degenerate, since the two functions $e^{ \pm i q n}$ associated to opposite Fourier momenta $(\pm q)$ are two independent valid solutions,
 and the spectral decomposition of Eq. \ref{Gspectral} involves an integral over $ q \in ]-\pi,+\pi[$.

(ii) for the biased random walk on
 the half-line $n \in ]0,+\infty[$ with an absorbing boundary condition at $n=-1$
the eigenvalue $\lambda (q) =\lambda(-q)$ is not degenerate anymore,
since one needs a linear combination of the two functions $e^{ \pm i q n}$ associated to opposite Fourier momenta $(\pm q)$ in order to build the sinus $\sin[ q ( n+1) ] $ that satisfies the boundary condition,
and the spectral decomposition of Eq. \ref{Ptimagesspectral} involves an integral over $ q \in ]0,+\pi[$.

%%%%%%%%%%%%%%%%%%%%%%%%%%%%%%%%%%%%%

\subsubsection{ Time-Laplace-transform ${\tilde P}^{Abs}_s(n \vert n_0) $ of the propagator $P^{Abs}_t(n \vert n_0)   $ }

The time-Laplace-transform of the first line of Eq. \ref{Ptimages} yields 
\begin{eqnarray}
{\tilde P}^{Abs}_s(n \vert n_0) && \equiv \sum_{t=0}^{+\infty} e^{-s t} P^{Abs}_t(n \vert n_0)
=  {\tilde P}^{\mathrm {Free}}_s(n - n_0) - \left(\frac{p}{1-p}\right)^{n_0+1} {\tilde P}^{\mathrm {Free}}_s(n +n_0+2)
 \label{PLaplaceimages}
\end{eqnarray}
in terms of $ {\tilde P}^{\mathrm {Free}}_s(m) $ of Eq. \ref{GLaplace},
so that its explicit expression reads in the two regions $n \geq n_0$ and $0 \leq n \leq n_0$  
\begin{eqnarray}
\text{for }  n \geq n_0 : \ \ {\tilde P}^{Abs}_s(n \vert n_0)   &&   
= \frac{r_+(s)+r_-(s)}{ r_+(s)-r_-(s) } [r_-(s)]^{n-n_0} - \left(\frac{p}{1-p}\right)^{n_0+1} \frac{r_+(s)+r_-(s)}{ r_+(s)-r_-(s) } [r_-(s)]^{n+n_0+2}
\nonumber \\
\text{for } 0 \leq n \leq n_0 : \ \ {\tilde P}^{Abs}_s(n \vert n_0)  
&&= \frac{r_+(s)+r_-(s)}{ r_+(s)-r_-(s) } \left[ r_+(s) \right]^{n-n_0} - \left(\frac{p}{1-p}\right)^{n_0+1} \frac{r_+(s)+r_-(s)}{ r_+(s)-r_-(s) } [r_-(s)]^{n+n_0+2}
\label{sPabs}
\end{eqnarray}

It is useful to use Eq. \ref{sumprod}
to replace $\left(\frac{p}{1-p}\right) = \frac{1}{r_+(s) r_-(s)} $ in order to 
write the solution of Eq. \ref{sPabs}
in terms of only the two roots $r_{\pm}(s)$
\begin{eqnarray}
\text{for }  n \geq n_0 : \ \ {\tilde P}^{Abs}_s(n \vert n_0)   &&   
= \frac{r_+(s)+r_-(s)}{ r_+(s)-r_-(s) } [r_-(s)]^{n+1} \bigg( [r_-(s)]^{-n_0-1}- [ r_+(s) ]^{-n_0-1} \bigg)
\nonumber \\
\text{for } 0 \leq n \leq n_0 : \ \ {\tilde P}^{Abs}_s(n \vert n_0)  &&
 = \frac{r_+(s)+r_-(s)}{ r_+(s)-r_-(s) } [r_+(s)]^{-n_0-1} \bigg( [r_+(s)]^{n+1}- [ r_-(s) ]^{n+1} \bigg)
 \nonumber \\
\label{sPabspm}
\end{eqnarray}

The time-Laplace-transform ${\tilde A}(s \vert n_0) $ of the 
absorption probability $A(t \vert n_0) = p P^{Abs}_{t-1}(0 \vert n_0) $ 
of Eq \ref{probaAbst} then reads using Eq. \ref{sumprod}
\begin{eqnarray}
{\tilde A}(s \vert n_0) && \equiv \sum_{t=1}^{+\infty} e^{-s t} A(t \vert n_0) 
= p \sum_{t=1}^{+\infty} e^{-s t} P^{Abs}_{t-1}(0 \vert n_0) 
 = p e^{-s} {\tilde P}^{Abs}_s(0 \vert n_0)
 =  p e^{-s} \frac{r_+(s)+r_-(s)}{ r_+(s)-r_-(s) } [r_+(s)]^{-n_0-1} (r_+-r_-)
\nonumber \\
&& = [ r_+(s)]^{-n_0-1}
\label{probaAbstLaplace}
\end{eqnarray}
while
the time-Laplace-transform ${\tilde S}(s \vert n_0) $ of the 
survival probability $S(T \vert n_0) $ of Eq. \ref{survivalproba}
is given by
\begin{eqnarray}
{\tilde S}(s \vert n_0) && \equiv \sum_{T=0}^{+\infty} e^{-s T} S(T \vert n_0) 
= \sum_{T=0}^{+\infty} e^{-s T} \left[ 1 - \sum_{t=1}^{T} A(t \vert n_0)  \right] 
= \frac{1- {\tilde A}(s \vert n_0) }{1-e^{-s}} 
\nonumber \\
&& = \frac{1-  [ r_+(s)]^{-n_0-1} }{1-e^{-s}} 
\label{survivalLaplace}
\end{eqnarray}

 The value of Eq. \ref{probaAbstLaplace}
 at $s=0$ gives the normalization of $A(t \vert n_0) $ 
complementary to the forever-survival probability $S(\infty \vert n_0) $
of Eq. \ref{probaAbstinfinitydef}
\begin{eqnarray}
\sum_{t=1}^{+\infty} A(t \vert n_0) \equiv 1- S(\infty \vert n_0) 
&& = {\tilde A}(s=0 \vert n_0)  = [ r_+(0)]^{-n_0-1}
=  \left( \frac{ 2 p  }{ 1 + \vert 1-2p \vert}  \right)^{n_0+1}
\nonumber \\
&& =  \begin{cases}
1 \ \ \ \ \ \ \ \ \ \ \ \ \ \ \ \ \ \ \ \ \ \ \ \ \ \ \text{ if } \frac{1}{2} \leq p \leq 1
 \\
 \left(  \frac{p }{ 1-p } \right)^{n_0+1} <1 \ \ \ \ \ \ \ \  \text{ if } 0 \leq p < \frac{1}{2} 
\end{cases}
\label{probaAbstinfinity}
\end{eqnarray}
with the following conclusion:

(i) in the region $ 0 \leq p < \frac{1}{2} $ where the bias is towards $(+\infty)$, the probability to survive forever 
is strictly positive 
\begin{eqnarray}
S(\infty \vert n_0) = 1 -  \left(  \frac{p }{ 1-p } \right)^{n_0+1} >0 \ \ \text{ if } 0 \leq p < \frac{1}{2} 
\label{foreversrurvivalfinite}
\end{eqnarray}

(ii) in the region $ \frac{1}{2} < p \leq 1$ where the bias is towards the absorbing condition at $n=-1$,
the forever survival probability vanishes $S(\infty \vert n_0) =0$, the probability distribution $A(t \vert n_0) $
of the absorption time $t$ is normalized with the asymptotic exponential decay of Eq. \ref{stirling}
\begin{eqnarray}
A(t +1\vert n_0) \opsimeq_{t \to + \infty}  
 \frac{  (n_0+1) p}{\sqrt \pi} \left(\frac{1-p}{p}\right)^{- \frac{n_0}{2}} [ 4 p (1-p)]^{\frac{t}{2}} \left( \frac{2}{t} \right)^{\frac{3}{2}} 
 \oppropto_{t \to + \infty}  \frac{e^{-t \frac{ \ln \left(\frac{1}{4p(1-p) } \right) }{2} }}{t^{3/2}}
\label{stirlingchaos}
\end{eqnarray}
and with the first moment 
\begin{eqnarray}
\sum_{t=1}^{+\infty} t A(t \vert n_0) =  - \frac{ d {\tilde A}(s \vert n_0) }{ds} \bigg \vert_{s=0}
= \frac{ n_0+1}{ 2p-1}  
\label{MFABs}
\end{eqnarray}
while the survival probability $S(T \vert n_0)$ is governed by the same exponential decay as 
the absorption probability of Eq. \ref{stirlingchaos}
\begin{eqnarray}
S(T \vert n_0) = \sum_{t=T}^{+\infty}  A(t +1\vert n_0) 
\opsimeq_{t \to + \infty}  
 \frac{  (n_0+1) p}{\sqrt \pi} \left(\frac{1-p}{p}\right)^{- \frac{n_0}{2}} \int_T^{+\infty} dt
  [ 4 p (1-p)]^{\frac{t}{2}} \left( \frac{2}{t} \right)^{\frac{3}{2}} 
 \oppropto_{t \to + \infty}  \frac{e^{-t \frac{ \ln \left(\frac{1}{4p(1-p) } \right) }{2} }}{t^{3/2}}
\label{survivalchaos}
\end{eqnarray}

(iii) at the critical point $p_c=\frac{1}{2}$ where the bias vanishes, the forever survival probability vanishes 
$S(\infty \vert n_0) =0$, the probability distribution $A(t \vert n_0) $
of the absorption time is normalized, but its asymptotic decay of Eq. \ref{stirling} contains only the power-law $t^{-3/2}$
\begin{eqnarray}
 A(t +1\vert n_0) \opsimeq_{t \to + \infty}  
\sqrt{\frac{2}{\pi} }  \frac{  (n_0+1) }{ t^{\frac{3}{2}} }
\label{stirlingcriti}
\end{eqnarray}
so that its first moment diverges in contrast to the finite value of Eq. \ref{MFABs} for $p>\frac{1}{2}$,
while the time-Laplace-transform of Eq. \ref{probaAbstLaplace}
displays the following singularity in $\sqrt{s}$ for $s \to 0$
\begin{eqnarray}
\text{ for } p_c=\frac{1}{2} : \ \  {\tilde A}(s \vert n_0) 
= \left( \frac{ e^{-s} }{ 1 + \sqrt{ 1 -  e^{-2s}}}  \right)^{n_0+1} = 1 - (n_0+1) \sqrt{2s} +O(s)
\label{probaAbstLaplacecriti}
\end{eqnarray}
The survival probability $S(T \vert n_0)$ obtained from Eq. \ref{stirlingcriti} decays only as $ T^{-\frac{1}{2}} $
\begin{eqnarray}
S(T \vert n_0) = \sum_{t=T}^{+\infty}  A(t +1\vert n_0) 
\opsimeq_{t \to + \infty}  \sqrt{\frac{2}{\pi} } (n_0+1) \int_{t=T}^{+\infty}  \frac{ dt  }{ t^{\frac{3}{2}} }
 \oppropto_{t \to + \infty}  2 \sqrt{\frac{2}{\pi} } \frac{   (n_0+1)}{ \sqrt{T} }
\label{survivalcriti}
\end{eqnarray}

%%%%%%%%%%%%%%%%%%%%%%%%%%%%%%%%%%%%%%%%%%%%%

\subsection{ Results when the initial condition $n_0$ is drawn with the probability distribution $2^{-1-n_0}$  }

For the discussions in the main text, it is also useful to write the various observables 
when the initial condition $n_0$ is drawn with the probability distribution $2^{-1-n_0}$.

\subsubsection{ Probability $P^{Abs}_t(n ) $ to be surviving at site $n$ at time $t$ }

When the initial condition $n_0$ is averaged over the probability distribution $2^{-1-n_0}$, 
 the probability $P^{Abs}_t(n ) $ to be surviving at time $t$ at site $n$
\begin{eqnarray}
P^{Abs}_t(n ) \equiv \sum_{n_0=0}^{+\infty} P^{Abs}_t(n\vert n_0) 2^{-1-n_0}
\label{Propagatorreset}
\end{eqnarray}
has for time-Laplace-transform using Eq. \ref{sPabspm}

\begin{eqnarray}
&& {\tilde P}^{Abs}_s(n )      \equiv \sum_{n_0=0}^{+\infty}{\tilde P}^{Abs}_s(n \vert n_0)  2^{-1-n_0}
\nonumber \\
&& = \frac{r_+(s)+r_-(s)}{ r_+(s)-r_-(s) }
\bigg[  \sum_{n_0=0}^{n} [r_-(s)]^{n+1} \bigg( [r_-(s)]^{-n_0-1}- [ r_+(s) ]^{-n_0-1} \bigg)2^{-1-n_0}
\nonumber \\
&& \ \ \ \ \ \ \ \ \ \ \ +  \sum_{n_0=n+1}^{+\infty}  [r_+(s)]^{-n_0-1} \bigg( [r_+(s)]^{n+1}- [ r_-(s) ]^{n+1} \bigg)2^{-1-n_0}
 \bigg] 
\nonumber \\
&&   =  
\frac{r_+(s)+r_-(s)}{ r_+(s)-r_-(s) } 
\bigg( [r_-(s)]^{n+1}  \sum_{n_0=0}^{n}  [2r_-(s)]^{-n_0-1} 
 +   [r_+(s)]^{n+1}\sum_{n_0=n+1}^{+\infty}  [2 r_+(s)]^{-n_0-1} 
-  [r_-(s)]^{n+1}  \sum_{n_0=0}^{+\infty}   [ 2 r_+(s) ]^{-n_0-1} 
\bigg)
  \nonumber \\
&&  = \frac{r_+(s)+r_-(s)}{ r_+(s)-r_-(s) } 
\bigg( \frac{2^{-1-n}  - [r_-(s)]^{n+1}}{1- 2 r_-(s)} 
 +   \frac{ 2^{-1-n}}{2 r_+(s)  -1 } 
 -   \frac{[r_-(s)]^{n+1}}{ 2  r_+(s)   -1} 
 \bigg)
 \nonumber \\
&&  = \frac{r_+(s)+r_-(s)}{ r_+(s)-r_-(s) } \left( \frac{1}{1- 2 r_-(s)} +   \frac{ 1}{2 r_+(s)  -1 } \right) 
\left( 2^{-1-n}  - [r_-(s)]^{n+1}\right)
 \nonumber \\
&&  = \frac{2 [ r_+(s)+r_-(s)]}{[1- 2 r_-(s)] [2 r_+(s)  -1 ] } 
\left( 2^{-1-n}  - [r_-(s)]^{n+1}\right)
\label{pabsreset}
\end{eqnarray}

In particular for $s=0$, one obtains that the total occupation time at position $n$ before absorption reads
\begin{eqnarray}
 {\tilde P}^{Abs}_{s=0}(n )     \equiv \sum_{t=0}^{+\infty} P^{Abs}_t(n )   
  =   \frac{2}{3p-2} \left[2^{-1-n}  - \left( \frac{1-p}{p}\right)^{n+1} \right]
\label{pabsresetzero}
\end{eqnarray}

%%%%%%%%%%%%%%%%%%%%%%

\subsubsection{ Probability distribution $ A(t )$ of the absorption time }

When the initial condition $n_0$ is averaged over the probability distribution $2^{-1-n_0}$, 
the absorption probability $A(t ) $ becomes
\begin{eqnarray}
A(t) \equiv \sum_{n_0=0}^{+\infty} A(t \vert n_0) 2^{-1-n_0}
\label{Abdistrireset}
\end{eqnarray}
with the corresponding survival probability of Eq. \ref{survivalproba}
\begin{eqnarray}
S(T) \equiv \sum_{n_0=0}^{+\infty}  S(T \vert n_0) 2^{-1-n_0} = 1 - \sum_{t=1}^{T} A(t ) 
\label{survivalprobareset}
\end{eqnarray}

The asymptotic behavior of Eq. \ref{Abdistrireset} using Eq. \ref{stirling} 
\begin{eqnarray}
A(t +1) && \opsimeq_{t \to + \infty}  
 \frac{   p}{\sqrt \pi}  [ 4 p (1-p)]^{\frac{t}{2}} \left( \frac{2}{t} \right)^{\frac{3}{2}}  \left(\frac{1-p}{p}\right)^{ \frac{1}{2}}
 \left[ \sum_{n_0=0}^{+\infty} (n_0+1)  \left( 2^{-1} \sqrt{ \frac{p}{1-p} } \right)^{ n_0+1} \right]
\nonumber \\
&& =   \frac{p 4 \sqrt{2}  [ 4 p (1-p)]^{\frac{t}{2}} }{ \sqrt \pi\left( 2- \sqrt{ \frac{p}{1-p} } \right)^2 t^{\frac{3}{2}}}
\label{stirlingreset}
\end{eqnarray}
 is still dominated by the exponential decay $[ 4 p (1-p)]^{\frac{t}{2}} $ for $p \ne \frac{1}{2}$,
and by the power-law $t^{\frac{3}{2}} $ for the critical case $p_c=\frac{1}{2}$.

The time-Laplace-transform of $A(t)$ reads using Eq. \ref{probaAbstLaplace}
\begin{eqnarray}
{\tilde A}(s ) && \equiv \sum_{t=1}^{+\infty} e^{ -s T}A(t ) = \sum_{n_0=0}^{+\infty} {\tilde A}(s \vert n_0) 2^{-1-n_0}
 =\sum_{n_0=0}^{+\infty}   [ 2 r_+(s) ]^{-n_0-1}
  = \frac{1}{ 2 r_+(s)-1}
\label{Absinireset}
\end{eqnarray}
with the corresponding time-Laplace-transform ${\tilde S}(s ) $ of the 
survival probability $S(T ) $ as in Eq. \ref{survivalLaplace}
\begin{eqnarray}
{\tilde S}(s ) && \equiv \sum_{T=0}^{+\infty} e^{-s T} S(T ) 
= \frac{1- {\tilde A}(s ) }{1-e^{-s}} 
\label{survivalLaplacereset}
\end{eqnarray}

The value of Eq. \ref{Absinireset} for $s=0$ gives the normalization of the probability $A(t)$ 
that is complementary to the survival probability $S(\infty)$
\begin{eqnarray}
1-S(\infty) = {\tilde A}(s=0 ) \equiv \sum_{t=1}^{+\infty} A(t ) =\frac{ 1}
{  \frac{ 1 + \vert 1-2p \vert  }{p }  -1 } 
 =  \begin{cases}
1 \ \ \ \ \ \ \ \ \text{ if } \frac{1}{2} \leq p \leq 1
 \\
   \frac{ p }{ 2 - 3 p }  \ \ \ \text{ if } 0 \leq p < \frac{1}{2} 
\end{cases}
\label{probaAbstnormareset}
\end{eqnarray}
that can be also understood as the averaged of Eq. \ref{probaAbstinfinity} over $n_0$ with the probability distribution $2^{-1-n_0}$ with the similar discussion:

(i) in the region $ 0 \leq p < \frac{1}{2} $ where the bias is towards $(+\infty)$, the forever-survival probability  
is strictly positive 
\begin{eqnarray}
S( \infty) \equiv 1- {\tilde A}(s=0 ) = 1-  \frac{ p }{ 2 - 3 p } =   \frac{ 2(1-2p)  }{ 2 - 3 p } >0
\label{Survivalinfinityreset}
\end{eqnarray}

(ii) in the region $ \frac{1}{2} < p \leq 1$ where the bias is towards the absorbing condition at $n=-1$,
the forever-survival probability vanishes $S( \infty) =0$, the probability distribution $A(t ) $
of the absorption time is normalized, with the asymptotic exponential decay of Eq. \ref{stirlingreset}
\begin{eqnarray}
A(t +1) && \opsimeq_{t \to + \infty}  
     \frac{p 4 \sqrt{2}  [ 4 p (1-p)]^{\frac{t}{2}} }{ \sqrt \pi\left( 2- \sqrt{ \frac{p}{1-p} } \right)^2 t^{\frac{3}{2}}}
      \oppropto_{t \to + \infty}  \frac{e^{-t \frac{ \ln \left(\frac{1}{4p(1-p) } \right) }{2} }}{t^{3/2}}
\label{stirlingresetchaos}
\end{eqnarray}
and with the first moment 
\begin{eqnarray}
\sum_{t=1}^{+\infty} t A(t ) =  - {\tilde A}'(s=0)
= \frac{ 2}{ 2p-1}  
\label{MFABsreset}
\end{eqnarray}
that can be also understood as the averaged of Eq. \ref{MFABs} over $n_0$ with the probability distribution $2^{-1-n_0}$,
while the survival probability $S(T )$ is governed by the same exponential decay as 
the absorption probability of Eq. \ref{stirlingresetchaos}
\begin{eqnarray}
S(T ) = \sum_{t=T}^{+\infty}  A(t +1) 
\opsimeq_{T \to + \infty}  
 \int_T^{+\infty} dt   \frac{p 4 \sqrt{2}  [ 4 p (1-p)]^{\frac{t}{2}} }{ \sqrt \pi\left( 2- \sqrt{ \frac{p}{1-p} } \right)^2 t^{\frac{3}{2}}}
 \opsimeq_{T \to + \infty}  
   \frac{p 8 \sqrt{2}   e^{-T \frac{ \ln \left(\frac{1}{4p(1-p) } \right) }{2} } }{ \sqrt \pi\left( 2- \sqrt{ \frac{p}{1-p} } \right)^2 \left[ \ln \left(\frac{1}{4p(1-p) } \right) \right] T^{3/2}}
\label{survivalresetchaos}
\end{eqnarray}

(iii) at the critical point $p_c=\frac{1}{2}$ where the bias vanishes, the forever survival probability vanishes $S(\infty) =0$, the probability distribution $A(t ) $
of the absorption time is normalized, but its asymptotic decay of Eq. \ref{stirlingreset} contains only the power-law $t^{-3/2}$
\begin{eqnarray}
 A(t +1) \opsimeq_{t \to + \infty}  
 \frac{  1 }{\sqrt{\pi} } \left( \frac{2}{t} \right)^{\frac{3}{2}}
\label{stirlingcritireset}
\end{eqnarray}
so that the first moment diverges,
while the time-Laplace-transform of Eq. \ref{Absinireset}
displays the following singularity in $\sqrt{s}$ for $s \to 0$
\begin{eqnarray}
   {\tilde A}(s )  
= \frac{1}{  2 e^s  \left[ 1 + \sqrt{ 1 -  e^{-2s}} \right]  -1}
= 1 - 2 \sqrt{2s} +O(s)
\label{Absiniresetcriti}
\end{eqnarray}
The survival probability $S(T )$ obtained from Eq. \ref{stirlingcritireset} decays only as $ T^{-\frac{1}{2}} $
\begin{eqnarray}
S(T ) = \sum_{t=T}^{+\infty}  A(t +1) 
\opsimeq_{t \to + \infty}   \frac{  1 }{\sqrt{\pi} } \left( \frac{2}{t} \right)^{\frac{3}{2}}
 \opsimeq_{t \to + \infty}  4 \sqrt{\frac{2}{\pi T} } 
\label{survivalresetcriti}
\end{eqnarray}

%%%%%%%%%%%%%%%%%%%%%%%%%%%%%%%%%%%%%%%%%%%

\section{ Spectral decomposition of the propagator $ \pi_t (n \vert  n_0) $ }

\label{app_spectral}

This Appendix is devoted to the spectral decomposition of Eq. \ref{spectralreset}
the propagator $ \pi_t (n \vert  n_0) $, where the steady state 
$ \pi_*(n)  $ existing only in the chaotic region $\frac{1}{2} <p \leq 1$ 
has already been discussed in detail in the main text.
So we will focus here on the continuum spectrum of eigenvalues $\lambda(q)= \sqrt{ 4 p (1-p) } \cos q$
already encountered in Eq. \ref{lambdaq} and in Eq. \ref{Ptimagesspectral}
in order to write the appropriate right and left eigenvectors that will replace 
the right and left eigenvectors of Eq. \ref{eigenabs} concerning the biased random walk with an absorbing boundary condition at $n=-1$.

%%%%%%%%%%%%%%%%%%%%%%%%%%%%%%

\subsection{ Right eigenvector $R_q(n)$ associated to the momentum $q \in ]0,\pi[$ 
and eigenvalue $\lambda (q)= \sqrt{ 4 p (1-p) } \cos q $ }

The eigenvalue equation for the right eigenvector $R_q(n)=\langle n \vert R_q \rangle $ associated to the eigenvalue $\lambda (q)= \sqrt{ 4 p (1-p) } \cos q $ of the Markov matrix $M$ of Eq. \ref{MarkovMatrixM}
\begin{eqnarray}
\lambda_q  R_q(  n) && =  \sum_{m=0}^{+\infty}  \ M(n \vert m)  R_q(m) 
\nonumber \\
&& =  p    R_q(n+1) 
 +  (1-p)    R_q(n-1)    \theta( n \geq 1 )
 + p 2^{- n -1}   R_q(0)  
\label{eigenright}
\end{eqnarray}
is a recurrence very similar to Eq. \ref{piLaplacerecinhomo} of Appendix \ref{app_laplace}.
Using the correspondence 
\begin{eqnarray}
e^s \to \lambda(q)=\sqrt{ 4 p (1-p) } \cos q = \sqrt{ p (1-p) } (e^{iq}+e^{-iq} )
\label{replacement}
\end{eqnarray}
with its consequences for the two roots $r_{\pm}(s) $ of Eq. \ref{2dsol} that become the two complex-conjugate roots $z_{\pm}(q) $
\begin{eqnarray}
r_{\pm}(s) \to   \sqrt{ \frac{1-p}{p} } e^{\pm iq} \equiv z_{\pm}(q)
\label{2dsolcomplex}
\end{eqnarray}
one obtains that the solution of Eq. \ref{eigenright} 
reads 
\begin{eqnarray}
 R_q(  n)  
&& =   \frac{1}{\sqrt 2} \left[    z_+^{n+1} - z_-^{n+1} 
+  \frac{ (  z_+  -    z_-)   }{    ( z_+ -1  ) (1-2 z_-)}  (2^{-n-1}    - z_-^{n+1}  ) \right]
\nonumber \\
&& 
=   \frac{1}{\sqrt 2} \left[    z_+^{n+1} -  \frac{(z_--1)(2 z_+-1)}{(z_+-1)(2 z_- -1 )}  z_-^{n+1} 
-  \frac{ (  z_+  -    z_-)   }{    ( z_+ -1  ) (2 z_- -1 )}  2^{-n-1}    \right]
 \label{Rqn}
\end{eqnarray}
whose translation in terms of the parameter $p$ and the momentum $q$ via Eq. \ref{2dsolcomplex}
is given in Eq. \ref{Rqnq}
of the main text.

%%%%%%%%%%%%%%%%%%%%%%%%%%%%%%%%%%%%%%%%%%%%%

\subsection{ Left eigenvector $L_q(  n_0)$ associated to the momentum $q \in ]0,\pi[$ 
and eigenvalue $\lambda (q)= \sqrt{ 4 p (1-p) } \cos q $}

The eigenvalue equation for the left eigenvectors $L_q(  n_0)=\langle L_q \vert n_0 \rangle $ associated to the eigenvalue $\lambda (q)= \sqrt{ 4 p (1-p) } \cos q $ of the Markov matrix $M$ of Eq. \ref{MarkovMatrixM}
 \begin{eqnarray}
\lambda_q  L_q(  n_0) &&  =  \sum_{m}^{+\infty}  L_q(  m) M(m \vert n_0) 
\nonumber \\
&&=  p  L_q(  n_0-1)  \theta( n_0 \geq 1 )
  + (1-p)   L_q(  n_0+1)  
   +\delta_{n_0,0} \left[ p \sum_{m=0}^{+\infty}  L_q(  m)   2^{-m-1} \right]
\label{eigenLeft}
\end{eqnarray}
corresponds for $n_0 \geq 1$ to an homogeneous recurrence with constant coefficients,
whose general solution can be written in terms of the two roots $z_{\pm}(q)$ introduced in Eq. \ref{2dsolcomplex}
as 
\begin{eqnarray}
L_q(  n_0)   =  \frac{1}{\sqrt 2} \left[   z_+^{-n_0-1} + \gamma(q) z_-^{-n_0-1} \right]
 \label{Lqnsol}
\end{eqnarray}
The constant $\gamma(q)$ is determined by Eq. \ref{eigenLeft}
at the boundary site $n_0=0$ 
 \begin{eqnarray}
0 && = - \lambda_q  L_q(  0) +  (1-p)   L_q(  1)  
   + p \sum_{m=0}^{+\infty}  L_q(  m)   2^{-m-1} 
   = -p   L_q(  -1)  
   + p \sum_{m=0}^{+\infty}  L_q(  m)   2^{-m-1} 
\nonumber \\
&&= - \frac{p}{\sqrt 2}    \left[1 + \gamma \right]
   + \frac{p}{\sqrt 2}  \sum_{m=0}^{+\infty} \left[z_+^{-m-1} + \gamma z_-^{-m-1} \right]    2^{-m-1} 
%   \nonumber \\ &&
=  \frac{p}{\sqrt 2}    \left[-1 - \gamma    +  \frac{1}{2 z_+-1} + \gamma  \frac{1}{2 z_- -1} \right]
   \nonumber \\
&&= \frac{p}{\sqrt 2}    \left[
     \frac{2(1-z_+)}{2 z_+-1} + \gamma  \frac{2(1-z_-)}{2 z_- -1}   \right]
\label{eigenLeft0}
\end{eqnarray}
leading to the value
 \begin{eqnarray}
\gamma=    -  \frac{(z_+-1)(2 z_- -1 )}{(z_--1)(2 z_+-1)} 
\label{dephasage}
\end{eqnarray}
that can be plugged into Eq. \ref{Lqnsol} to obtain the left eigenvector
\begin{eqnarray}
L_q(  n_0)   =    \frac{1}{\sqrt 2} \left[   z_+^{-n_0-1} -  \frac{(z_+-1)(2 z_- -1 )}{(z_--1)(2 z_+-1)}  z_-^{-n_0-1} \right]
 \label{Lqn}
\end{eqnarray}
whose translation in terms of the parameter $p$ and the momentum $q$ via Eq. \ref{2dsolcomplex}
is given in Eq. \ref{Lqnq}
of the main text.

%%%%%%%%%%%%%%%%%%%%%%%%%%%%%%%%%%%%%%%%%%%

\end{document}